\documentclass[twocolumn]{aastex63}
\usepackage{lineno}
\pdfoutput=1 
\usepackage{amsmath,amstext,amssymb}
\usepackage[T1]{fontenc}
\usepackage[figure,figure*]{hypcap}
\usepackage[caption=false]{subfig}
\graphicspath{{./}{figures/}}

\received{?}
\revised{?}
\accepted{?}
\submitjournal{AAS}

\shorttitle{TOI-1136}
\shortauthors{Dai et al.}

\begin{document}

\title{TOI-1136 is a Young, Coplanar, Aligned Planetary System in a Pristine Resonant Chain}

\author[0000-0002-8958-0683]{Fei Dai} 
\affiliation{Division of Geological and Planetary Sciences,
1200 E California Blvd, Pasadena, CA, 91125, USA}
\affiliation{Department of Astronomy, California Institute of Technology, Pasadena, CA 91125, USA}
\affiliation{NASA Sagan Fellow}

\author[0000-0003-1298-9699]{Kento Masuda}
\affiliation{Department of Earth and Space Science, Osaka University, Osaka 560-0043, Japan}

\author[0000-0001-7708-2364]{Corey Beard}
\affiliation{Department of Physics \& Astronomy, University of California Irvine, Irvine, CA 92697, USA}

\author[0000-0003-0149-9678]{Paul Robertson}
\affiliation{Department of Physics \& Astronomy, University of California Irvine, Irvine, CA 92697, USA}

\author[0000-0003-3868-3663]{Max Goldberg}
\affiliation{Department of Astronomy, California Institute of Technology, Pasadena, CA 91125, USA}

\author[0000-0002-7094-7908]{Konstantin Batygin}
\affiliation{Division of Geological and Planetary Sciences,
1200 E California Blvd, Pasadena, CA, 91125, USA}

\author{Luke Bouma}
\affiliation{Department of Astronomy, California Institute of Technology, Pasadena, CA 91125, USA}
\affiliation{51 Pegasi b Fellow}

\author[0000-0001-6513-1659]{Jack J. Lissauer}
\affiliation{Space Science \& Astrobiology Division, MS 245-3, NASA Ames Research Center
 Moffett Field, CA 94035, USA}

\author[0000-0001-7880-594X]{Emil Knudstrup}
\affiliation{Stellar Astrophysics Centre, Department of Physics and Astronomy, Aarhus University, Ny Munkegade 120, DK-8000 Aarhus C, Denmark}

\author[0000-0003-1762-8235]{Simon Albrecht}
\affiliation{Stellar Astrophysics Centre, Department of Physics and Astronomy, Aarhus University, Ny Munkegade 120, DK-8000 Aarhus C, Denmark}

\author[0000-0001-8638-0320]{Andrew W. Howard}
\affiliation{Department of Astronomy, California Institute of Technology, Pasadena, CA 91125, USA}

\author[0000-0002-5375-4725]{Heather A. Knutson}
\affiliation{Division of Geological and Planetary Sciences,
1200 E California Blvd, Pasadena, CA, 91125, USA}

\author[0000-0003-0967-2893]{Erik A. Petigura}
\affiliation{Department of Physics \& Astronomy, University of California Los Angeles, Los Angeles, CA 90095, USA}

\author[0000-0002-3725-3058]{Lauren M. Weiss}
\affiliation{Department of Physics and Astronomy, University of Notre Dame, Notre Dame, IN 46556, USA}

\author[0000-0002-0531-1073]{Howard Isaacson}
\affiliation{501 Campbell Hall, University of California at Berkeley, Berkeley, CA 94720, USA}
\affiliation{Centre for Astrophysics, University of Southern Queensland, Toowoomba, QLD, Australia}

\author[0000-0002-2607-138X]{Martti Holst Kristiansen}
\affiliation{Brorfelde Observatory, Observator Gyldenkernes Vej 7, DK-4340 Tølløse, Denmark}

\author[0000-0002-4047-4724]{Hugh Osborn}
\affiliation{NCCR/PlanetS, Centre for Space \& Habitability, University of Bern, Bern, Switzerland}
\affiliation{Department of Physics and Kavli Institute for Astrophysics and Space Research, Massachusetts Institute of Technology, Cambridge, MA 02139, USA}

\author[0000-0002-7846-6981]{Songhu Wang} 
\affiliation{Department of Astronomy, Indiana University, Bloomington, IN 47405, USA}

\author[0000-0002-0376-6365]{Xian-Yu Wang} 
\affiliation{National Astronomical Observatories, Chinese Academy of Sciences, Beijing 10010, China}
\affiliation{University of the Chinese Academy of Sciences, Beijing, 100049, China}

\author[0000-0003-0012-9093]{Aida Behmard}
\affiliation{Division of Geological and Planetary Sciences,
1200 E California Blvd, Pasadena, CA, 91125, USA}

\author[0000-0002-0371-1647]{Michael Greklek-McKeon}
\affiliation{Division of Geological and Planetary Sciences,
1200 E California Blvd, Pasadena, CA, 91125, USA}

\author[0000-0003-2527-1475]{Shreyas Vissapragada}
\affiliation{Division of Geological and Planetary Sciences,
1200 E California Blvd, Pasadena, CA, 91125, USA}

\author[0000-0002-7030-9519]{Natalie M. Batalha}
\affiliation{Department of Astronomy and Astrophysics, University of California, Santa Cruz, CA 95060, USA}

\author[0000-0002-4480-310X]{Casey L. Brinkman}
\affiliation{Institute for Astronomy, University of Hawai'i, 2680 Woodlawn Drive, Honolulu, HI 96822 USA}

\author[0000-0003-1125-2564]{Ashley Chontos}
\altaffiliation{Henry Norris Russell Fellow}
\affiliation{Department of Astrophysical Sciences, Princeton University, 4 Ivy Lane, Princeton, NJ, 08544, USA}
\affiliation{Institute for Astronomy, University of Hawai`i, 2680 Woodlawn Drive, Honolulu, HI 96822, USA}

\author{Ian Crossfield}
\affiliation{Department of Physics and Astronomy, University of Kansas, Lawrence, KS, USA}

\author{Courtney Dressing}
\affiliation{501 Campbell Hall, University of California at Berkeley, Berkeley, CA 94720, USA}

\author[0000-0002-3551-279X]{Tara Fetherolf}
\altaffiliation{UC Chancellor's Fellow}
\affiliation{Department of Earth and Planetary Sciences, University of California, Riverside, CA 92521, USA}

\author[0000-0003-3504-5316]{Benjamin Fulton}
\affiliation{NASA Exoplanet Science Institute/Caltech-IPAC, MC 314-6, 1200 E California Blvd, Pasadena, CA 91125, USA}

\author[0000-0002-0139-4756]{Michelle L. Hill}
\affiliation{Department of Earth and Planetary Sciences, University of California, Riverside, CA 92521, USA}

\author[0000-0001-8832-4488]{Daniel Huber}
\affiliation{Institute for Astronomy, University of Hawai`i, 2680 Woodlawn Drive, Honolulu, HI 96822, USA}

\author[0000-0002-7084-0529]{Stephen R. Kane}
\affiliation{Department of Earth and Planetary Sciences, University of California, Riverside, CA 92521, USA}

\author[0000-0001-8342-7736]{Jack Lubin}
\affiliation{Department of Physics \& Astronomy, University of California Irvine, Irvine, CA 92697, USA}

\author[0000-0003-2562-9043]{Mason MacDougall}
\affiliation{Astronomy Department, 475 Portola Plaza, University of California, Los Angeles, CA 90095, USA}

\author[0000-0002-7216-2135]{Andrew Mayo}
\affiliation{501 Campbell Hall, University of California at Berkeley, Berkeley, CA 94720, USA}

\author[0000-0003-4603-556X]{Teo Mo\v{c}nik}
\affiliation{Gemini Observatory/NSF's NOIRLab, 670 N. A'ohoku Place, Hilo, HI 96720, USA}

\author[0000-0001-8898-8284]{Joseph M. Akana Murphy}
\altaffiliation{NSF Graduate Research Fellow}
\affiliation{Department of Astronomy and Astrophysics, University of California, Santa Cruz, CA 95060, USA}

\author[0000-0003-3856-3143]{Ryan A. Rubenzahl}
\altaffiliation{NSF Graduate Research Fellow}
\affiliation{Department of Astronomy, California Institute of Technology, Pasadena, CA 91125, USA}

\author[0000-0003-3623-7280]{Nicholas Scarsdale}
\affiliation{Department of Astronomy and Astrophysics, University of California, Santa Cruz, CA 95060, USA}

\author{Dakotah Tyler}
\affiliation{Department of Physics \& Astronomy, University of California Los Angeles, Los Angeles, CA 90095, USA}

\author[0000-0002-4290-6826]{Judah Van Zandt}
\affil{Department of Physics \& Astronomy, University of California Los Angeles, Los Angeles, CA 90095, USA}

\author[0000-0001-7047-8681]{Alex S. Polanski}
\affil{Department of Physics and Astronomy, University of Kansas, Lawrence, KS 66045, USA}

\author[0000-0002-1637-2189]{Hans Martin Schwengeler}
\affiliation{Citizen scientist, c/o Zooniverse, Department of Physics, University of Oxford, Denys Wilkinson Building, Keble Road, Oxford, OX1 3RH, UK}

\author[0000-0002-0654-4442]{Ivan A. Terentev}
\affiliation{Citizen Scientist, Petrozavodsk, Russia}

\author{Paul Benni}
\affiliation{Acton Sky Portal private observatory, Acton, MA, USA}

\author[0000-0001-6637-5401]{Allyson Bieryla}
\affiliation{Center for Astrophysics \textbar \ Harvard \& Smithsonian, 60 Garden Street, Cambridge, MA 02138, USA}

\author[0000-0002-5741-3047]{David Ciardi}
\affiliation{NASA Exoplanet Science Institute/Caltech-IPAC, MC 314-6, 1200 E California Blvd, Pasadena, CA 91125, USA}

\author{Ben Falk}
\affiliation{Space Telescope Science Institute, 3700 San Martin Drive, Baltimore, MD, 21218, USA}

\author[0000-0001-9800-6248]{E. Furlan}
\affiliation{NASA Exoplanet Science Institute, Caltech/IPAC, Mail Code 100-22, 1200 E. California Blvd., Pasadena, CA 91125, USA}

\author{Eric Girardin}
\affiliation{Grand Pra Observatory, 1984 Les Haudères, Switzerland}

\author{Pere Guerra}
\affiliation{Observatori Astronòmic Albanyà, Camí de Bassegoda S/N, Albanyà 17733, Girona, Spain}

\author[0000-0002-2135-9018]{Katharine~M.~Hesse}
\affiliation{Department of Physics and Kavli Institute for Astrophysics and Space Research, Massachusetts Institute of Technology, Cambridge, MA 02139, USA}

\author[0000-0002-2532-2853]{Steve~B.~Howell}
\affil{NASA Ames Research Center, Moffett Field, CA 94035, USA}

\author[0000-0003-3742-1987]{J.~Lillo-Box}
\affil{Centro de Astrobiolog\'ia (CAB, CSIC-INTA), Depto. de Astrof\'isica, ESAC campus, 28692, Villanueva de la Ca\~nada (Madrid), Spain}

\author[0000-0003-0593-1560]{Elisabeth C. Matthews}
\affiliation{Observatoire de l'Universit\`e de Gen\`eve, Chemin Pegasi 51, 1290 Versoix, Switzerland}

\author[0000-0002-6778-7552]{Joseph D. Twicken}
\affiliation{SETI Institute, Mountain View, CA  94043, USA}
\affiliation{NASA Ames Research Center, Moffett Field, CA  94035, USA}

\author{Joel Villase{\~ n}or}
\affiliation{Department of Physics and Kavli Institute for Astrophysics and Space Research, Massachusetts Institute of Technology, Cambridge, MA 02139, USA}

\author[0000-0001-9911-7388]{David W. Latham}
\affiliation{Center for Astrophysics | Harvard \& Smithsonian, 60 Garden St, Cambridge, MA 02138, USA}

\author[0000-0002-4715-9460]{Jon M. Jenkins}
\affiliation{NASA Ames Research Center, Moffett Field, CA 94035, USA}

\author[0000-0003-2058-6662]{George R. Ricker}
\affiliation{Department of Physics and Kavli Institute for Astrophysics and Space Research, Massachusetts Institute of Technology, Cambridge, MA 02139, USA}

\author[0000-0002-6892-6948]{Sara Seager}
\affiliation{Department of Physics and Kavli Institute for Astrophysics and Space Research, Massachusetts Institute of Technology, Cambridge, MA
02139, USA}
\affiliation{Department of Earth, Atmospheric and Planetary Sciences, Massachusetts Institute of Technology, Cambridge, MA 02139, USA}
\affiliation{Department of Aeronautics and Astronautics, MIT, 77 Massachusetts Avenue, Cambridge, MA 02139, USA}

\author[0000-0001-6763-6562]{Roland Vanderspek}
\affiliation{Department of Physics and Kavli Institute for Astrophysics and Space Research, Massachusetts Institute of Technology, Cambridge, MA 02139, USA}

\author[0000-0002-4265-047X]{Joshua N. Winn}
\affiliation{Department of Astrophysical Sciences, Princeton University, 4 Ivy Lane, Princeton, NJ 08544, USA}



\begin{abstract}
\noindent Convergent disk migration has long been suspected to be responsible for forming planetary systems with a chain of mean-motion resonances (MMR). Dynamical evolution over time could disrupt the delicate resonant configuration. We present TOI-1136, a $700\pm 150$-Myr-old G star hosting at least 6 transiting planets between $\sim$2 and 5 $R_\oplus$. The orbital period ratios deviate from exact commensurability by only $10^{-4}$, smaller than the $\sim$\,$10^{-2}$ deviations seen in typical Kepler near-resonant systems. A transit-timing analysis measured the masses of the planets (3-8$M_\oplus$) and demonstrated that the planets
in TOI-1136 are in true resonances with librating resonant angles. Based on a Rossiter-McLaughlin measurement of planet d, the star's rotation appears to be aligned with the planetary orbital planes. The well-aligned planetary system and the lack of detected binary companion together suggests that TOI-1136's resonant chain formed in an isolated, quiescent disk with no stellar fly-by, disk warp or significant axial asymmetry. With period ratios near 3:2, 2:1, 3:2, 7:5, and 3:2, TOI-1136 is the first known resonant chain involving a second-order MMR (7:5) between two first-order MMR. The formation of the delicate 7:5 resonance places strong constraints on the system's migration history. Short-scale (starting from $\sim$0.1 AU) Type-I migration with an inner disk edge is most consistent with the formation of TOI-1136. A low disk surface density ($\Sigma_{\rm 1AU}\lesssim10^3$g~cm$^{-2}$; lower than the minimum-mass solar nebula) and the resultant slower migration rate likely facilitated the formation of the 7:5 second-order MMR. TOI-1136's deep resonance suggests that it has not undergone much resonant repulsion during its 700-Myr lifetime. One can rule out rapid tidal dissipation within a rocky planet b or obliquity tides within the largest planets d and f. TOI-1136 is a pristine example of the orbital architecture produced by convergent disk migration, and may be a precursor of the mature Kepler multi-planet systems.
\end{abstract}

\keywords{planets and satellites: composition; planets and satellites: formation; planets and satellites: interiors}

\section{Introduction}
Disk migration is predicted to be a common stage of planet formation: in most scenarios the net effect is migration towards the central star \citep{Goldreich1979,Ward,Lin1986,McNeil,Terquem_2007,Nelson2018}. A pair of planets may become locked into a mean-motion resonance (MMR) if the migration is slow (adiabatic) and convergent (outer planets catching up with the inner planet). This process can be extended to capture multiple planets in a chain of resonance \citep[see ][and references therein]{Kley_2012}. Different studies using adiabatic perturbation theory \citep{Henrard,Batygin2015_capture}, modified N-body integration \citep[e.g.][]{Lee2002,Terquem_2007} and hydrodynamic simulations \citep[e.g.,][]{Kley_2005,McNeil,Ogihara,Cresswell,Ataiee} all came to the same conclusion that convergent disk migration consistently generates compact, first-order resonant chains of planets. This process of resonant capture is considered to be so effective and robust that it is difficult to understand why only a few percent of {\it Kepler} multi-planet systems are near first-order MMR \citep{Fabrycky2014}. Upon closer examination, most of these systems still show 1--2\% positive deviation from perfect period commensurability. Transit-timing-variation (TTV) modeling \citep[e.g.,][]{Hadden2017} has shown that most of these systems are near-resonant (with circulating resonant angles) rather than being truly resonant (librating resonant angles).

Planetesimal scattering \citep{Chatterjee_2015}, tidal dissipation \citep{Lithwick_repulsion,Batygin_repulsion}, secular chaos \citep{Petrovich}, and orbital instability \citep{PuWU,Izidoro,Goldberg2022} are some of the possible mechanisms for breaking migration-induced resonances as planetary systems mature. Some of these processes may take as long as billions of years to manifest. One might expect, therefore, that when the Kepler multi-planet systems were younger, they were also closer to resonance or truly resonant. In this paper, we present a young system that is deep in resonance ( observed orbital period ratios are close to small integer ratios; relevant resonant angles are also librating). TOI-1136 has a resonant chain of at least 6 transiting planets, all of which display TTVs. The planets' orbital period ratios deviate from perfect integer period ratio by $10^{-4}$. With an age of only 700 Myr, TOI-1136 may still record a pristine orbital architecture produced by convergent disk migration, before subsequent dynamical evolution have had the chance to disrupt the resonance. We present in this paper a series of observations and dynamical modeling to characterize the system and explore how the system formed and dynamically evolved.

The paper is organized as follows. Section \ref{sec:stellar_para} characterizes the host star, establishes its youth and puts limits of the presence of a stellar companion. Section \ref{sec:obliquity} presents a Rossiter-McLaughlin measurement of planet d. Section \ref{sec:transits} and \ref{sec:ttv} contain our analyses of the transit signal and transit timing variations. Section \ref{sec:dynamical} describes a series of dynamical models to investigate the dynamical stability, resonant configuration, disk migration, and resonant repulsion of TOI-1136. Section \ref{sec:discussion} discusses the implications for the
formation and evolution of TOI-1136 in relation to other multi-planet systems. Section \ref{sec:summary} is a brief summary of the paper.

\section{Host Star Properties}\label{sec:stellar_para}
\subsection{Spectroscopic and Stellar Parameters}
We obtained three high-resolution, high-signal-to-noise-ratio (SNR), iodine-free spectra of TOI-1136 with the High Resolution Echelle Spectrometer on the 10m Keck I telescope \citep[Keck/HIRES,][]{HIRES}. We employed {\tt SpecMatch-Syn}\footnote{\url{https://github.com/petigura/specmatch-syn}} \citep[for details see][]{CKS1} to extract the spectroscopic parameters ($T_{\rm eff}$, log$~g$ and [Fe/H]) of the host star. The results are listed in Table \ref{tab:stellar_para}. The cross correlation function of our HIRES spectra ruled out a spectroscopic binary that contributes more than 1\% of the observed flux.

To derive the stellar parameters, including the mass and radius of the host star, we fitted the measured spectroscopic parameters with Gaia parallax information
\citep{Gaia} in the {\tt Isoclassify} package \citep{Huber}. Our procedure was similar to that presented in \citet{CKS7}. We summarize the stellar parameters in Table \ref{tab:stellar_para}. \citet{Tayar} showed that between different theoretical model grids, the systematic uncertainties from  {\tt Isoclassify} could potentially amount to $\sim2\%$ in $T_{\rm eff}$, $\sim4\%$ in $M_{\star}$ and $\sim5\%$ in $R_{\star}$. We caution the readers that these systematic uncertainties are not explicitly included in Table \ref{tab:stellar_para}.

\begin{deluxetable*}{lcc}
\tablecaption{Stellar Parameters of TOI-1136\label{tab:stellar_para}} 
\tablehead{
\colhead{Parameters} & \colhead{Value and 68.3\% Credible Interval} & \colhead{Reference}}
\startdata
TIC ID  & 142276270 & A\\
R.A.  & 12:48:44.38  & A\\
Dec.  & +64:51:18.99& A\\
V (mag) & 9.534 $\pm$0.003& A\\
K (mag) & 8.034 $\pm$0.021& A\\
Distance (pc) & 84.5362$\pm$0.158& A\\
Effective Temperature $T_{\text{eff}} ~$(K) & $5770\pm50$ & B \\
Surface Gravity $\log~g~(\text{dex})$ &$4.47 \pm 0.04$& B \\
Iron Abundance $[\text{Fe/H}]~(\text{dex})$ &$0.07 \pm 0.06$& B \\
Rotational Broadening $v~\text{sin}~i_\star$ ~(km~s$^{-1}$) &$6.7\pm0.6    $& B \\
Stellar Radius $R_{\star} ~(~R_{\odot})$ &$0.968\pm0.036$& B \\
Stellar Mass $M_{\star} ~(M_{\odot})$ &$1.022\pm0.027$& B \\
Stellar Density $\rho_\star$ ($\rho_\odot$) &$1.11\pm0.12$&B \\
Limb Darkening q$_1$ & $0.38\pm0.16$& B\\
Limb Darkening q$_2$ & $0.24\pm0.11$& B\\
Activity Indicator $S_{\rm HK}$ &$0.32\pm0.03$&B\\
Activity Indicator log$R^\prime_{\rm HK}$ &$-4.49\pm0.05$&B\\
Age (Myr) from Gyrochronology, Activity Indicator, and Lithium &$700\pm150$&B\\
\enddata
\tablecomments{A:TICv8 \citep{Stassun}; B: this work.}
\end{deluxetable*}

\subsection{Rotation Period}

\begin{figure}
\includegraphics[width = 1.\columnwidth]{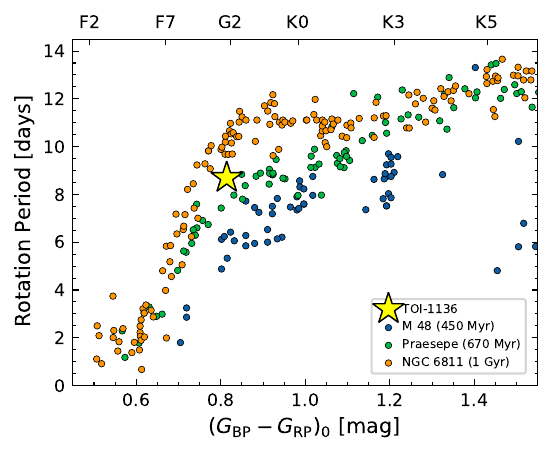}
\caption{The rotation period and de-reddened {\it Gaia} $G_{\rm BP}-G_{\rm RP}$ color of TOI-1136 (yellow star) and stars within selected young clusters. Based on this `gyrochronal' comparison, TOI-1136 is likely younger than NGC\,6811\citep[1 Gyr old,][]{curtis_2019}, and older than M48 \citep[450 Myr old,][]{barnes_2015}. The current rotation period of TOI-1136 ($8.7\pm0.1$ day) suggests an age similar or slightly older than Praesepe \citep[670 Myr old,][]{douglas_2017}.}
\label{fig: prot_color}
\end{figure}

We measured the rotation period of TOI-1136 from the rotational modulation seen in the {\it TESS} light curve. With a Lomb-Scargle periodogram \citep{Lomb1976,Scargle1982}, we measured a period of $P_{\rm rot}$ = $8.7\pm0.1$ days for the strongest peak in the periodogram. The corresponding flux variation has an amplitude of about 1\% (see Fig.~\ref{fig:tess_raw_light_curve1}).

We estimated the age of the system using gyrochronology. Given a $8.7\pm0.1$ day rotation period for a star like TOI-1136, the gyrochronal relation from \citet{Schlaufman2010} yields an age of $610\pm15$ Myr. Alternatively, if one follows \citet{Mamajek}, the estimated age is $700\pm20$ Myr. To leverage the latest empirical results, we put TOI-1136 on a rotation versus de-reddened color diagram to compare against young clusters with precise rotation period measurements (Fig.~\ref{fig: prot_color}).  Given $G_{\rm BP}-G_{\rm RP}=0.81$ and ignoring reddening due to the $\sim$85\,pc distance, TOI-1136 rotates at roughly the same rate as stars with comparable color in Praesepe \citep[670 Myr old,][]{douglas_2017}.
It rotates slower than any comparable stars in M48 \citep[450 Myr old,][]{barnes_2015}, and faster than any comparable stars in NGC\,6811 \citep[1 Gyr old,][]{curtis_2019}. Given TOI-1136's overlap with the stars in the Praesepe
cluster, we conclude that the age of TOI-1136 is $\approx$700 Myr.
This estimate is tied to Praesepe's age, which could be as high as 800\,Myr \citep{brandt_age_2015}.

\subsection{Lithium Absorption}
We modeled the lithium absorption in our Keck/HIRES spectra of TOI-1136 to corroborate the youth of system. We modeled the \ion{Li}{1} doublet at 6708 \AA~as well as the nearby \ion{Fe}{1} line simultaneously. Following the procedure of \citet{Bouma_li}, we estimated an equivalent width (EW) of 67.9$\pm$1.0 m\AA. This Li EW is again consistent with the Praesepe cluster \citep[see Fig  7 of ][]{Bouma_li} and is higher than that of most field stars \citep[see also Fig  5 of ][]{Berger_li}.

\subsection{Ca HK Emission \& Adopted Age}
Chromospheric emission lines can provide further constraint on the youth of TOI-1136. We analyzed the Ca II H\&K lines in our HIRES spectra and extracted the $S_{\rm HK}$ and $\log R^\prime_{\rm HK}$ values using the method of \citet{Isaacson}. TOI-1136 has enhanced stellar activity compared to field stars: we obtained a mean $S_{\rm HK} = 0.32\pm0.03$ and $\log R^\prime_{\rm HK}$ = -4.49$\pm$0.05 \citep[field stars of similar spectral type typically have $\log R^\prime_{\rm HK}=-5.0$][]{Isaacson}.  We converted the $\log R^\prime_{\rm HK}$ to an estimate of the age of the host star. We followed the relation linking $B-V$ color, $\log R^\prime_{\rm HK}$, and age calibrated by \citet{Mamajek}. The age of TOI-1136 was estimated to be 570$\pm$ 200 Myr, consistent with the age from gyrochronology and Li absorption. We combined the various age indicators by taking a weighted average, and we enlarged the formal uncertainty to reflect the systematic uncertainties in the different methods to arrive at an age for TOI-1136 of $700\pm150$ Myr.

\subsection{Cluster Membership}
Given its youth, TOI-1136 may be part of a young comoving group. We checked the proper motion of TOI-1136 for comoving groups against {\tt Banyan-$\Sigma$} \citep{Gagne} as well as the more recent compilation of open clusters and moving groups by \citet{Bouma_2022}. No match was found. We also used the {\tt Python} package {\tt COMOVE} \citep{Tofflemire} to search for comoving stars. We limited our search to a radius of 25 pc in spatial separation. {\tt COMOVE} returned 11 stars with tangential velocity difference $<2$ km/s within this search volume.  The closest had a 3-D separation of about 17 pc. These separations could not establish a firm kinematic connection between these stars and TOI-1136.

\subsection{High Resolution Imaging}
\begin{figure}
\vspace{-.5in}
\hspace{-.5in}
\includegraphics[width = 1.2\columnwidth]{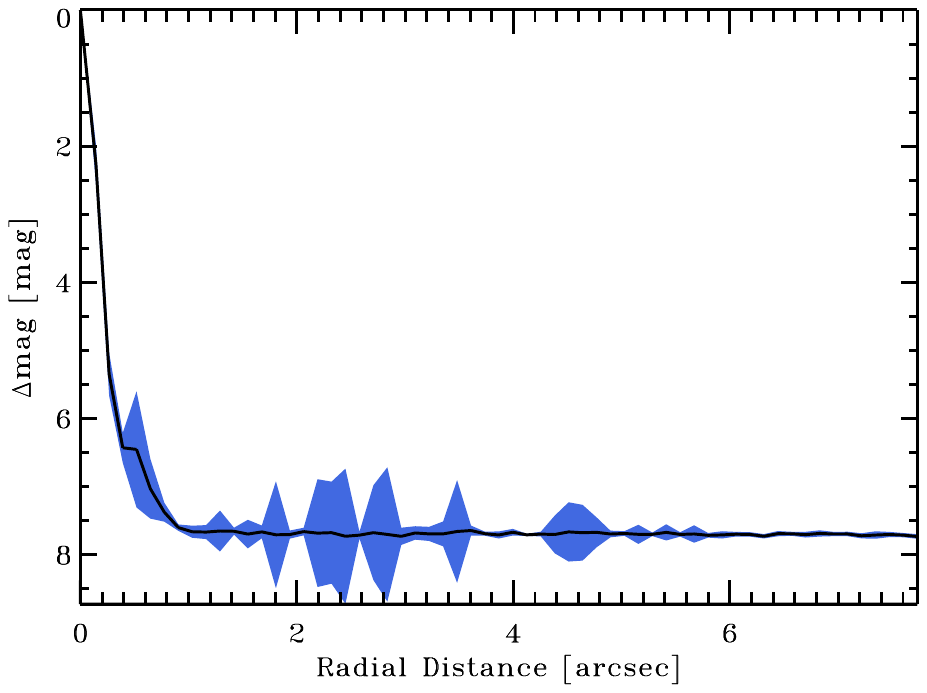}
\vspace{-2.5in}
\caption{The contrast curve as function of radial separation for TOI-1136 using the AO imaging from Gemini/NIRI in K-band. No stellar companion was identified.}
\label{fig:AO}
\end{figure}

To rule out nearby stellar companion, we performed a series of high resolution imaging on TOI-1136 (see Appendix). We highlight here the Adaptive Optics (AO) imaging observation on Gemini/NIRI \citep{Hodapp} on UT Dec 06 2019. We obtained 9 frames, each with exposure time 1.8 sec in the Br$\gamma$-band. We dithered the frames by 2'' in a 2-D grid. The data were reduced with a custom IDL routine that removes bad pixels, subtracts sky background, flattens the field, and co-adds the frames. No stellar companion was seen anywhere in the combined image (total FoV $\sim$26''$\times$26''). We also performed an injection/recovery test to quantify the sensitivity of the AO observation. The resultant sensitivity curve is shown in Fig. \ref{fig:AO}. We can rule out companions with $\Delta$mag of 6.4 at separations larger than 0.5''.

\subsection{A Single Star}

TOI-1136 has no reported visual or comoving companion on SIMBAD, VIZIER or Gaia DR3 \citep{Gaia_dr3_companion}. Gaia DR3 astrometry provides additional information on the possibility of inner companions that may have gone undetected by either Gaia or the high resolution imaging. The Gaia Renormalised Unit Weight Error (RUWE) is a metric, similar to a reduced chi-square, where values that are $\lesssim 1.4$  indicate that the Gaia astrometric solution is consistent with the star being single whereas RUWE values $\gtrsim 1.4$ may indicate an astrometric excess noise, possibly caused the presence of an unseen companion \citep[e.g., ][]{ziegler2020}.  TOI~1136 has a Gaia DR3 RUWE value of 0.99 indicating that the astrometric fit is consistent with a single-star model. The lack of a spectroscopic (spectra), blended (AO), visual (SIMBAD), and comoving (Gaia) companion indicate that TOI-1136 is likely a single star.

\section{Rossiter-McLaughlin Observation}\label{sec:obliquity}

TOI-1136 is amenable to a Rossiter-McLaughlin (RM) measurement  given its large rotational broadening, bright V-band magnitude, and relatively long transit duration. Moreover, it provides a rare chance to obtain a stellar obliquity measurement for a young planetary system with a resonant chain of planets. We observed a total of 52 spectra of TOI-1136 with the High Resolution Echelle Spectrometer on the 10m Keck I telescope \citep[Keck/HIRES][]{HIRES} on the night of UTC 2022 March 11 during a transit of TOI-1136 d as part of the Tess Keck Survey \citep[TKS, see][]{Chontos}. We obtained the spectra with the iodine cell in the light path. The dense and well-measured molecular lines serve to anchor the wavelength solution and the model of the line spread function. Each exposure lasted about 500\,sec and reached a median signal-to-noise ratio (SNR) of 200 per reduced pixel near 5500 Å. We had previously obtained a series of iodine-free spectra which were used to create a high-SNR template stellar spectrum for radial velocity extraction. The radial velocities were extracted using our standard HIRES forward-modeling pipeline \citep{Howard}. The extracted RVs and uncertainties are reported in Table \ref{tab:rv}.

\begin{figure*}
\center
\includegraphics[width = 1.5\columnwidth]{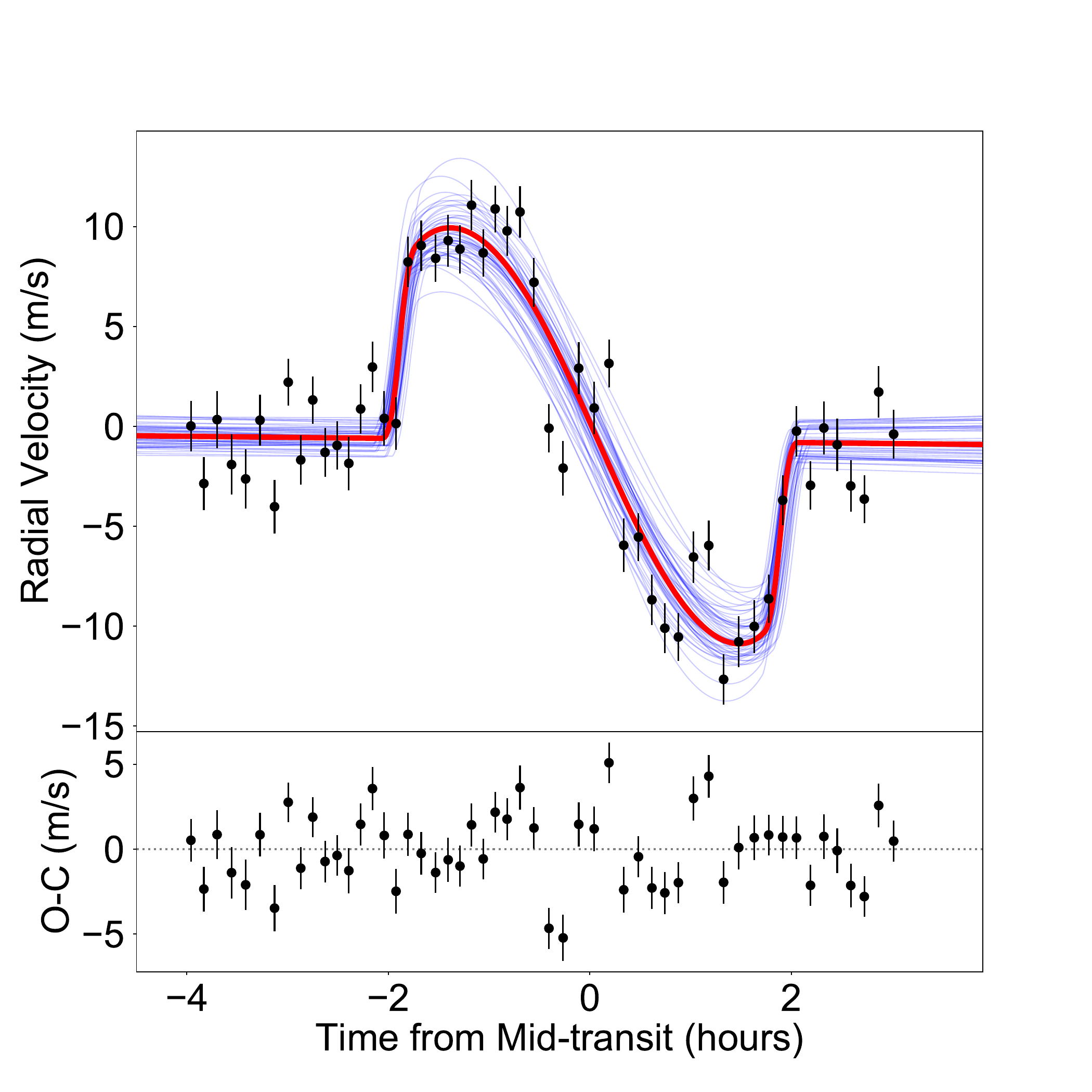}
\caption{The measured Rossiter-McLaughlin effect of TOI-1136 d suggests a well-aligned orbit with a sky-projected obliquity of $\lambda = 5\pm5^\circ$. Taking into account the stellar rotation period, the stellar radius and $v\sin i_\star$ measurements, the stellar obliquity of TOI-1136 d is consistent with being 0$^\circ$ with a 95\% upper limit of 28$^\circ$. The black points are our HIRES measurements. The red curve is the best-fit model; blue curves are random posterior draws. The mid-transit times from the RM measurement confirmed and followed the trend of the TTV seen in the {\it TESS} data (Fig.~\ref{fig:ttv}).}
\label{fig:rm}
\end{figure*}

We used our best-fit transit model from the {\it TESS} light curves (Section \ref{sec:transits}) to assist in the modeling of the RM effect. Specifically, we modeled the phase-folded, transit-timing-variation-adjusted {\it TESS} transits of planet d simultaneously with the RM effect. The model for the RM effect included the time of conjunction as a free parameter to account for the large transit-timing-variations (TTV). Reassuringly, the best-fit mid-transit time of the RM measurement confirmed the TTV of planet d and followed the trend that we expected from the {\it TESS} data (Fig.~\ref{fig:ttv}). Our RM model follows the prescription of \citet{Hirano} closely. In addition to the usual transit parameters modeled in Section \ref{sec:transits}, the RM model also requires the following parameters: the sky-projected obliquity $\lambda$, the projected rotational velocity $v\sin i_\star$, a linear function of time to describe the local radial velocity (RV) trend with an offset $\gamma$ and the local gradient $\dot{\gamma}$. An RV jitter term was also included to subsume any additional astrophysical or instrumental noise. No clear sign of a red noise component was seen in the RM residuals (Fig.~\ref{fig:rm}); we therefore adopted a simple $\chi^2$ likelihood function with a penalty term for the jitter parameter \citep[e.g.,][]{Howard78}.  We found the best-fit model using the {\tt Levenberg-Marquardt} method implemented in {\tt Python} package {\tt lmfit} \citep{LM}.

To sample the posterior distribution, we used the Markov Chain Monte Carlo (MCMC) technique
implemented in {\tt emcee} \citep{emcee}. We launched 128 walkers near the best-fit model, and ran them for 10000 links. We used the Gelman-Rubin convergence statistic \citep{BB13945229} to assess convergence of our MCMC process. The statistic was below 1.01 for each parameter by the end of the process, indicating good convergence. The results are summarized in Table \ref{tab:planet_para}. In short, we found that TOI-1136 d has a sky-projected obliquity $\lambda$ of $5\pm5^\circ$, consistent with zero. Moreover, the RM modeling provided a consistent but tighter constraint on the rotational broadening $v\sin i_\star$=6.7 $\pm 0.6$ km/s compared to the spectroscopic value ($v\sin i_\star =5.3 \pm 1.3$ km/s). Combining the $v\sin i_\star$, the stellar radius, and stellar rotation period from {\it TESS}, we placed a constraint on the stellar inclination $\sin i_\star$ \citep{Masuda_vsini}. Following the procedure outlined by \citet{Albrecht2021}, we found that the stellar obliquity $\Psi$ is consistent with being zero, with an upper limit of 28$^\circ$ at a 95\% credible level. We also performed an independent RM measurement of TOI-1136 d on HARPS-N that yielded consistent result. The details are outlined in the Appendix.

\section{TESS Observations}\label{sec:transits}
TOI-1136  (TIC 142276270) was observed by the {\it TESS} mission \citep{Ricker} in Sectors 14, 15, 21, 22, 41, and 48 from UT Jul 18 2019 to Feb 25 2022. Our analysis was based on the 2-min cadence light curve reduced by the TESS Science Processing Operations Center \citep[SPOC][]{jenkinsSPOC2016} available on the Mikulski Archive for Space
Telescopes website\footnote{\url{https://archive.stsci.edu}}. It can be accessed via \dataset[DOI]{doi:10.17909/t9-nmc8-f686}. We experimented with both the Simple Aperture Photometry \citep[SAP][]{Twicken2010,Morris2020} and the Presearch Data Conditioning Simple Aperture Photometry \citep[PDCSAP][]{Stumpe2012,Stumpe2014,Smith2012} versions of the light curves. We chose to present the results based on the SAP light curve in this paper. SAP light curve preserves the stellar variability much better, while both versions produced nearly identical transit fits. We minimized the influence of anomalous data by excluding cadences with non-zero Quality flags.

\subsection{Transit Modeling}

The TESS team reported four transiting planet candidates \citep{Guerrero} with orbital periods of 6.3 (TOI 1136.02), 12.5 (TOI 1136.01), 18.8 (TOI 1136.04), and 26.3 (TOI 1136.03) days, based on Threshold Crossing Events produced in the SPOC transit search \citep[e.g.][]{Jenkins2020J}. 
The ExoFOP website\footnote{\url{https://exofop.ipac.caltech.edu}} reported two additional planets on 4.2 and 39.5-day orbits identified by the community ( ExoFOP website). We confirmed the detection of these candidates with an independent Box-Least-Square search \citep[BLS,][]{Kovac2002} previously used in \citet{Dai_1444}.

We realized that TOI-1136 may display large transit timing variations (TTV) given how close the orbital periods are to resonance (see Section \ref{sec:commensurability}). We employed the {\tt Python} package {\tt Batman} \citep{Kreidberg2015} to model the transit light curves. The precise stellar density derived in Section \ref{sec:stellar_para} served as a prior in our transit modeling. A precise stellar density prior assists transit modeling by mitigating the degeneracy in semi-major axis, impact parameter, and orbital eccentricity \citep{Seager}. We adopted a quadratic limb darkening profile in the reparameterization of $q_1$ and $q_2$ by \citet{Kipping} for efficient sampling. We imposed a Gaussian prior (width = 0.3) on the limb darkening coefficients centered on the theoretical values from {\tt EXOFAST} \citep{Eastman2013}. The mean stellar density and the limb darkening coefficients are the three global parameters shared by all planets in TOI-1136. Each planet has its usual transits parameters: the orbital period $P_{\text{orb}}$, the time of conjunction $T_{\text{c}}$, the planet-to-star radius ratio $R_{\text{p}}/R_\star$, the scaled orbital distance $a/R_\star$, the transit impact parameter $b$, the orbital eccentricity $e$, and the argument of pericenter $\omega$. 

 The first step in our transit modeling was to remove any stellar variability and instrumental flux variation by fitting a cubic spline of length 0.5 day to the {\it TESS} light curve. Before fitting the spline, we removed any data points within 2 times the transit duration $T_{\rm 14}$ around each transit (and TTVs were accounted for in subsequent iterations of this process). The original light curve (with transits) was then divided by the spline fit. Fig.~\ref{fig:tess_raw_light_curve1} shows the original {\it TESS} light curve, the spline fit, and the detrended light curve. Visual inspection confirmed that the detrending procedure was
 successful, with no obvious distortions of the transit light curve.

The next step was to fit the transits of each planet assuming a constant orbital period. We obtained the best-fit model with the {\tt Levenberg-Marquardt} method implemented in {\tt Python} package {\tt lmfit} \citep{LM}. The best-fit model served as a template when we fitted for the mid-transit time of each individual transit. During the fit for each transit, the only free parameters were the mid-transit time and three parameters of a quadratic function of time that accounts for any residual out-of-transit flux variations. In TOI-1136, there are often cases where transits of different planets partially overlap with each other. In those cases, we fitted the involved planets simultaneously. The loss of light was assumed
to be the sum of the losses due to each planet, without
accounting for possible planet-planet eclipses \citep[e.g.,][]{Hirano_planet_planet}. We delay a thorough investigation of possible planet-planet eclipses to a future work (Beard et al., in preparation) that employs a full photodynamical model \citep[e.g.,][]{Carter,Mills2017}.

\begin{figure*}
\begin{center}
\vspace{-0.25in}
\subfloat[planet b]{%
  \includegraphics[width=0.8\columnwidth]{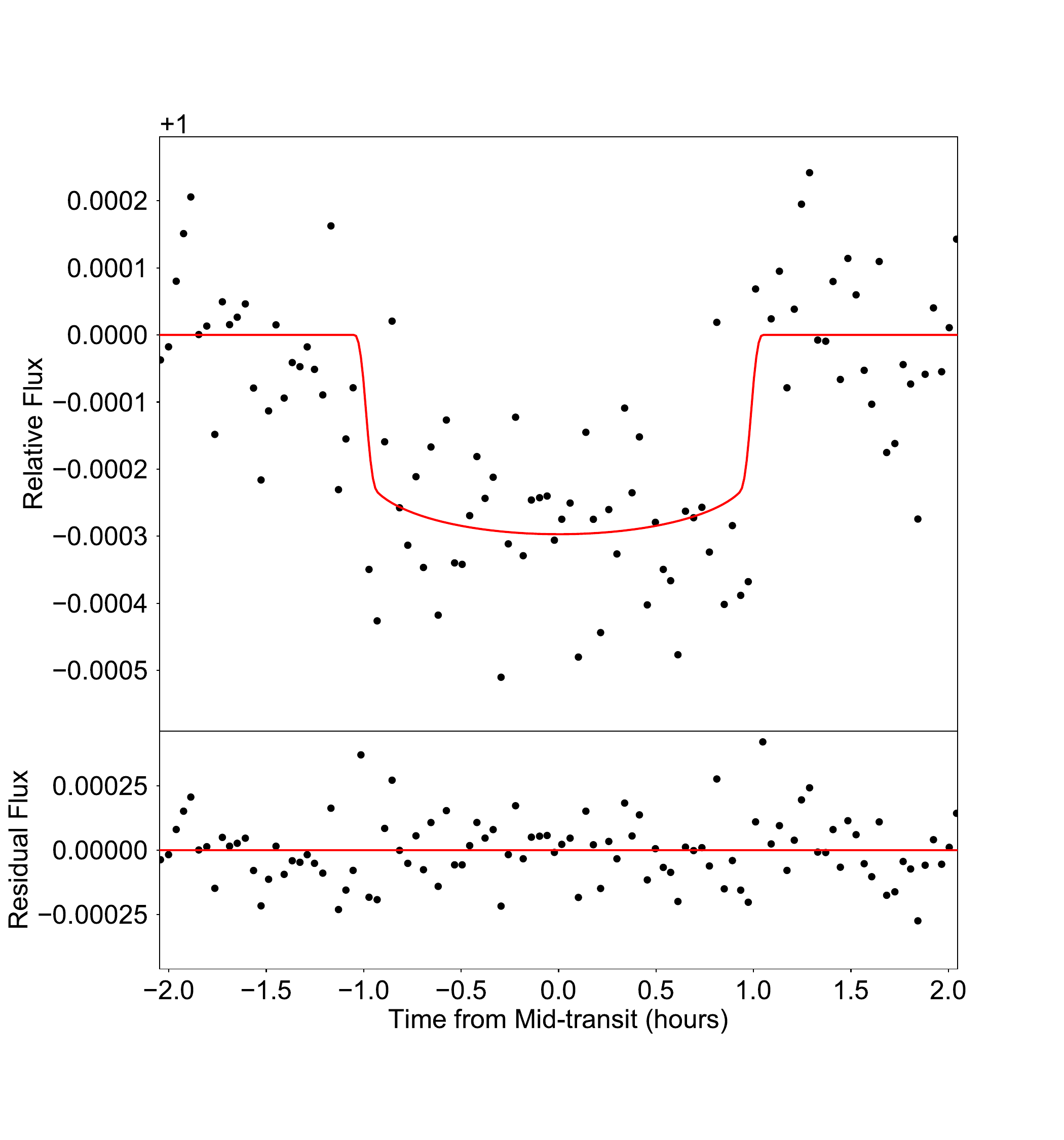}%
}\qquad
\subfloat[planet c]{%
  \includegraphics[width=0.8\columnwidth]{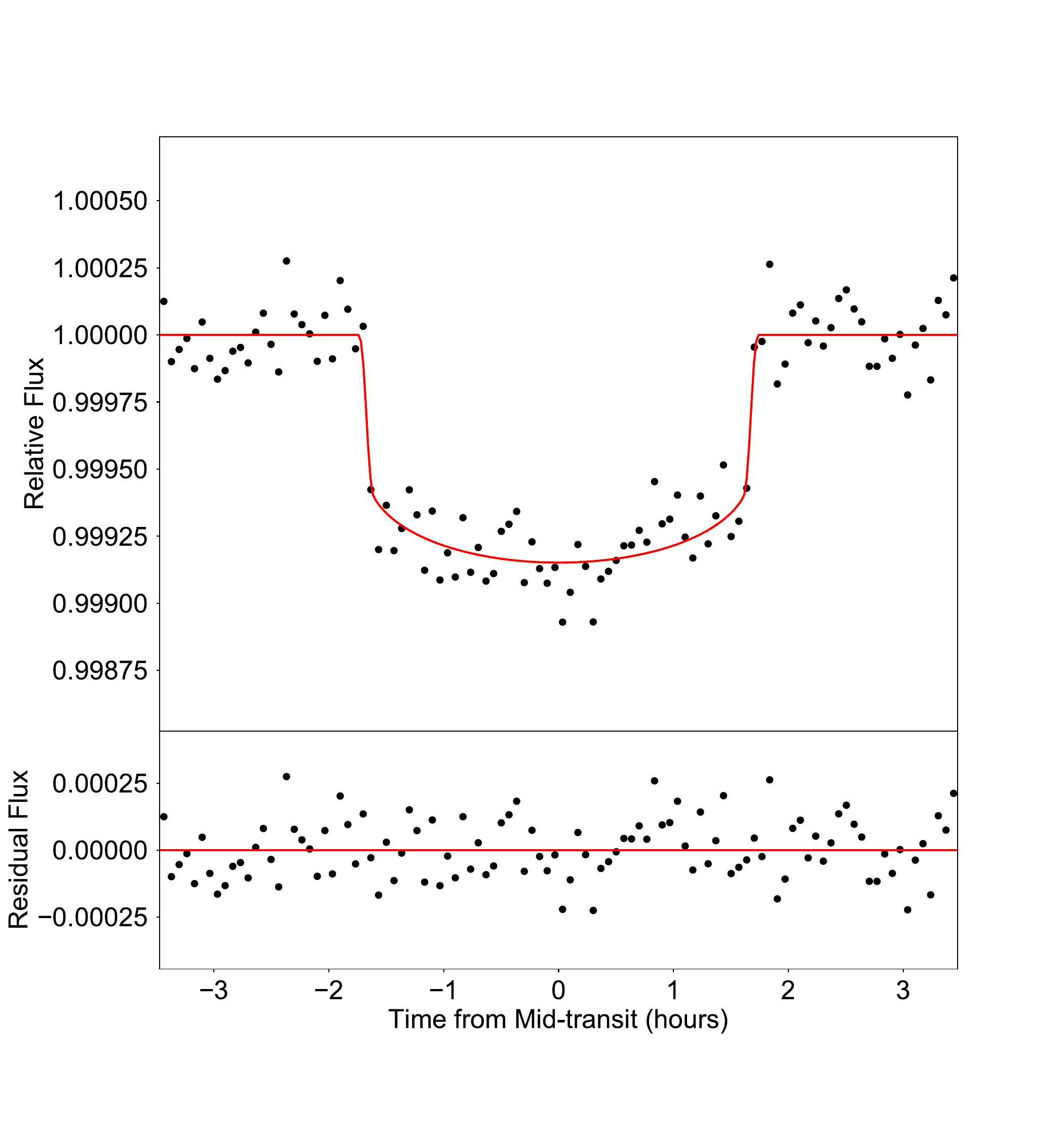}%
}\qquad
\subfloat[planet d]{%
  \includegraphics[width=0.8\columnwidth]{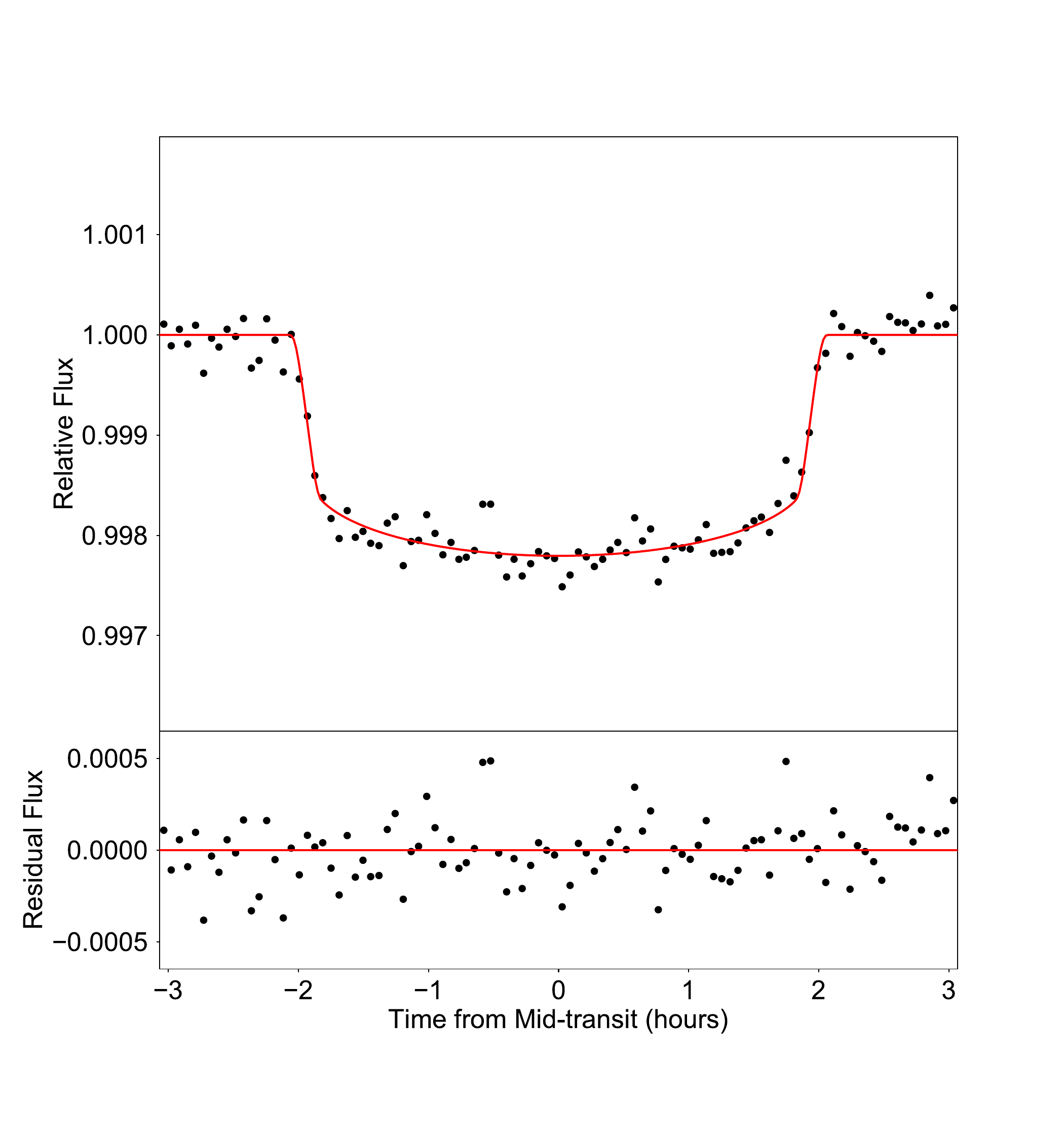}%
}\qquad
\subfloat[planet e]{%
  \includegraphics[width=0.8\columnwidth]{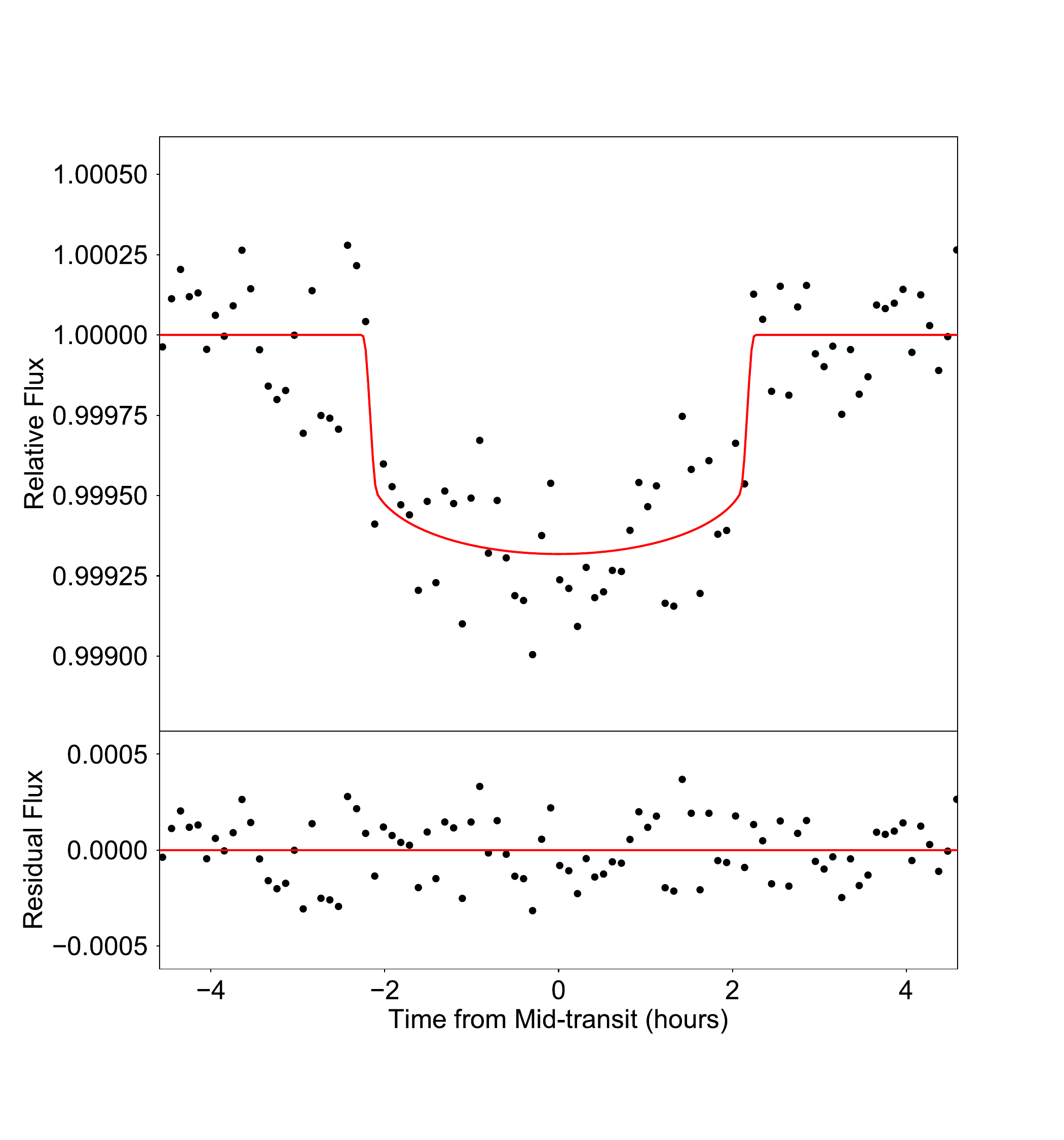}%
}\qquad
\subfloat[planet f]{%
  \includegraphics[width=0.8\columnwidth]{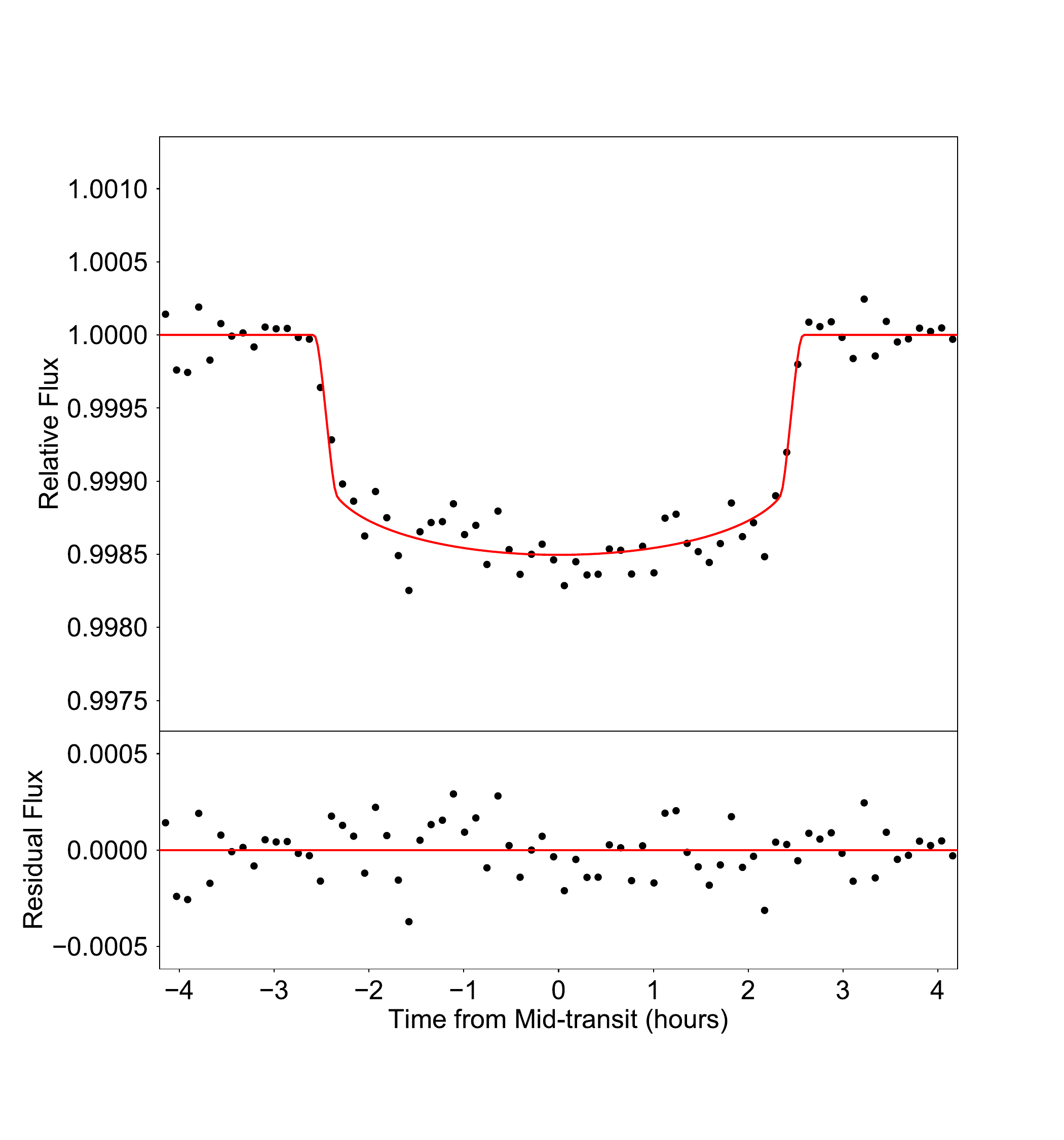}%
}\qquad
\subfloat[planet g]{%
  \includegraphics[width=0.8\columnwidth]{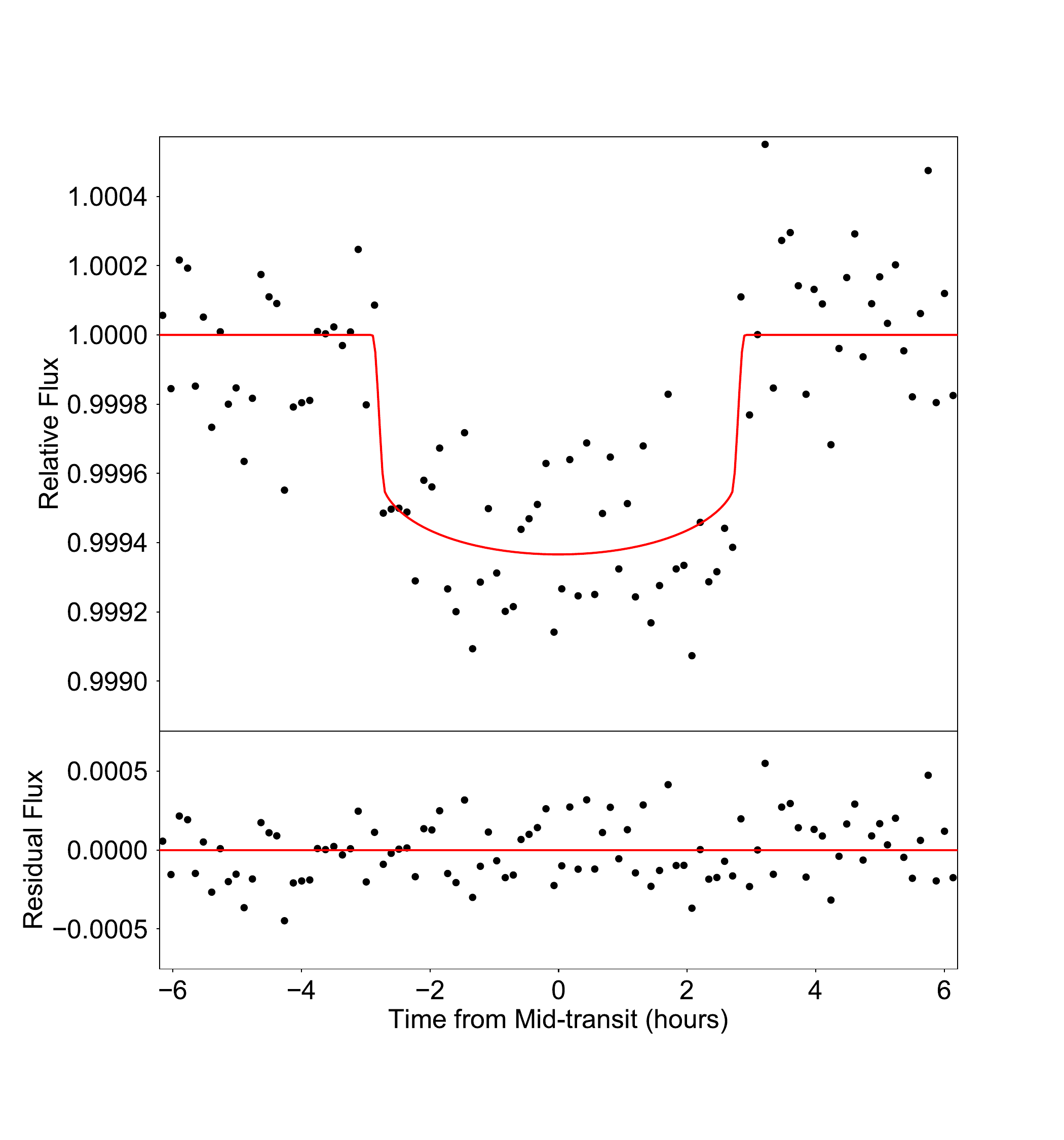}%
}\qquad
\caption{The {\it TESS} light curve phase-folded and binned after removing the measured TTV in TOI-1136. The red curves are our best fit transit models. Simultaneous transits (where two planets transit the host star) were removed before making this plot.}
\label{fig:transit}
\end{center}
\end{figure*}

\begin{figure*}
\begin{center}
\includegraphics[width = 1.\columnwidth]{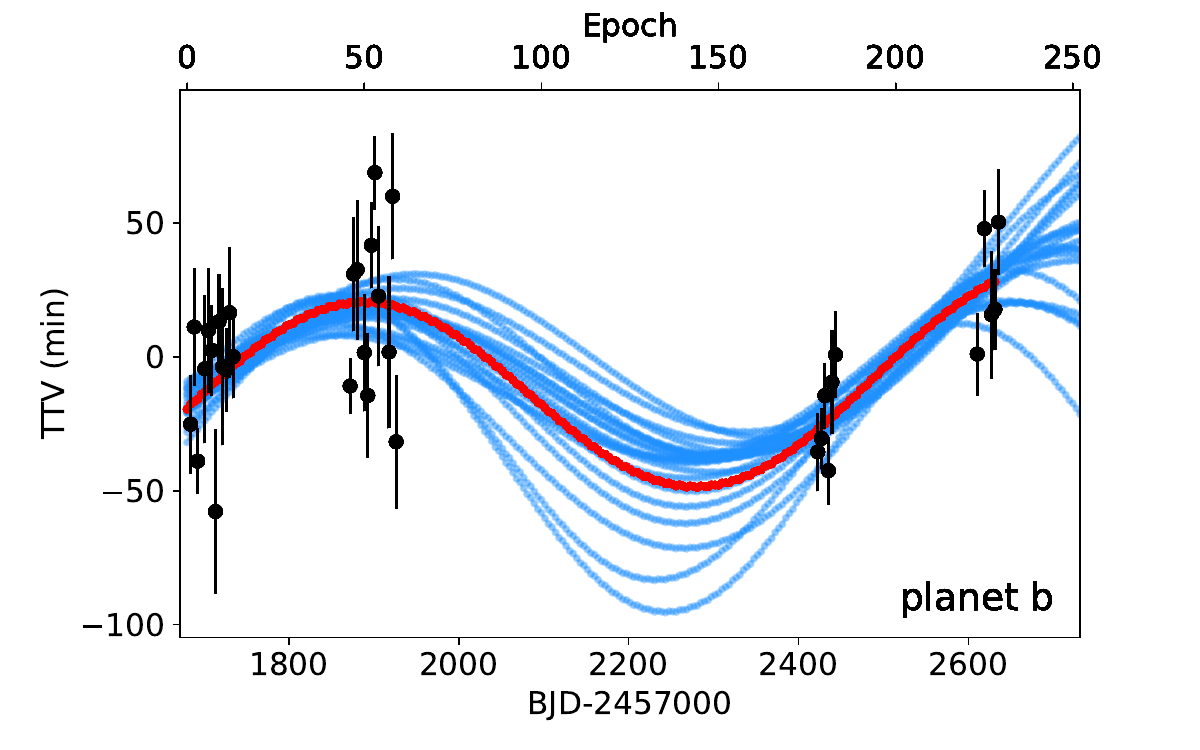}
\includegraphics[width = 1.\columnwidth]{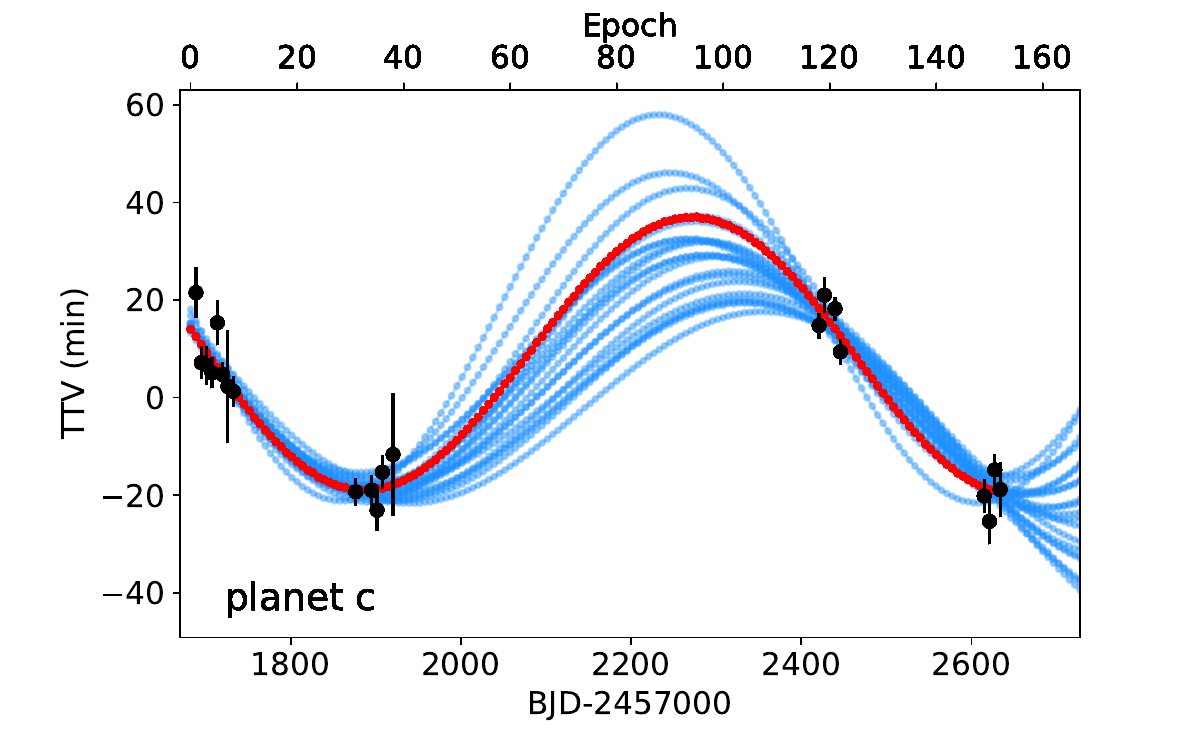}
\includegraphics[width = 1.\columnwidth]{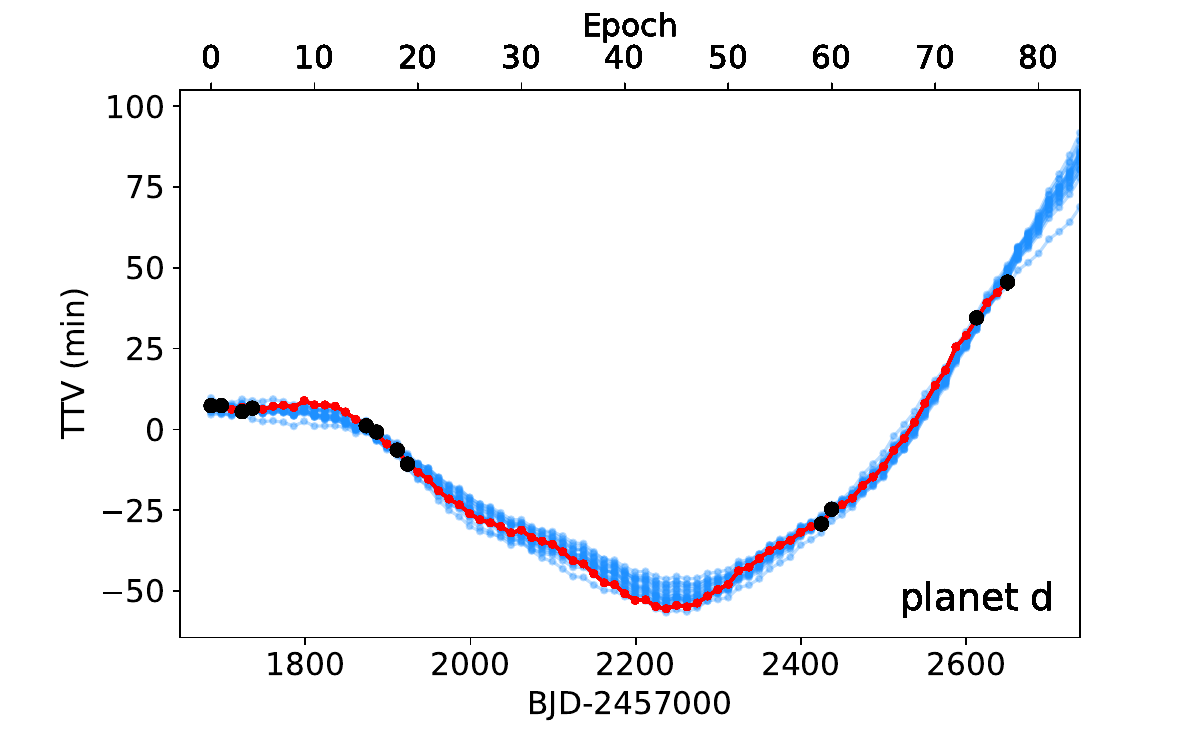}
\includegraphics[width = 1.\columnwidth]{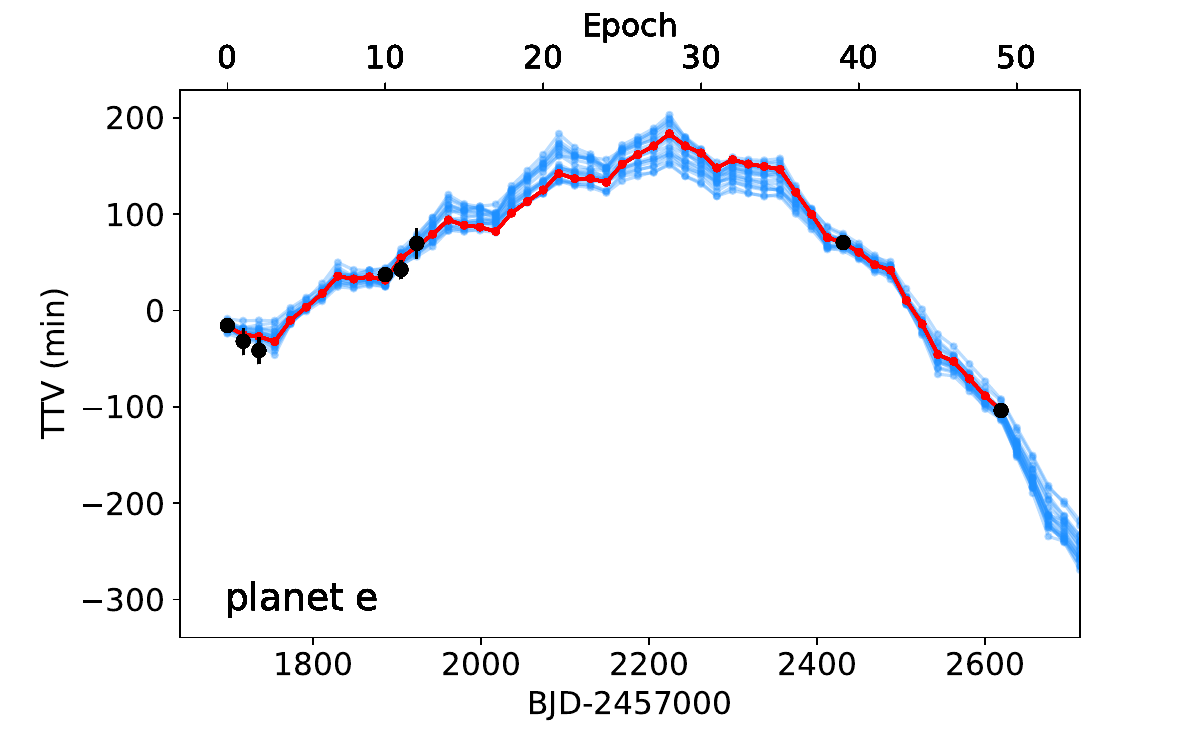}
\includegraphics[width = 1.\columnwidth]{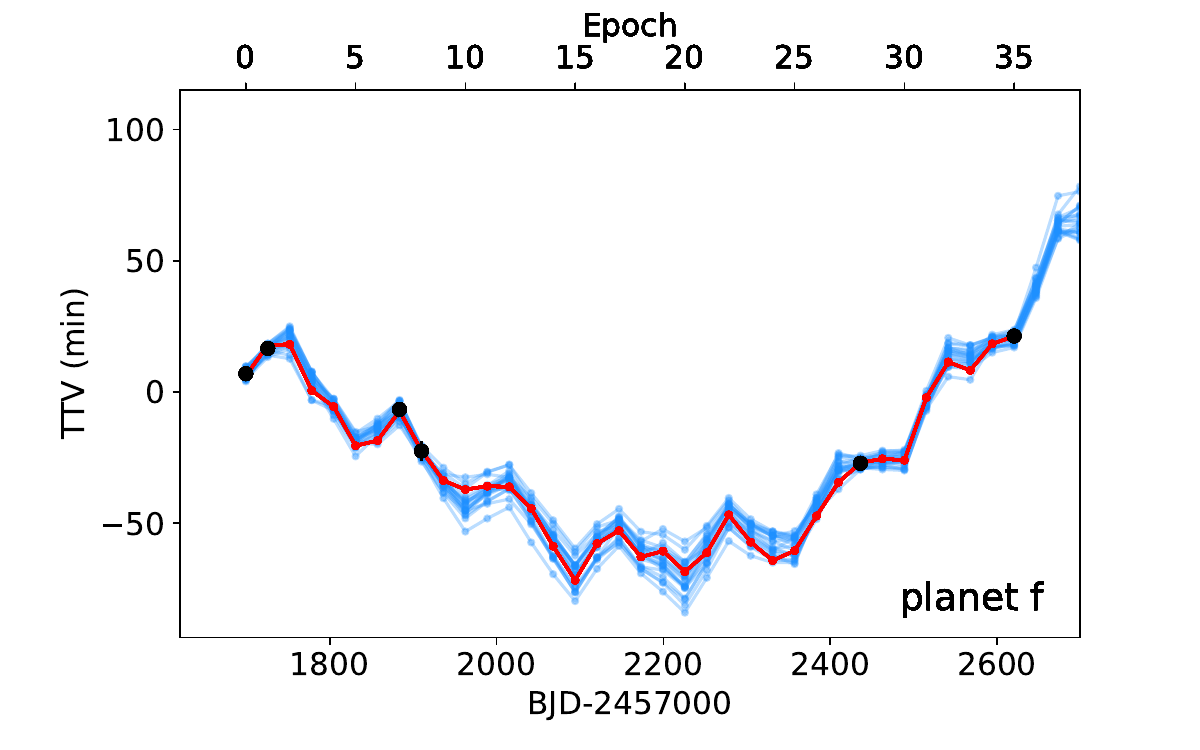}
\includegraphics[width = 1.\columnwidth]{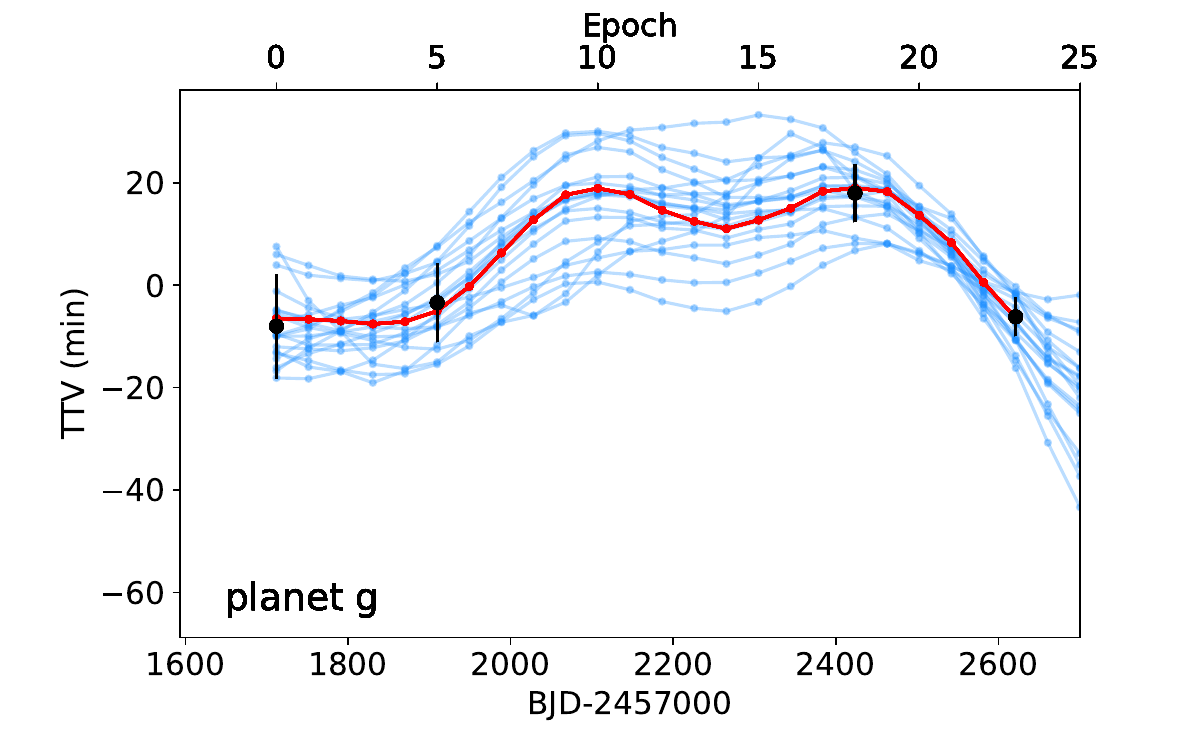}
\caption{The observed transit timing variations of the planets in TOI-1136, the best fit TTV model (red curve), and 20 dynamically stable posterior samples (blue curves). All data came from {\it TESS} observations except for the last transit of planet d, which came from our RM measurement. TTVs from neighboring planets are anti-correlated. The super-periods are estimated to be $\gtrsim10000$ days, which is much longer than the current observational baseline. Instead, the TTVs are driven by the libration of the resonant angles \citep{Nesvorny}. The libration periods were estimated $P_l \approx P_{\rm orb} (\frac{m_1+m_2}{m_\star})^{-2/3}$ \citep{Agol2005,Nesvorny,Goldberg_stability} to be between $\sim700$ days and $\sim 5000$-days, with the shortest period for the bc pair. The observed TTV show variations on similar timescales.}
\label{fig:ttv}
\end{center}
\end{figure*}

After performing these steps, TTVs were detected
(see Fig.~\ref{fig:ttv}). We phase-folded the individual transits after taking into account their TTVs. Fig.~\ref{fig:transit} shows the phase folded and binned transit light curves of each planet. Without accounting for TTVs, the phase-folded transits would have appeared V-shaped as opposed to U-shaped, and would have led to inaccurate transit parameters. We fit all the planets simultaneously with {\tt emcee} \citep{emcee}. We initialized 128 walkers near the best-fit model from  {\tt lmfit}. We ran the MCMC for 50000 links and assessed convergence using the Gelman-Rubin potential scale reduction factor \citep{BB13945229}. It dropped to below 1.02, indicating good convergence. The resultant posterior distribution is summarized in Table \ref{tab:planet_para}, while Fig.~\ref{fig:transit} shows the best-fit transit models.

We note that the initial detrending of the light curve and isolation of transit windows depends crucially on both a good knowledge of the TTVs and the transit durations. We therefore iterated the whole process outlined in this section twice to ensure convergence. In the Appendix, we present a search for additional transiting planets in this system.

\section{Transit Timing Variations}\label{sec:ttv}

We modeled the observed TTVs with full $N$-body integrations of the orbits. As we will show below, at least some of the planets of TOI-1136 are likely locked in mean motion resonances. In this case, the TTV signal cannot be adequately described by the combination of well-known analytic formulae for the near-resonant \citep{Lithwick_ttv} and individual conjunction \citep[chopping,][]{Deck_chopping} TTVs based on perturbation theory. The TTVs in a fully resonant system showing nonlinear dynamics cannot be treated in the same way \citep{Agol2005, Nesvorny}.

We integrated the orbits using a symplectic integrator \citep{1991AJ....102.1528W, 2014ApJ...787..132D} with the constant time step of $0.1\,\mathrm{days}$, considering only the Newtonian gravitational interactions between the six planets and the central star all treated as point masses. Over the observational baseline of a few years, any relativistic precession should be negligible and hence ignored. The model transit times were computed as described in \citet{2010arXiv1006.3834F} by finding the minima of the sky-projected star--planet distances. During this iteration for finding transit times, the system was integrated using a fourth-order Hermite integrator \citep{2004PASJ...56..861K}. The system was initialized using values for the planet-to-star mass ratio, orbital period $P$, eccentricity $e$, argument of pericenter $\omega$, and time $T_c$ of inferior conjunction nearest to the epoch $\mathrm{BJD}=2458680$, which was converted to the time of pericenter passage $\tau$ via 
$2\pi (T_c-\tau)/P = E_0 - e\sin E_0$ with $E_0=2\arctan\left[\sqrt{{1-e}\over{1+e}}\tan\left({\pi \over 4}-{\omega \over 2}\right)\right]$.
The orbital inclinations and the longitudes of ascending nodes were held fixed at $\pi/2$ and $0$, respectively. The mass ratios and osculating orbital elements were converted to Jacobi coordinates using the interior mass in Kepler's Third law as in \citet{2015MNRAS.452..376R} (see their Section~2.2),\footnote{We note that this conversion is different from what is adopted in the {\tt TTVFast} code \citep{2014ApJ...787..132D}, which performs the conversion following the Hamiltonian splitting defined by \citet{1991AJ....102.1528W}. The difference comes from the arbitrariness of how to split the motion into non-perturbed (i.e., Keplerian) and perturbed parts, and the resulting mappings between the coordinates and orbital elements differ slightly by an amount on the order of magnitude of the planet-to-star mass ratio. This difference is well below the stated uncertainties of any of the parameters, but it matters when one tries to reproduce the TTV signal.}
and the \citet{2006AJ....131.2294W} correction for the difference between real and mapping Hamiltonian was applied once at the beginning of integration as in \citet{2014ApJ...787..132D}. 
The sky plane was chosen to be the reference plane, with respect to which arguments of pericenters and the line of nodes were defined. The ascending node was defined with respect to the $+Z$-axis chosen to point toward the observer.
The transit timing code was implemented in {\tt JAX} \citep{jax2018github} to enable automatic differentiation with respect to the input parameters \citep[see also][]{2021MNRAS.507.1582A}, and is available through GitHub.\footnote{\url{https://github.com/kemasuda/jnkepler}; in this work, we used commit 6cac1c2.}.

The $N$-body transit time model $m(\theta)$ as described above was used to sample from the posterior probability distribution for the model parameters $\theta$ conditioned on the observed transit times $D=\{t_i\}$, $p(\theta|D) \propto p(D|\theta)\,p(\theta)$. 
We adopted the following log-likelihood function:
\begin{equation}
    \ln p(D|\theta)
    = -{1\over 2} \sum_i 
    \left\{ 
    {\left[t_i-m_i(\theta)\right]^2 \over \sigma_i^2} + \ln\left(2\pi\sigma_i^2\right)\right\}
\end{equation}
which is based on the assumption that the observed transit times are drawn from the independent and identical Gaussian distributions around the model values, with variances  $\sigma_i^2$ estimated from the modeling of transit and Rossiter-McLaughlin data in Section~\ref{sec:obliquity} and \ref{sec:transits} (Table \ref{tab:ttv}). The residuals of transit time fitting did not show clear evidence for any non-Gaussianity in the tails of the distributions, as has been seen in some other works \citep{JontofHutter2016, Agol2021}. 

We adopted a prior probability distribution function $p(\theta)$ separable for each model parameter, as summarized in Table~\ref{tab:ttvpriors}.
The sampling was performed using Hamiltonian Monte Carlo and the No-U-Turn Sampler \citep{DUANE1987216, 2017arXiv170102434B} as implemented in {\tt NumPyro} \citep{bingham2018pyro, phan2019composable}. We ran four chains in parallel until we obtained at least 50 effective samples for each parameter and the resulting chains had the Gelman-Rubin statistic of $\hat{R}<1.05$ \citep{BB13945229}.

\begin{deluxetable}{lc}
\caption{Priors adopted in the TTV Modeling.}
\label{tab:ttvpriors}
\tablehead{
	\colhead{Parameter} & \colhead{Prior}
} 
\startdata
Planet/Star Mass Ratio & $\mathcal{U}(0,5\times 10^{-4})$\\
Orbital Period (days) & $\mathcal{U}(P_0-0.5, P_0+0.5)$\\
Orbital Eccentricity & $\mathcal{U}(0,0.4)$\\
Argument of Pericenter & $\mathcal{U}(0,2\pi)$\\
Time of First Inferior Conjunction (days) & $\mathcal{U}(T_0-0.1, T_0+0.1)$
\enddata
\tablecomments{$\mathcal{U}(a,b)$ is the uniform distribution between $a$ and $b$. The symbols $P_0$ and $T_0$ denote the linear ephemeris computed from observed transit times for each planet. The argument of pericenter was wrapped at $2\pi$. 
}
\end{deluxetable}

During the TTV posterior sampling, we did not impose any requirement for long-term dynamical stability.  Instead, we imposed a stability requirement in post-processing, as will be described in the next Section. The planetary parameters reported in Table \ref{tab:planet_para} will be based on the stable TTV posterior samples.

\section{Dynamical Modeling}\label{sec:dynamical}

\subsection{Stability Analyses}\label{sec:stability}

After examining the posterior distribution of our TTV analysis, we realized that many of the posterior samples would experience orbital instability on relatively short timescales. Since TOI-1136 is about 700 Myr old, it should be stable on similar timescales. However, with the TTV data in hand, the TTV analysis alone may not be able to pin down the system's configuration (with $>30$ parameters) to the island of stability that the real system resides. Near MMR the system is dynamically rich, a small change of system parameters may lead to very different dynamical behavior. This is especially true considering the fine structure of second-order resonance, and the relatively short TTV baseline of the current {\it TESS} data.

We therefore proceeded to trim down the posterior samples by removing TTV solutions that go unstable quickly. We employed the {\tt Python} package {\tt REBOUND} \citep{Rein}. We used the built-in {\tt mercurius} integrator, which is a hybrid integrator similar to {\tt Mercury} by \citet{Chambers}. {\tt mercurius} makes use of the symplectic Wisdom-Holman integrator {\tt WHFast} \citep{1991AJ....102.1528W} when planets are far away from each other, and switches to the high-order integrator {\tt IAS15} \citep{IAS15} whenever it detects a close encounter within a user-defined distance. We switched the integrator when any two planets are less than 4 mutual Hill radii from each other. 

We integrated all the posterior samples from Section \ref{sec:ttv} for 1 Myr. We acknowledge that this is much shorter than the system's age of $\sim$700 Myr. The choice of 1 Myr was a compromise between computation time and gauging the long-term stability of the TTV solutions. We did not include tidal effects which may begin to manifest on timescales longer than 1 Myr. We removed posterior samples that were flagged as unstable by {\tt REBOUND}. Planets in these systems experienced collisions or became unbound.

To quantify the stability of the remaining  posterior samples, we further examined the orbital architectures after 1-Myr integration. Using the orbital period of the innermost planet b as a proxy, we show in Fig.~\ref{fig:stability} that some posterior samples underwent substantial changes in orbital architecture even though the system remained technically stable. In some cases, the orbital periods of planet b underwent order-of-unity changes from its initial value, Moreover, the orbital period ratio between the innermost planets $P_c/P_b$ moved significantly off resonance (Fig.~\ref{fig:stability}). These system later experienced orbital instability when we integrated them to 10 Myr. To maximize the long-term stability of our posterior samples, we kept only posterior samples in which 1) $P_b$ changed by $<1\%$ from its initial value and 2) $P_c/P_b$  changed by $<2\%$ from its initial value of 3:2 MMR after 1-Myr of N-body integration. These criteria are the orange box in Fig.~\ref{fig:stability}.

About 48\% of the original posterior samples remained after the selections just described. All of our subsequent analyses were based on this ``stable'' posterior sample. Table \ref{tab:planet_para} summarizes this stable posterior distributions and reports the osculating Keplerian elements at the time of reference BJD=2458680. We note that the osculating orbital period ratios should not be used to predict future transits or gauge the depth of resonance in this system. The osculating orbital periods suffer from large uncertainty as they vary rapidly after a close encounter between planets. Instead, we report the orbital period ratios by averaging the osculating orbital period of the stable solutions over a time interval of 50000 days (longer than the libration periods of the system, see Section \ref{sec:resonance}). The period ratios are extremely close to their respective resonance, with deviations $\Delta \equiv \frac{P_{out}/P_{in}}{p/q}-1$ of $6.9\pm1.9\times10^{-5}$ for bc, $2.01\pm0.97\times10^{-4}$ for cd, $4.4\pm1.3\times10^{-4}$ for de, $4.5\pm1.6\times10^{-4}$ for ef, and $8.4\pm2.9\times10^{-4}$ for fg. We compare this resonant structure to other known planetary systems in Section \ref{sec:commensurability}.

Given the limited TTV data and measurement uncertainty, we most likely have not located the true island of stability that is stable for 700 Myr. Resonant interaction involving several planets leads to a finely-structured and complex dependence of the system's dynamical evolution on the initial parameters. A small change of the system configuration may lead to very different dynamical behavior. A similar situation was encountered by \citet{Gillon} in their early analysis  of TRAPPIST-1. Most of their TTV solutions went unstable on a very short timescale  ($\sim$0.5 Myr). Only years later, when TTVs were observed over a longer timespan, did \citet{Agol2021} find solutions for the orbital architecture of TRAPPIST-1 that are stable for at least 50 Myr. With this in mind, we encourage follow-up transit observations of TOI-1136.

\begin{figure}
\center
\includegraphics[width = .9\columnwidth]{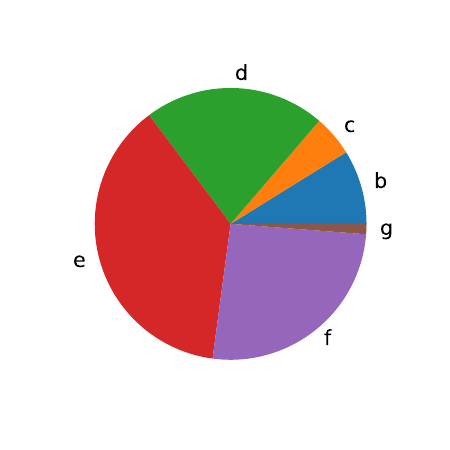}
\caption{The relative fractions of TOI-1136 planets that became dislodged from resonance first in our dynamical integration. Dynamical instability often ensues after breaking resonance. Planets e and f (the only second-order MMR in TOI-1136) are usually the first to become dislodged due to the weaker strength of second-order MMR. Together, they departed from resonance first in more 68\% of the posterior samples that went unstable within 1 Myr.} 
\label{fig:unstable_pie}
\end{figure}

\begin{figure*}
\center
\includegraphics[width = 1.\columnwidth]{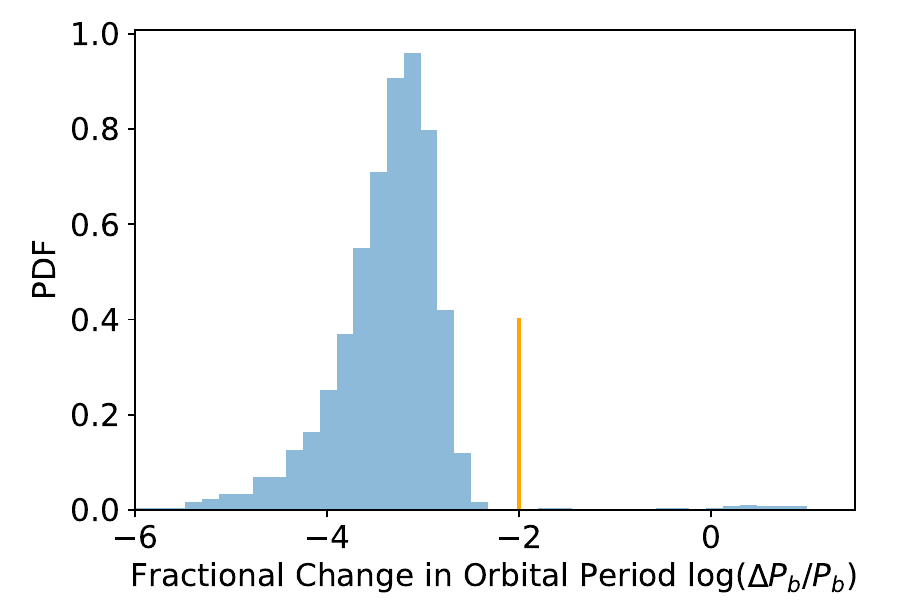}
\includegraphics[width = 1.\columnwidth]{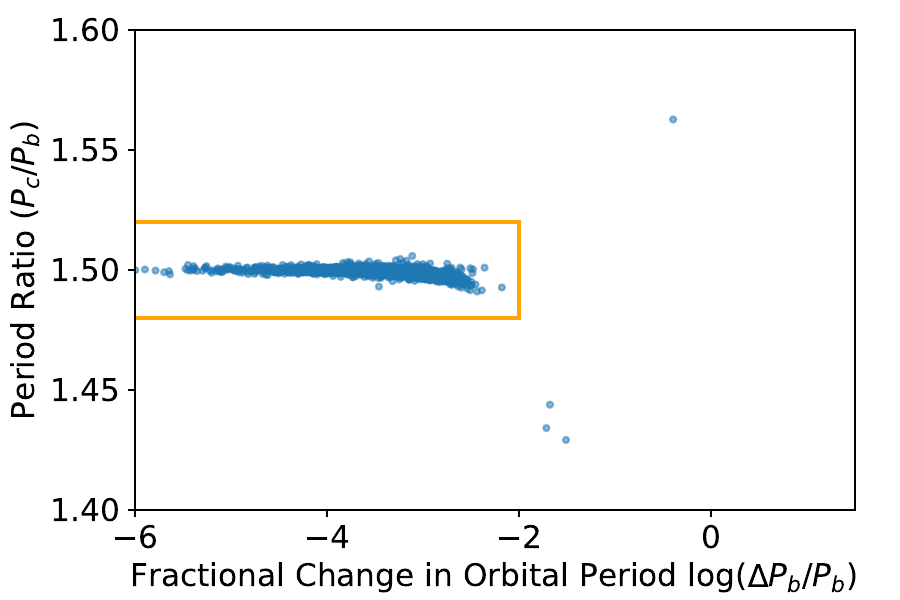}
\caption{The fractional change of orbital period for planet b (left) and the orbital period ratio between planet b and c (right) for our TTV posterior samples after 1-Myr N-body integration (Section \ref{sec:stability}). Posterior samples in which the orbital period of planet b moved more than 1\% from its initial value are also those that broke away from the resonant configuration ($P_c/P_b$ deviated from 3:2 MMR). We removed these systems (outside the orange box) as they quickly went unstable upon longer integration.}
\label{fig:stability}
\end{figure*}

We also tracked which of the TOI-1136 planets were dislodged from resonance first. As shown in Fig.~\ref{fig:unstable_pie}, planets e and f (7:5 second-order MMR) seems to be a weak link in the resonant chain: they were the first to be removed from resonance in more than 68\% of the unstable solutions. This is theoretically expected because second-order resonant interactions are weaker than first-order interactions by another factor of orbital eccentricity \citep[$e^{k}$ where k is the order of the MMR][]{Murray} and have thinner libration widths in semi-major axis (see Fig. \ref{fig:resonant_structure}). It has also been suggested that many second-order resonances formed by convergent disk migration may in fact be overstable \citep{Goldreich2014,Xu0217} and easily disrupted.

\subsection{Generalized Laplace Resonance} \label{sec:resonance}
We investigate in this section if TOI-1136 planets are indeed in mean-motion-resonance (MMR) rather than being near resonance by chance. The hallmark of true MMR is the libration of the relevant resonant angles. For a planetary system near resonance, one can decompose the Hamiltonian into the Keplerian, resonant, and secular terms \citep{Murray}.  The generalized coordinate for the resonant interaction is the resonant angle. For two-planet systems, the resonant angle  $\phi$ takes the form:

\begin{equation}\label{eqn:2-body}
    \phi_{\rm 12} = q\lambda_{1}-p\lambda_{2}+(p-q)\varpi_{1,2}
\end{equation}
where $p$ and $q$ are positive co-prime integers, $|p-q|$ is the order of the resonance. The mean longitude $\lambda$ is the sum of the mean anomaly $M$, the longitude $\Omega$ of the ascending node, and the argument of pericenter $\omega$. The angle $\varpi$ 
is defined as $\Omega + \omega$. Following D'Alembert's rule, $\varpi_{1,2}$ can be an integer combination of $\varpi_{1}$ and $\varpi_{2}$ such that the sum of the coefficients is $p-q$. The strength of the MMR is proportional to $e^{|p-q|}$. For a system in true 2-body MMR, $\phi_{\rm 12}$ librates around a libration center with limited amplitude, as opposed to circulating between 0 to 2$\pi$.

Several combinations of $\varpi_{1}$ and $\varpi_{2}$ are allowed by D'Alembert's rule, especially for higher-order MMR \citep{Murray}. Exploring all of them can be cumbersome and redundant. \citet{Sessin} suggested a canonical transformation such that 2-body resonance can be described by a single mixed pericenter angle \citep[see also ][]{Henrard1986,Wisdom1986,Batygin_resonance,Hadden_resonance}:

\begin{equation}\label{eqn:mixed_angle}
    \hat{\varpi}_{12}=\arctan{\left[\frac{f~e_1~\sin{\varpi_1}+g~e_2~\sin{\varpi_2}}{f~e_1~\cos{\varpi_1}+g~e_2~\cos{\varpi_2}}\right]}
\end{equation}
where $f$ and $g$ are the coefficients of the disturbing function \citep[see the tabulated values in e.g.,][]{Lithwick_ttv}. \citet{Petit} used this mixed angle to investigate 2-body MMR and found it useful for probing the resonant angles in K2-19: a system with high eccentricities and limited TTV data. We adopt this mixed pericenter angle formulation to analyze the 2-body resonances in TOI-1136.

When more than two planets are involved in MMR, one can generalize the resonant angle. One can simply subtract the 2-body resonant angles (Eqn.~\ref{eqn:2-body}) of neighboring pairs to remove any dependence on $\varpi$. For a concrete example, consider TOI-1136 b, c, and d:

\begin{equation}
\phi_{\rm bc} = 2\lambda_{b}-3\lambda_{c}+\varpi_c
\end{equation}
\begin{equation}
\phi_{\rm cd} = \lambda_{c}-2\lambda_{d}+\varpi_c
\end{equation}
\begin{equation}
\phi_{\rm bcd} =\phi_{\rm bc} -\phi_{\rm cd}  = 2\lambda_{b}-4\lambda_{c}+2\lambda_{d}
\end{equation}

A perceptive reader might point out
that the coefficients are no longer co-prime and that we should divide by 2. We chose not to do so following the suggestion of \citet{Siegel}. The benefit of keeping the original coefficients is that the preferred libration centers for 3-body MMR are now near 180$^\circ$ in this formulation. For example, in Kepler-60, \citet{Gozdziewski} defined the 3-body resonant angle $\phi_{\rm bcd} = \lambda_b-2\lambda_c+\lambda_d$. \citet{Gozdziewski} reported a libration center of $\sim 45^\circ$. The underlying 2-body MMR are 5:4 and 4:3; $\phi_{\rm bcd}$ should have been $\phi_{\rm bcd} = 4\lambda_b-8\lambda_c+4\lambda_d$ in the formulation of \citet{Siegel}. Correspondingly, $\phi_{\rm bcd}$ the libration center should have been 180$^\circ$. The significance of a libration center of 180$^\circ$ is perhaps best understood in the most famous example of Laplace's Resonance between the inner three Galilean moons Io, Europa, and Ganymede \citep[e.g.,][]{Sinclair}. The libration of $\phi_{\rm IEG}=\lambda_I-3\lambda_E+2\lambda_G$ around 180$^\circ$ ensures that whenever two satellites have a close encounter, the third satellite is far away, by either 90$^\circ$ or 180$^\circ$. Such a resonant configuration minimizes three-body conjunctions and chaotic interactions, and hence enhances the overall stability of the system. This geometric/phase relation holds true even for systems that have experienced long-range deviation from 2-body orbital period commensurability \citep[e.g. Kepler-221][]{Goldberg2021}.

One can extend this process to construct resonant angles when more planets are involved. In Table \ref{tab:libration}, we list the various resonant angles for TOI-1136. Before describing the results, we highlight an effect that can shift the libration centers. For a chain of planets in resonance, their mutual interactions change the topology of the Hamiltonian, especially when there is a non-adjacent first-order MMR. New libration centers can emerge that
are shifted away from 180$^\circ$ \citep[e.g.,][]{Siegel}. A system can be captured in one of the possible libration centers depending on the order of which planets are captured into resonance \citep{Delisle2017}. For example, Kepler-223 is in a 3:4:6:8 resonant chain \citep{MillsNature}. The bd pair (6:3$\equiv$2:1) and the ce pair (8:4$\equiv$2:1) are both examples of non-adjacent first-order MMR. The 3-body libration centers were hence shifted to 168$^\circ$ and 130$^\circ$ in that system \citep{Siegel}. \citet{Delisle2017} suggested that the observed configuration is perhaps most consistent with Kepler-223 c and d having been captured into MMR before e and b. Fortunately (or sadly), there is no non-adjacent first-order MMR in TOI-1136, so one need not worry about (or cannot take advantage of) this effect.

We integrated the stable TTV solutions from Section \ref{sec:stability} forward in time for 50000 days with {\tt REBOUND}. We recorded the various resonant angles of TOI-1136 listed in Table \ref{tab:libration}.  We identified systems in which the resonant angles are clearly circulating ($\phi$ varied by much more than 2$\pi$). Then, to identify the librating solutions, we calculated the mean of the resonant angles during this 50000-day period. We also computed the libration amplitude using the formula in \citet{Siegel} and \citet{Millholland_2018_resonance}:

\begin{equation}
A=\sqrt{\frac{2}{N}\sum{(\phi-\langle\phi\rangle)^2}}   
\end{equation}
where $\langle\phi\rangle$ is mean of the resonant angle. $N$ is the number of resonant angles sampled. If the libration of resonant angle is sinusoidal in shape and sampled regularly in time, then $A$ corresponds to the amplitude of that sinusoid. We adopted a generous definition of libration: a system is in libration if the amplitude is less than 90$^\circ$. We can see in Table \ref{tab:libration} that most libration amplitudes are much smaller than this threshold. 

\begin{figure*}
\center
\includegraphics[width = 2\columnwidth]{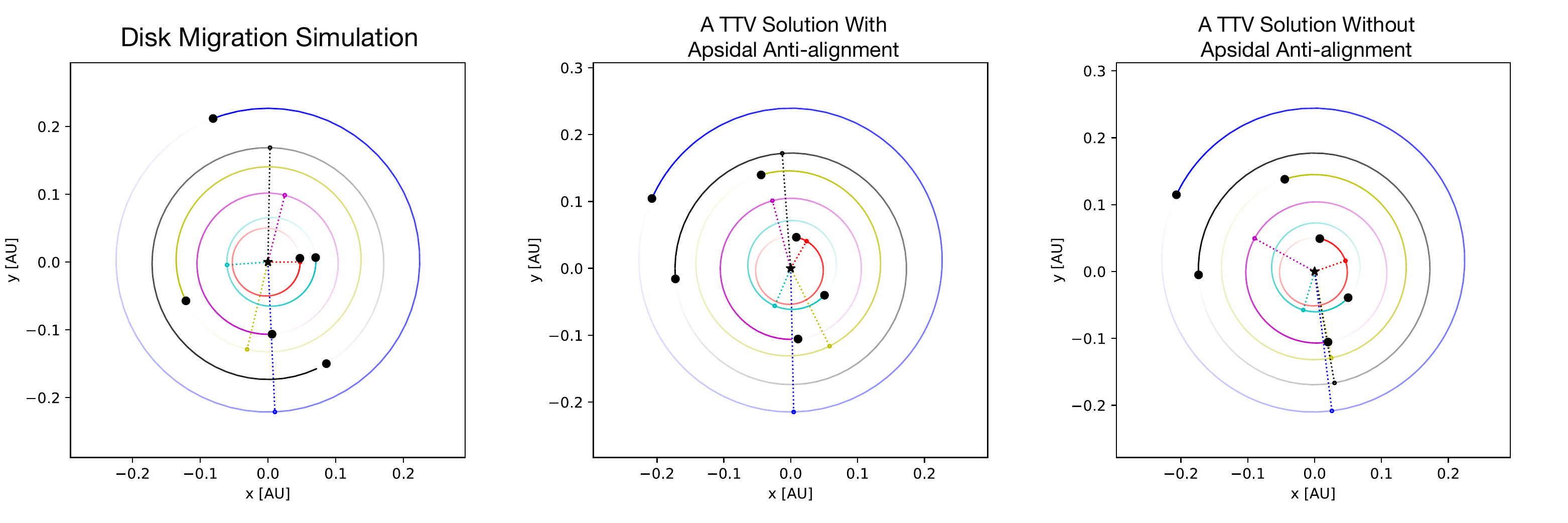}
\includegraphics[width = 2\columnwidth]{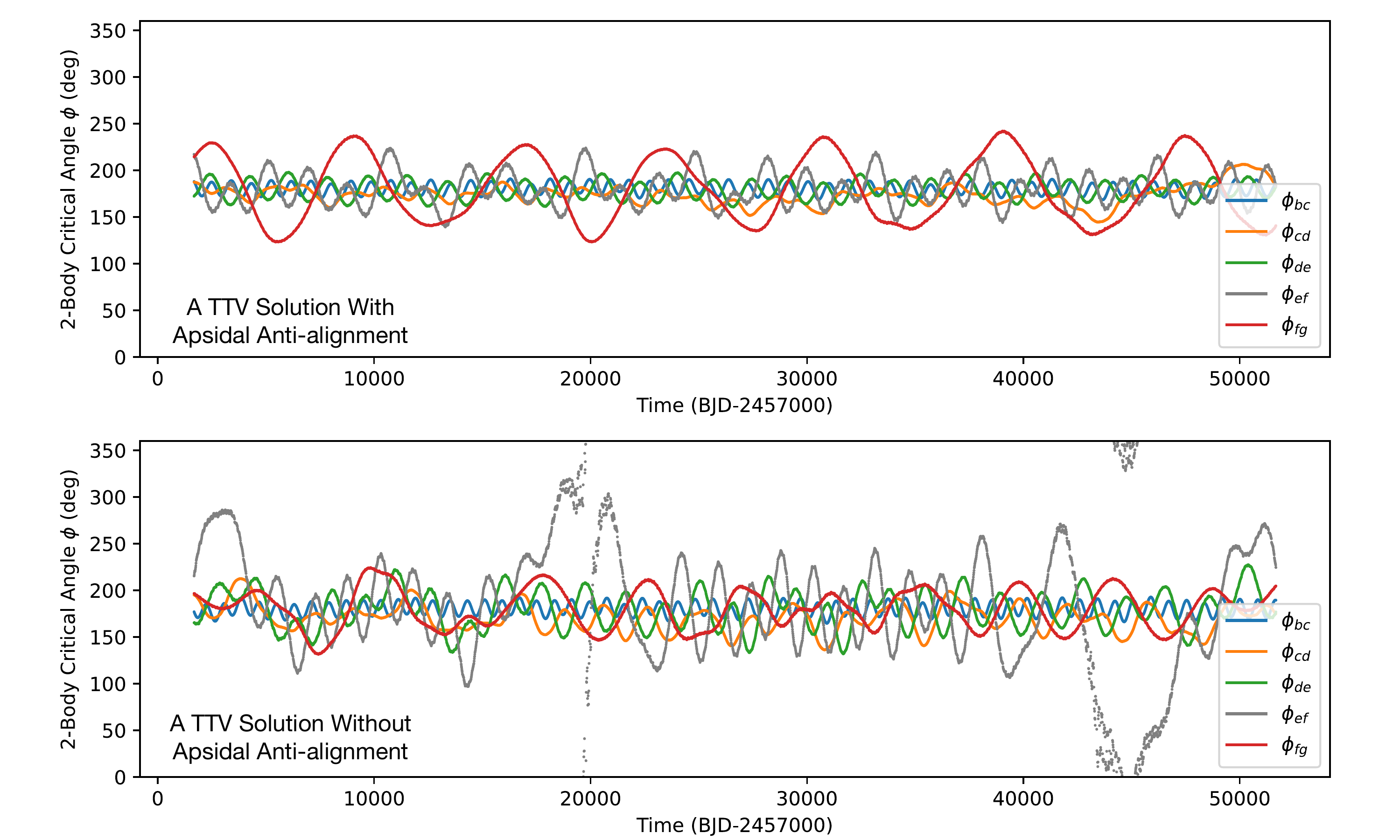}
\caption{
Top Row: The orbital configurations of TOI-1136 just after our convergent disk migration simulation of TOI-1136 (top left), a TTV solution with apsidal anti-alignment between neighboring planets (top center), and a TTV solution without apsidal anti-alignment (top right). The dotted lines indicate the pericenters of each planet. A classical prediction of convergent disk migration \citep[e.g.][]{Batygin2015_capture} is that neighboring planets should have anti-aligned pericenters (except the Asymmetric Libration of cd in 2:1 MMR, see text). A significant fraction of our TTV solutions conform to this prediction (see Fig. \ref{fig: e_vector}). The evolution of 2-body resonant angles of these solutions librate near 180$^\circ$ over the next 50000 days (middle panel). However, other TTV solutions are far from apsidal anti-alignment. Planet e and f (7:5 second-order resonance) in these solutions often show chaotic behavior where their 2-body resonant angle $\phi_{\rm ef}$ can oscillate between a state of libration and circulation (grey line in the bottom panel).}
\label{fig:ttv_resonant_angle}
\end{figure*}

\begin{figure*}
\center
\includegraphics[width = 1.2\columnwidth]{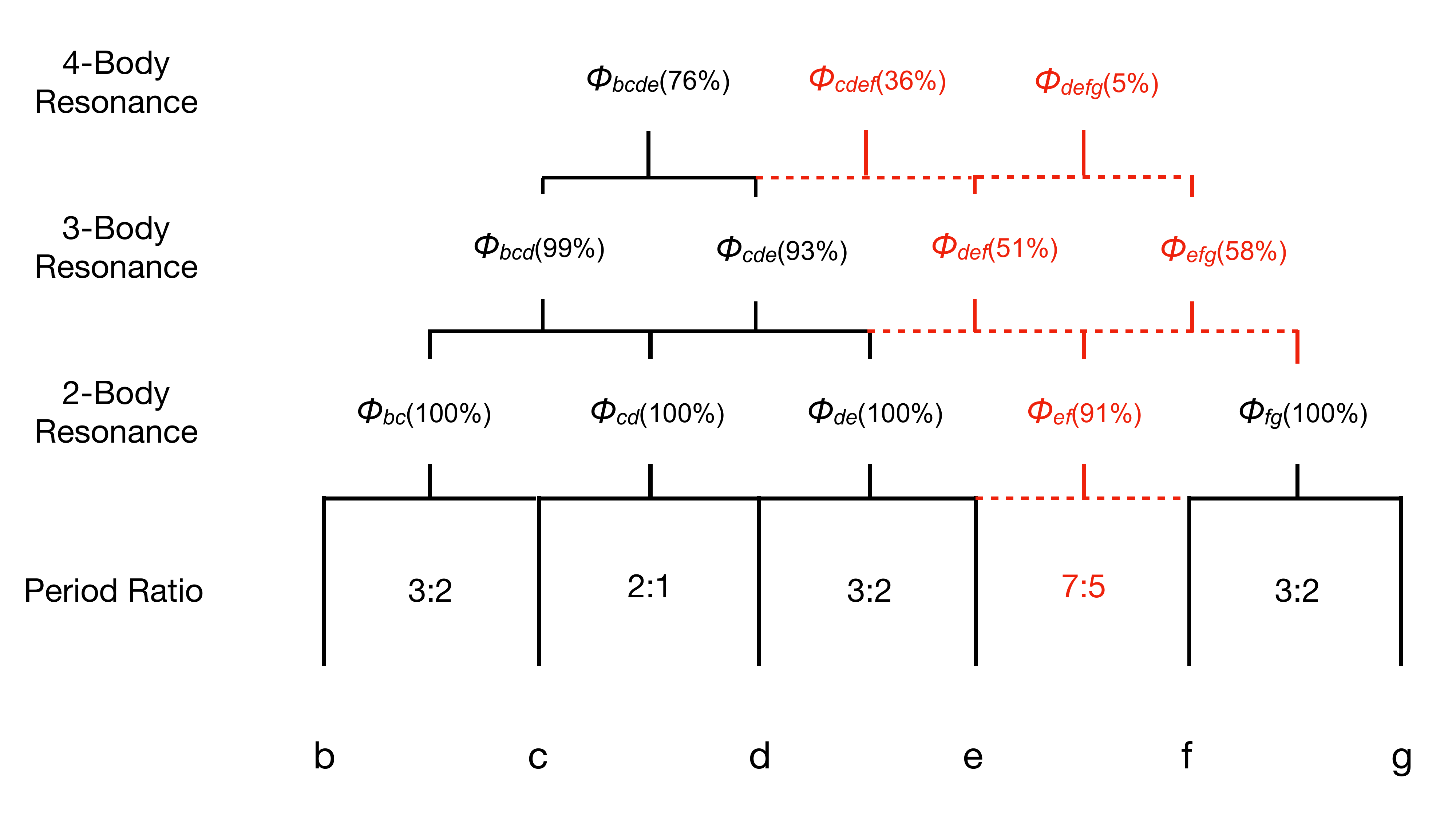}
\includegraphics[width = 0.8\columnwidth]{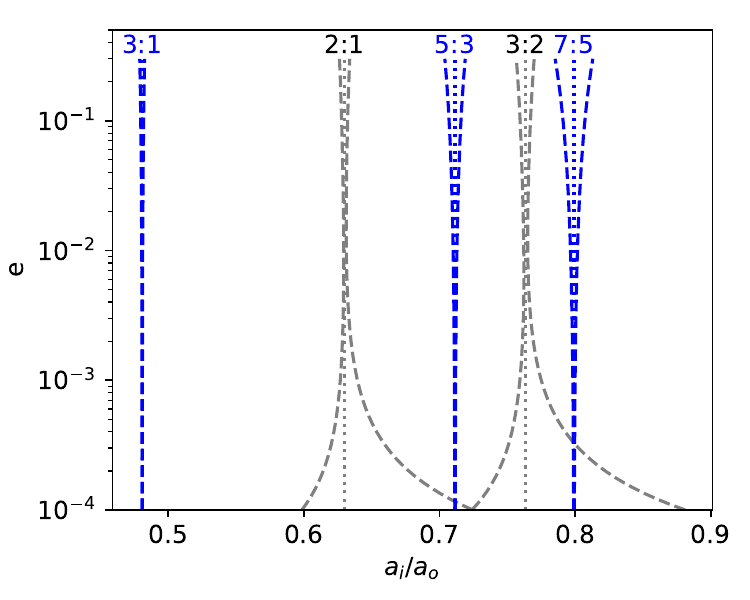}
\caption{Left: the ladder of the resonant angles involving increasingly more planets. We recorded the resonant angles (Table \ref{tab:libration}) in the stable TTV solutions for 50000 days. The fraction of the TTV posterior sample in which the specific resonant angle librates is shown in the bracket. The resonant angles that involve the second-order resonance of planet e and f (7:5 MMR) have a significantly reduced probability of libration. Second-order resonances have narrower libration width compared to first order resonance particularly towards low eccentricity (right panel, calculated with Eqn 8.76 in \citet{Murray}). Our TTV analyses, with measurement uncertainty, likely has not located the solution to the true island of stability that real system resides.}
\label{fig:resonant_structure}
\end{figure*}

Fig.~\ref{fig:resonant_structure} summarizes the relationships between the various resonant angles and the fraction of librating solutions for each angle. We found that the various resonant
angles involving only first-order resonance have a high probability of libration in our stable TTV solutions. The fraction is close to unity for the 2-body angles, and steadily drops as we move up the resonance ladder from 2-body resonance to multi-body resonance. The inner four planets bcde ($\phi_{\rm bcde}$) have a $76\%$ probability of being a resonant chain. Moreover, the libration centers are almost always near 0 or 180$^\circ$ (Table \ref{tab:libration}) as found by \citet{Siegel}. The only exceptions are the resonant angles involving planets c and d (2:1 MMR). \citet{Beauge} showed that the topology of the phase space of the 2-body 2:1, 3:1, n:1 MMR permits two libration centers that are shifted from 180$^\circ$ \citep[Asymmetric Libration][]{Beauge2006}. The shifts increase with orbital eccentricity. A planetary system may adopt one of these libration centers, or chaotically shift between them if the libration amplitude is large enough. This was confirmed in our convergent disk migration simulations (Section \ref{sec:migration} and the first panel of Fig.~\ref{fig:ttv_resonant_angle}): resonant angles involving TOI-1136 c and d are shifted from 180$^\circ$ by Asymmetric Libration.

In contrast to the first-order MMRs, the resonant angles that involve the only second-order MMR (planet e and f, 7:5) have significantly reduced probabilities of libration. Second-order MMR, by nature, is much weaker and much more localized in phase space than first-order MMR \citep[see Fig. \ref{fig:resonant_structure} and ][]{Murray}. In about 9\% of our TTV solution, the second-order resonant angle $\phi_{\rm ef}$ alternates between circulation and libration (Fig. \ref{fig:ttv_resonant_angle} lower panel). Alternation between libration and circulation is a hallmark of chaos and has been previously identified in Kepler-36 \citep{Carter}. However, we strongly suspect that e and f are indeed in a 7:5 second-order MMR.  In our stable TTV solutions, planets e and f do have a $\sim$91\% chance of being in 2-body libration. The observed orbital period ratio differs from 7:5 by only $4.5\pm1.6\times10^{-4}$; it seems very unlikely to be coincidental. See \citet{Bailey} for a dynamical exploration for the observed and expected period ratio of pairs of planets locked in second-order MMR. Our current TTV solutions of TOI-1136 are often chaotic on short timescales, with some Lyapunov times of the order 10$^5$ days. Again, we suspect that with the current TTV data, we have not located the true island of stability in phase space. The measurement uncertainty is particularly obvious for the second-order MMR that has thinner libration width in phase space (right panel of Fig. \ref{fig:resonant_structure}).

We examined the dominant periodicities of the observed TTV. For a near-resonant, circulating system, the TTV occurs on the timescale of the ``super-period'' $P_s = 1/|p/P_2-q/P_1|$ \citep{Lithwick_ttv}. In contrast,  for truly resonant systems, the TTV should vary on the timescale of the libration period $P_l \approx P_{\rm orb} (\frac{m_1+m_2}{m_\star})^{-2/3}$ for 2-body resonance \citep{Nesvorny,Goldberg_stability}. We estimated both  $P_s$ and $P_l$ in TOI-1136. Since the period ratios are so close to ratios of small
integers (Section \ref{sec:stability}), the super-periods $P_s$ are typically longer than 10$^4$ days for TOI-1136. On the other hand, the estimated libration periods $P_l$ are between
about 800 and 5000 days (from bc to fg) based on Eqn.~2 of \citet{Goldberg_stability}. The existing TTV data clearly show variations on the shorter timescales of $P_l$ (Fig.~\ref{fig:ttv}). For a more empirical test, we applied a Lomb-Scargle periodogram to the 2-body resonant angles $\phi_{\rm bc}$ to $\phi_{\rm fg}$ in our TTV posterior solutions. $P_l$ indeed span a range of $700$ to $5000$ days. This is another evidence that TOI-1136 planets are in resonance rather than near resonance.

\begin{deluxetable*}{lccc}
\tablecaption{Resonant Angles in Stable TTV Posterior \label{tab:libration}} 
\tablehead{
\colhead{Resonant Angle} & \colhead{Fraction in Libration} & \colhead{Libration Center} & \colhead{Libration Amplitude$^1$}}
\startdata
2-body Resonant Angles\\
$\phi_{\rm bc}$ = $2\lambda_{b}-3\lambda_{c}+\hat{\varpi}_{bc}~^2$ & 100\%  & 179.1$\pm$1.5 $^\circ$  & 9.6$\pm$1.5 $^\circ$\\
$\phi_{\rm cd}$ = $\lambda_{c}-2\lambda_{d}+\hat{\varpi}_{cd}$& 100\%   & 176.7$\pm$6.8 $^\circ$  & 14.6$\pm$6.6 $^\circ$ \\
$\phi_{\rm de}$ = $2\lambda_{d}-3\lambda_{e}+\hat{\varpi}_{de}$ & 100\%  & 180.5$\pm$1.5 $^\circ$  & 17.3$\pm$7.7 $^\circ$ \\
$\phi_{\rm ef}$ = $5\lambda_{e}-7\lambda_{f}+2\hat{\varpi}_{ef}$& 91\%   & 182.1$\pm$7.4$^\circ$ & 36$\pm$13$^\circ$\\
$\phi_{\rm fg}$ = $2\lambda_{f}-3\lambda_{g}+\hat{\varpi}_{fg}$ & 100\%  & 180.3$\pm$1.0 $^\circ$  & 19$\pm$15 $^\circ$ \\
\hline
3-body Resonant Angles\\
$\phi_{\rm bcd}$ = $2\lambda_{b}-4\lambda_{c}+2\lambda_{d}$ $^3$& 99\%   & 196$\pm$15 $^\circ$  & 19$\pm$9 $^\circ$ \\
$\phi_{\rm cde}$ = $1\lambda_{c}-4\lambda_{d}+3\lambda_{e}$ & 93\%   & 163$\pm$30 $^\circ$ & 45$\pm$22 $^\circ$ \\
$\phi_{\rm def}$ = $4\lambda_{d}-11\lambda_{e}+7\lambda_{f}$ & 51\%  & 173$\pm$37$^\circ$  & 64$\pm$13$^\circ$ \\
$\phi_{\rm efg}$ = $5\lambda_{e}-11\lambda_{f}+6\lambda_{g}$& 58\%   & 143$\pm$51 $^\circ$  & 69$\pm$19 $^\circ$ \\
\hline
4-body Resonant Angles\\
$\phi_{\rm bcde}$ = $2\lambda_{b}-5\lambda_{c}+6\lambda_{d}-3\lambda_{e}$& 76\%   & 24$\pm$36 $^\circ$  & 44$\pm$19 $^\circ$ \\
$\phi_{\rm cdef}$ = $1\lambda_{c}-8\lambda_{d}+14\lambda_{e}-7\lambda_{f}$& 36\%   & -7$\pm$39$^\circ$ & 72$\pm$6$^\circ$ \\
$\phi_{\rm defg}$ = $4\lambda_{d}-16\lambda_{e}+18\lambda_{f}-6\lambda_{g}$ & 5\%  & -  & -\\
\enddata
\tablecomments{$1$: Libration amplitude is defined as $A=\sqrt{\frac{2}{N}\sum{(\phi-\langle\phi\rangle)^2}}$ \citep{Millholland_2018_resonance,Siegel}. 2: $\lambda$ are the mean longitudes of each planet. According to the D'Alembert Rule, the longitudes of pericenters $\varpi$ of both planets involved in a mean-motion resonance could contribute to the resonant angles. However, with a canonical transformation, the 2-body resonance is dependent on just the mixed angle: $\hat{\varpi}_{12}=\arctan{(fe_1\sin{\varpi_1}+ge_2\sin{\varpi_2})/(fe_1\cos{\varpi_1}+ge_2\cos{\varpi_2})}$ (see Section \ref{sec:resonance} for more detail). For 3-body resonances and above, the lowest-order resonant angles are independent of $\varpi$. $3$: We did not reduce the coefficients to be co-prime, following the suggestion by \citet{Siegel}; in this way, the 3-body resonant angles librate near 180$^\circ$.}
\end{deluxetable*}

\subsection{Convergent Disk Migration}\label{sec:migration}
Simulating the formation of resonant-chain planetary systems with disk migration can constrain the disk density and turbulence, as well as the order of planets that captured into resonances \citep[e.g.][]{Huhn}. Previous works \citep{Xu0217} have shown that it is more much challenging to form a second-order MMR than first-order MMR through disk migration. If the disk migration were turbulent or simply rapid, a planet pair could have easily skipped a second-order resonance and become locked in nearby first-order resonances. We can leverage this difficulty of forming the observed second-order 7:5 MMR of TOI-1136 ef to constrain the properties of TOI-1136's protoplanetary disk.

\begin{figure*}
\centering
\includegraphics[width = 0.66\columnwidth]{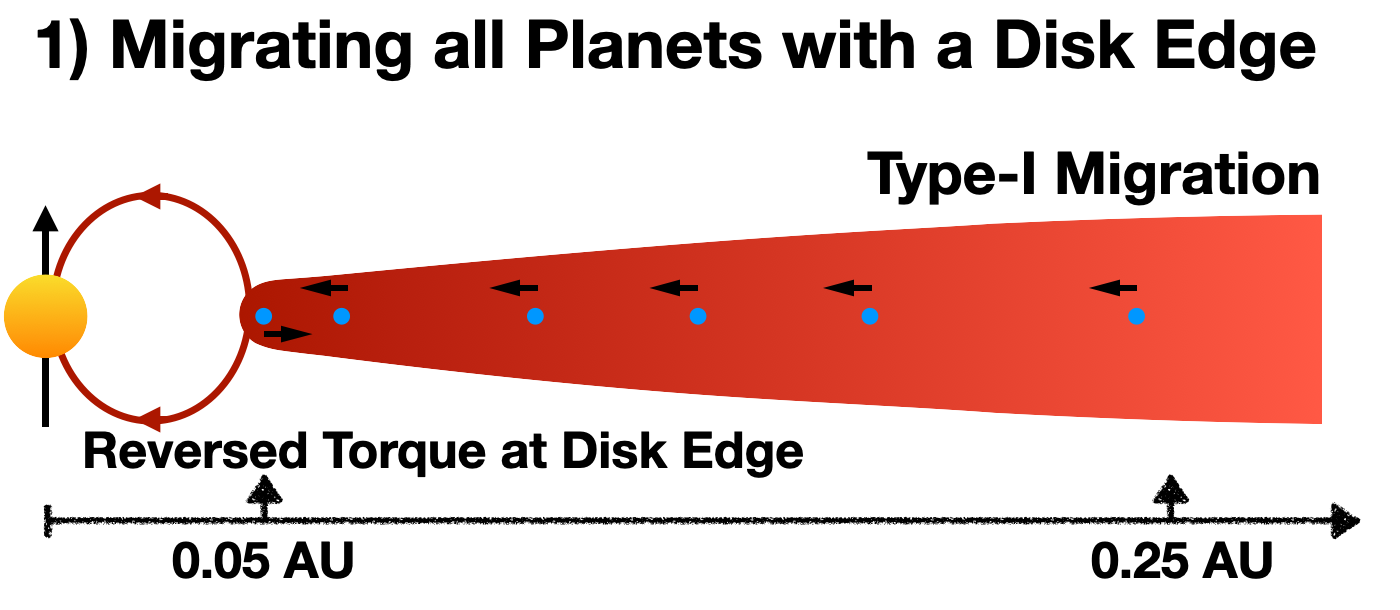}
\includegraphics[width = 0.66\columnwidth]{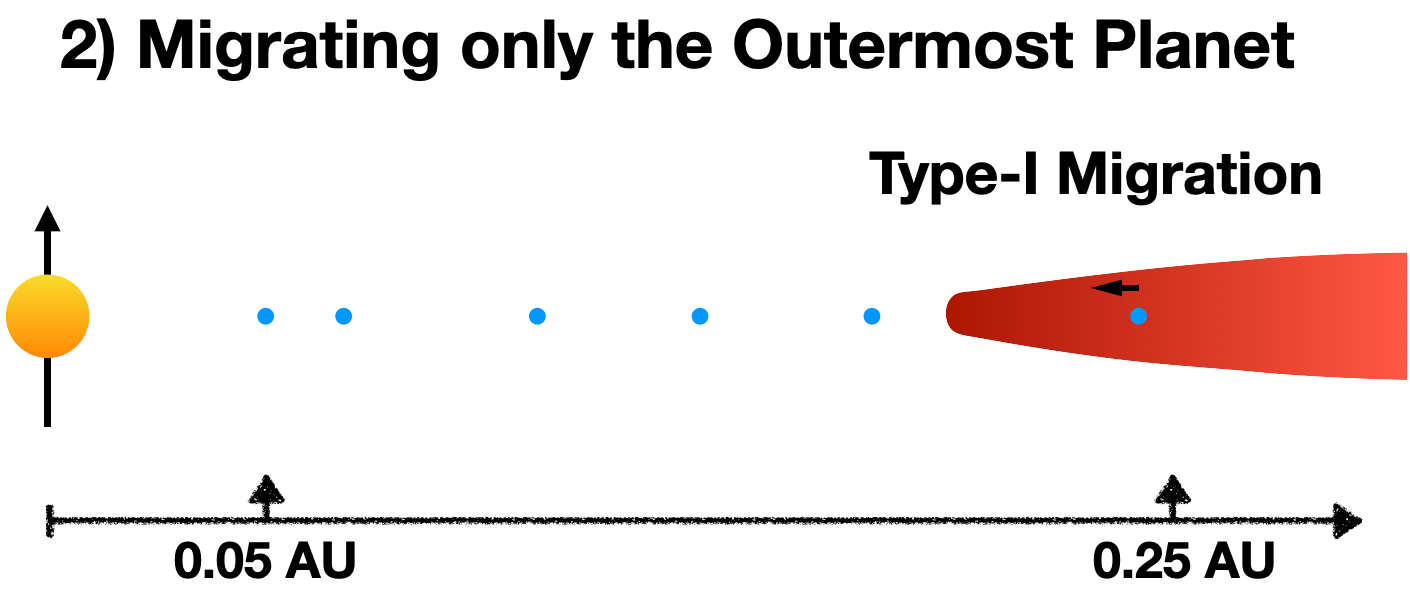}
\includegraphics[width = 0.66\columnwidth]{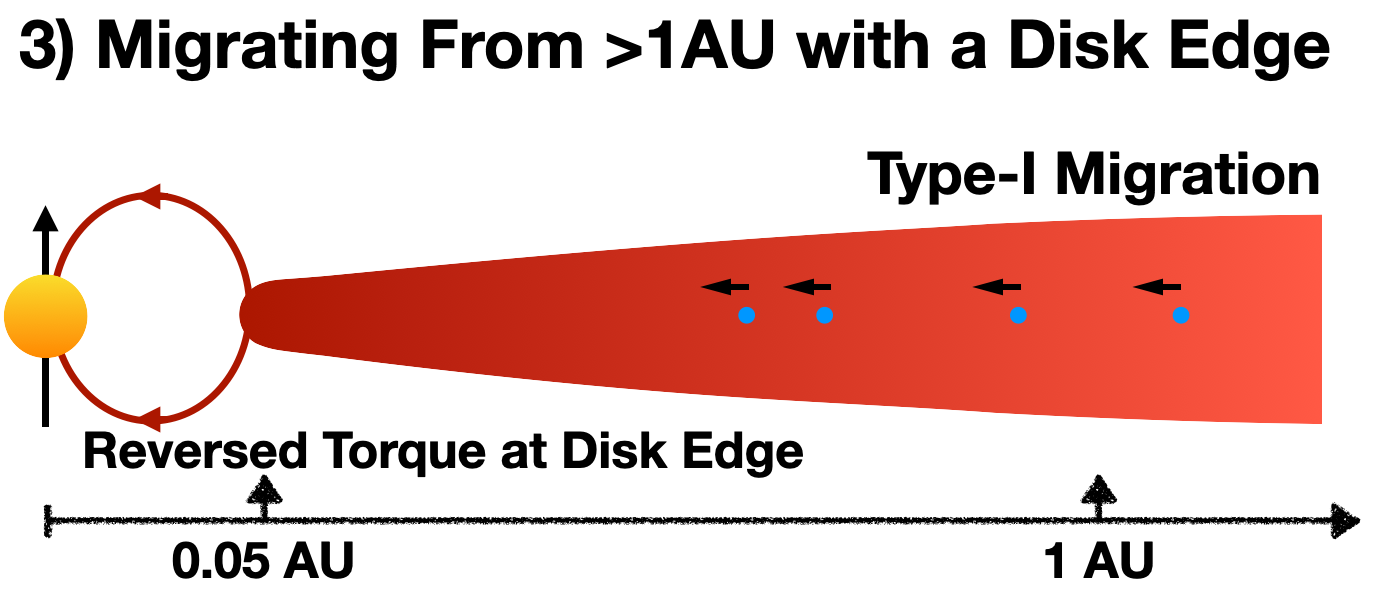}
\includegraphics[width = 0.66\columnwidth]{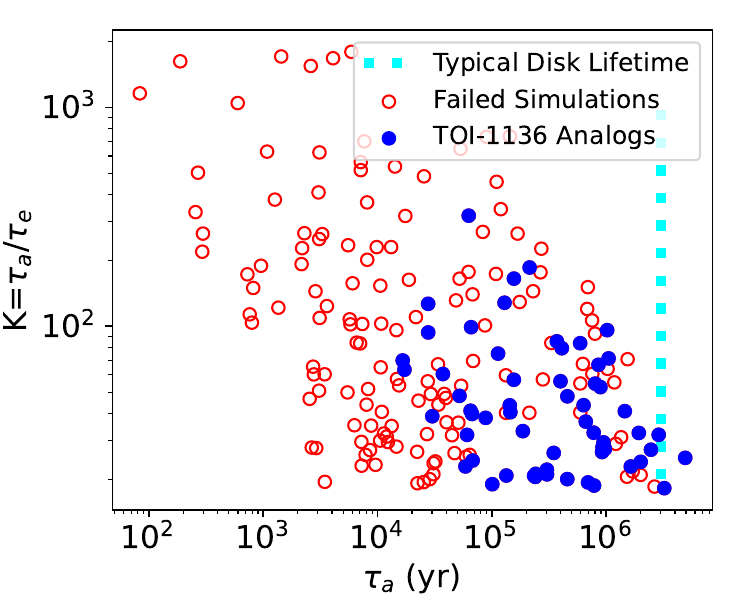}
\includegraphics[width = 0.66\columnwidth]{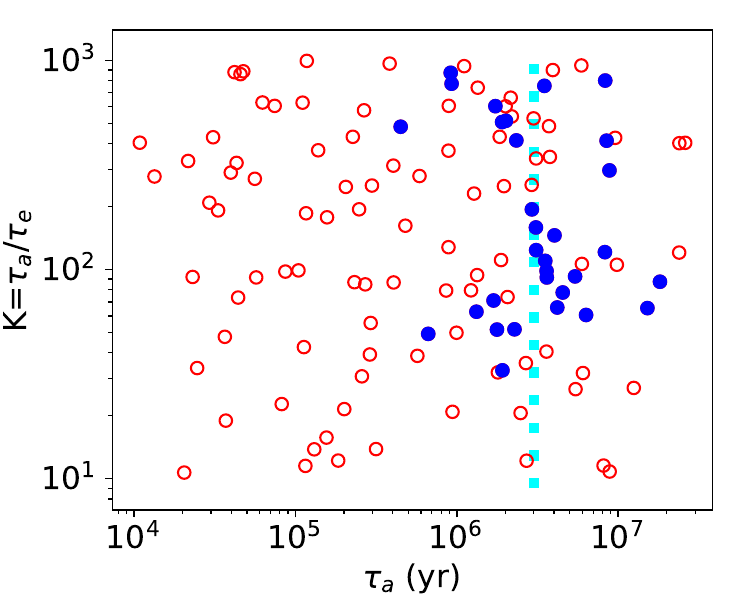}
\includegraphics[width = 0.66\columnwidth]{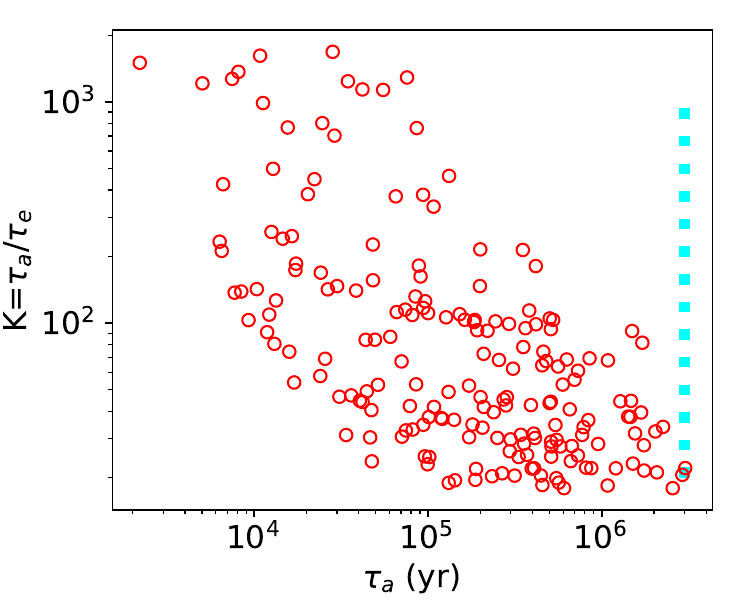}
\includegraphics[width = 0.66\columnwidth]{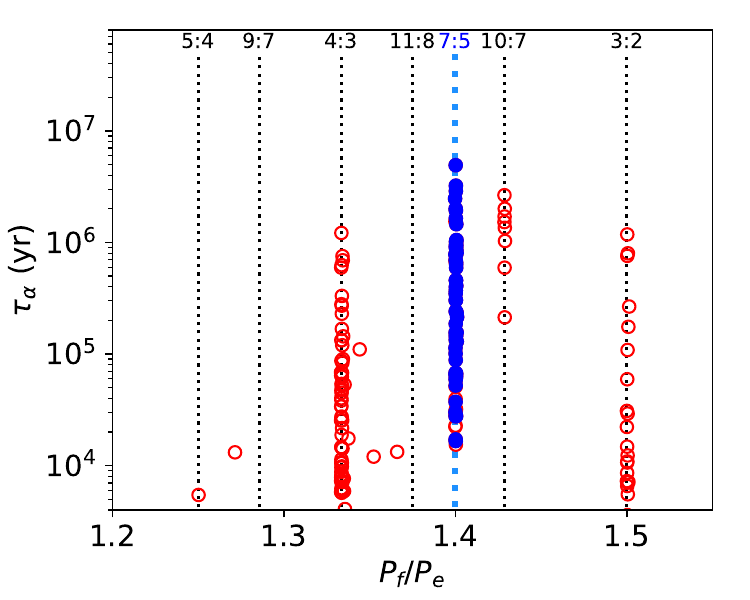}
\includegraphics[width = 0.66\columnwidth]{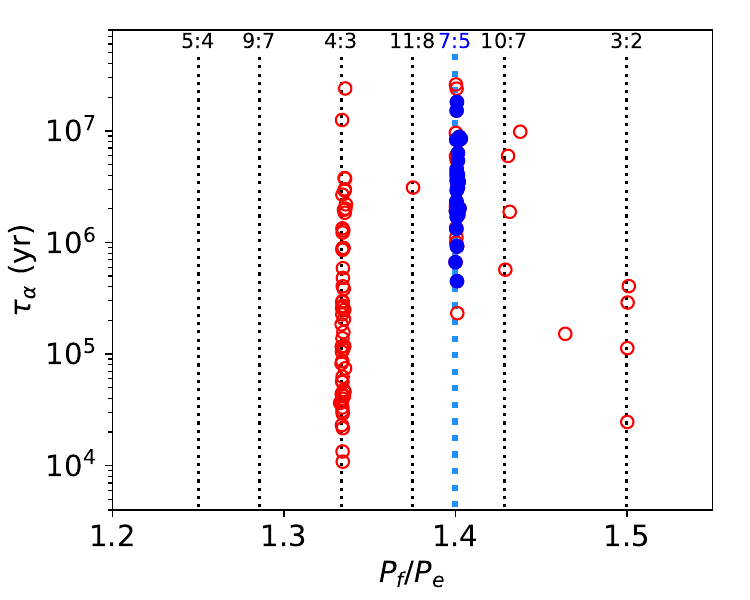}
\includegraphics[width = 0.66\columnwidth]{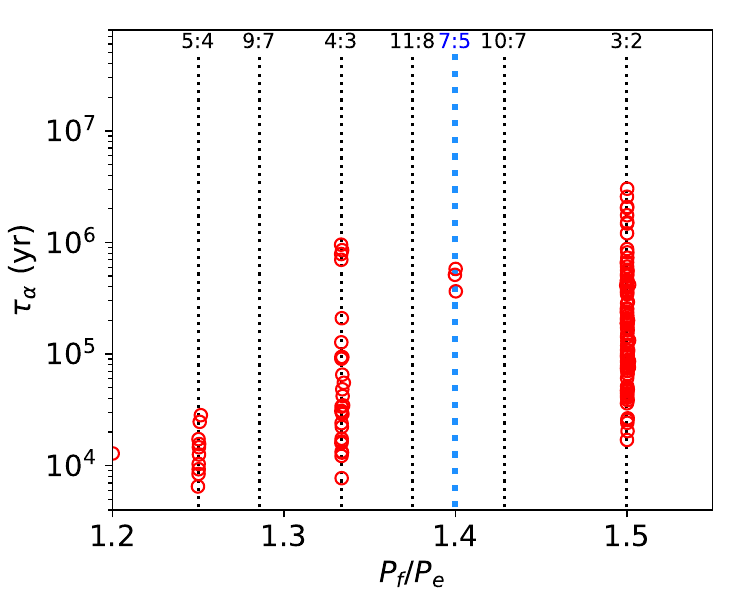}
\includegraphics[width = 0.66\columnwidth]{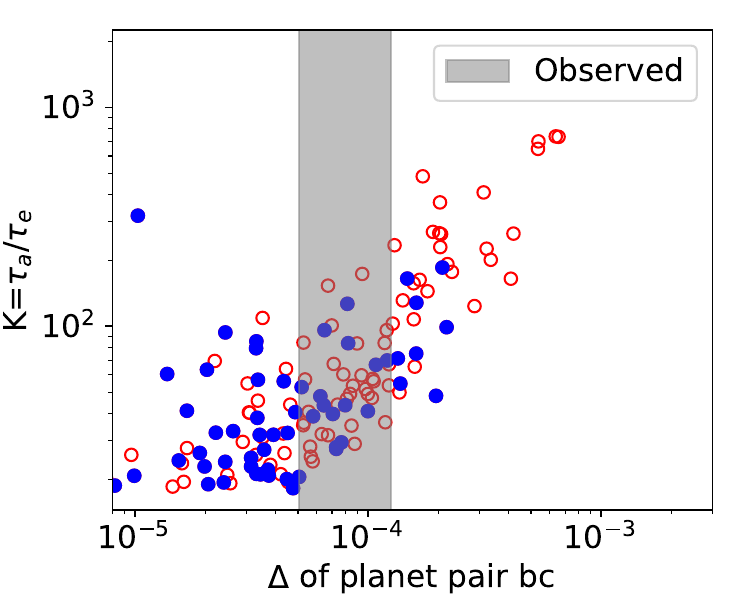}
\includegraphics[width = 0.66\columnwidth]{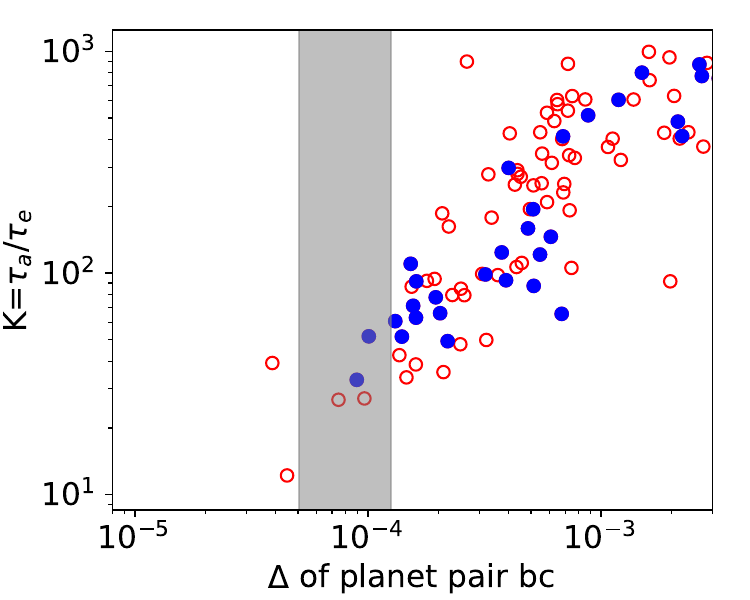}
\includegraphics[width = 0.66\columnwidth]{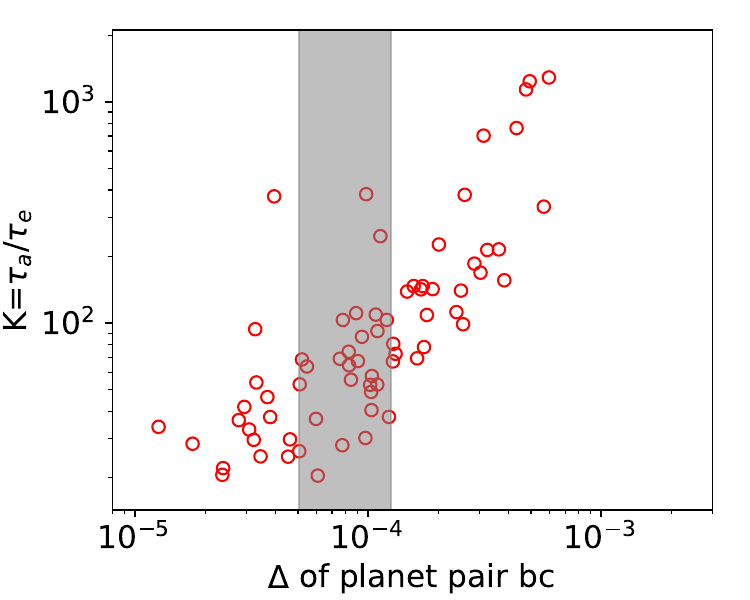}
\caption{Summary of our disk migration simulations (Section \ref{sec:migration}). We experimented with three prescriptions of disk migration: 1) we applied Type-I migration to all planets simultaneously with a disk edge (left column). 2) Type-I migration was only applied to the outermost planet (a scenario that may happen in transition disks, middle column). 3) Similar to the first case except that the planets migrated from beyond 1AU as opposed to the 0.1 AU in the previous two cases (right column). The top row shows the schematics for each mode of migration. The second row shows the results of the simulation in terms of migration timescales in $\tau_a$ and $K\equiv\tau_a/\tau_e$ compared with the typical disk lifetime \citep[$\sim$3Myr for sun-like stars,][]{Andrews_review}. The blue filled symbols are simulations that formed analogs of TOI-1136 where planets are in their observed resonances particularly with planet e and f in a second-order 7:5 MMR. The red hollow symbols are systems that have failed (usually e and f skipped 7:5 and became locked in a nearby first-order MMR). The third row shows the final orbital period ratio between planet e and f. The fourth row shows the depth of MMR produced in $\Delta$ at the end of the simulations. The gray area indicates the observed $\Delta$ between planet b and c. In general, the first prescription: short-scale (from 0.1AU) disk migration with a disk edge is the most robust at producing systems of TOI-1136. It can deposit systems deeper in MMR with $\Delta \lesssim 10^{-4}$ as was observed in TOI-1136. The migration process could be completed quickly within typical disk lifetime.}
\label{fig:migration_success}
\end{figure*}

\begin{figure*}
\begin{center}
\includegraphics[width = 1.7\columnwidth]{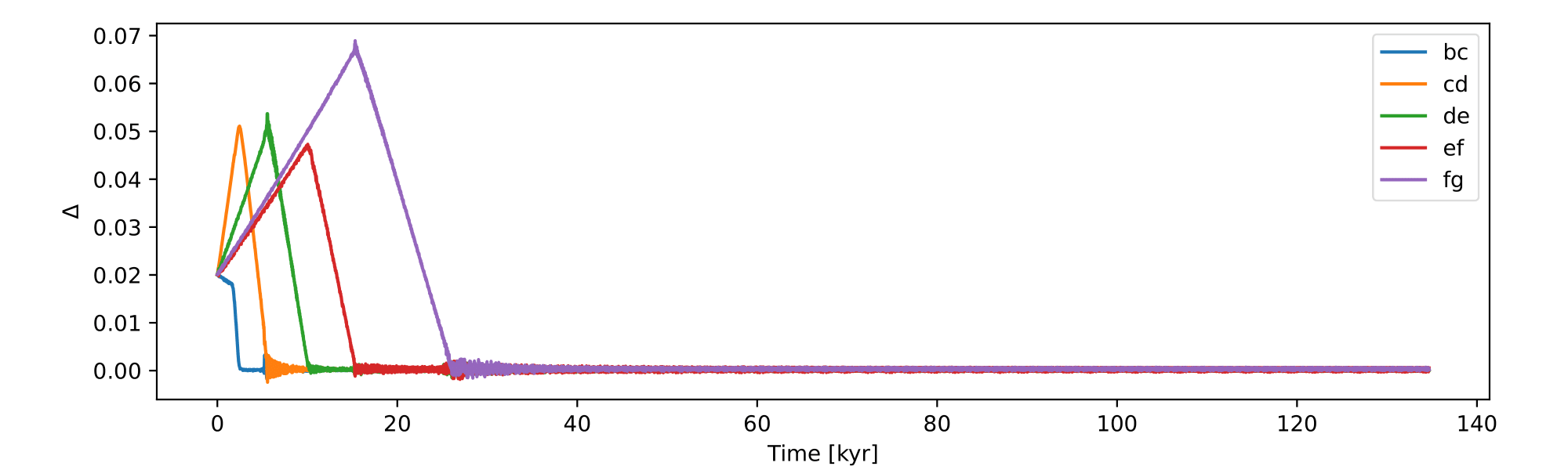}
\includegraphics[width = 1.7\columnwidth]{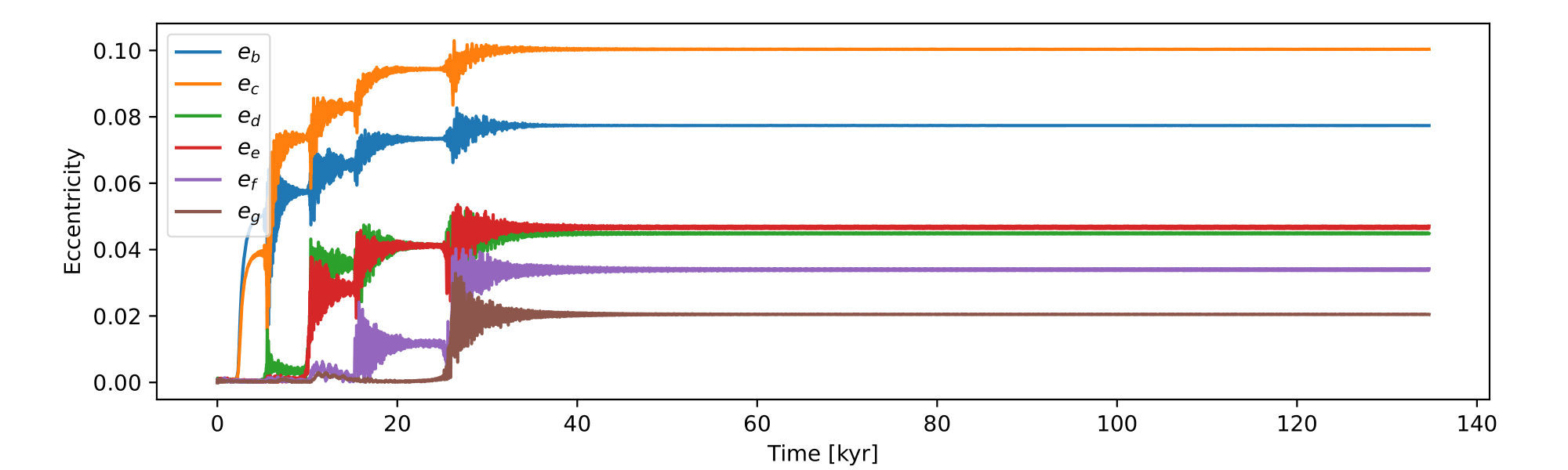}
\includegraphics[width = 1.7\columnwidth]{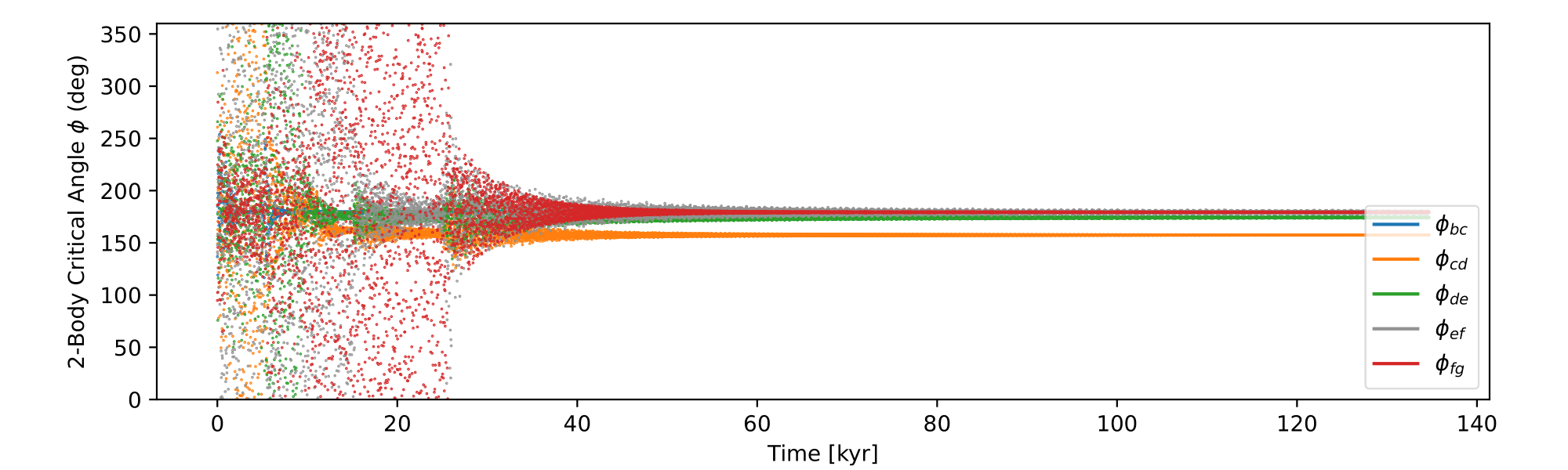}
\includegraphics[width = 1.7\columnwidth]{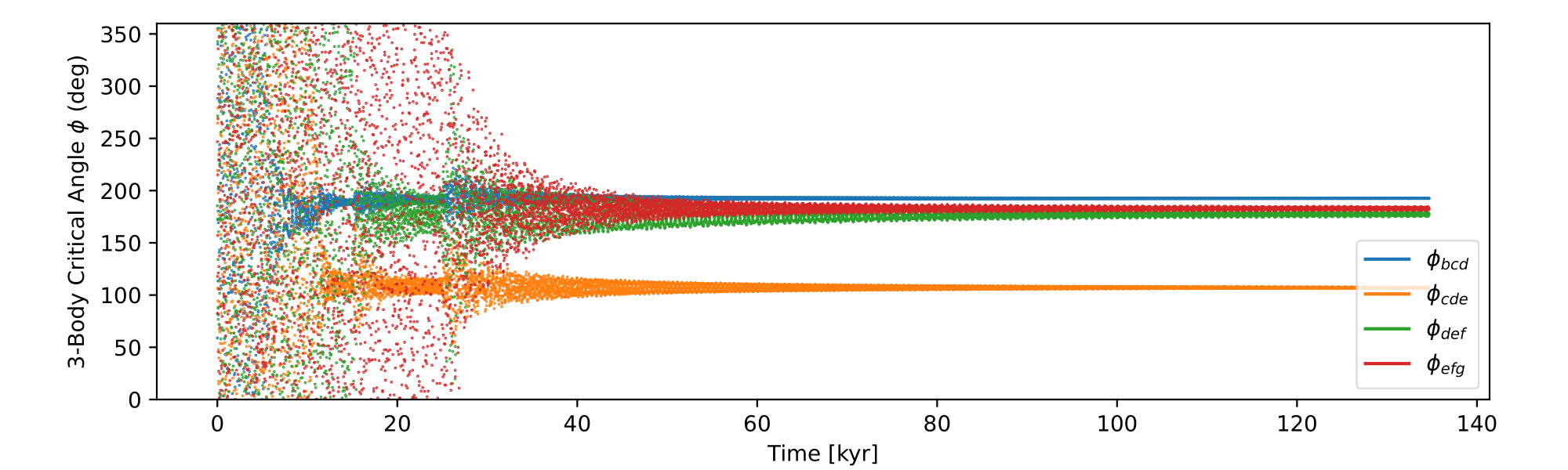}
\caption{Simulated Type-I migration where migration was applied to all planets and there is an inner edge of the disk at 0.05AU. The panels respectively shows the time evolution of the deviation from MMR $\Delta$, orbital eccentricities, 2-body resonant angles, and 3-body resonant angles. The inner disk edge halts the migration of the planets and turns initially divergent encounters into convergent encounters (first panel). As shown here, the system quickly captured into a resonant chain including the second-order resonance for planet e and f on a timescale of 10$^4$yr. Once in resonance, eccentricities are excited by resonant interaction, while the resonant angles start to librate.}
\label{fig:edge_migration}
\end{center}
\end{figure*}

We experimented three prescriptions of disk migration for TOI-1136 (see schematics in Fig. \ref{fig:migration_success}). Our first set of simulations follow the prescription of \citet{Cresswell2006}, \citet{Baruteau2014}, \citet{Pichierri} and \citet{Huhn}. Type-I migration was applied to all the planets simultaneously. The rate of migration on each planet was calculated based on the planetary properties and their current locations in the protoplanetary disk \citep[for details see Section 3 of][]{Pichierri}. This procedure was implemented in the {\tt type\_I\_migration} routine of {\tt REBOUNDx} \citep{Tamayo_x}. Crucially, to halt the migration and prevent planets from plunging into the host star, we included an inner edge of the disk in the simulations. The existence of an inner edge in a protoplanetary disk at the co-rotation radius is theoretically expected \citep[e.g.,][]{Ghosh,Ostriker}. Observationally, the inner edge may also be responsible for the decline of sub-Neptune occurrence inward of 10 days \citep[e.g.,][]{Terquem_2007,Lee}. The location of the inner edge was set to be 0.05 AU (near the current orbit of TOI-1136 b), with a transition region of 0.01 AU over which the migration torque is smoothly reversed to mimic the effect of the pressure bump. Planet b was initialized 5\% outside its currently observed orbit. The other planets were initialized with orbital separations such that each pair has a period ratio 2\% wider than their currently observed resonances. This is to represent {\it in-situ} formation of the planets followed by short-scale ($\sim0.1$AU) migration. The planetary masses were taken from the stable TTV posterior samples. The only exceptions are planets e and f, which were assigned a mass ratio $0.9<q<1.1$ and the same mass scale from TTV solutions. As suggested by \citet{Xu0217}, having a mass ratio near unity maximizes the chance of establishing and maintaining a second-order MMR. The planets had initially circular orbits and randomized arguments of pericenter and mean anomalies. The main tunable parameters in this simulation are the surface density of the protoplanetary disk at 1 AU ($\Sigma_{\rm 1AU}$) and the scale height $h\equiv H/R$. We assumed
$\Sigma = \Sigma_{\rm 1AU}~a^{-1.5}$ and varied $\Sigma_{\rm 1AU}$ between 10 to 10$^4$ g~cm$^{-2}$ uniformly in logarithmic space. Thus, the simulated disks have surface densities that between about $1/200$ and 10 times that of the minimum-mass solar nebula \citep[$\Sigma_{\rm 1AU}\approx$1700 g~cm$^{-2}$]{Weidenschilling,Hayashi} \footnote{see also the minimum-mass extrasolar nebulae, whose surface densities have substantial variation between different systems \citep{Chiang,Dai_mmen}}. $h$ was randomly chosen between 0.01 and 0.1, and was assumed to be a constant throughout the disk (no disk flaring). For easier comparison with the typical disk lifetime of sun-like star \citep[$\sim3$ Myr see e.g.][]{Andrews_review}, we converted [$\Sigma_{\rm 1AU}$, $h$] to [$\tau_a$, $\tau_e$] the decay timescale of the semi-major axis and orbital eccentricity using the equations in \citet{Pichierri}. The whole system was evolved for 3 $\tau_a$; visual inspection of the time evolution confirmed that all planets have had ample time to complete migration and settle into MMR (Fig. \ref{fig:edge_migration}).

Our second prescription of disk migration is widely used in the literature: e.g. \citet{Tamayo2017} employed this method to successfully simulate the formation of TRAPPIST-1 \citep{Gillon}. In this prescription, Type-I migration was only applied to the outermost planet. The benefit is that all encounter between the planets are now convergent: the inner planets do not migrate until they are captured in resonances with the outer planets. This prescription may seem contrived, however it may be the case in transition disks \citep{Espaillat} where the inner gas disk is starting to disperse (see schematic Fig. \ref{fig:migration_success}). There can be a time at which only the outermost planet is still embedded in a gas disk and experiences Type-I migration. We dynamically evolved the system using {\tt REBOUND} with the {\tt WHFAST} integrator \citep{Rein}. The effect of Type-I migration was implemented using the {\tt modify\_orbits\_forces} routine in {\tt REBOUNDx} \citep{Tamayo_x}. Since we are migrating just one planet, we directly varied $\tau_a$ uniformly in logarithmic space between 10$^4$ and 10$^7$ yr. Instead of varying $\tau_e$ directly, we varied $K\equiv\tau_a/\tau_e$ between 10 and 1000 (right panels of Fig.~\ref{fig:migration_success}). $K$ bears theoretical significance that will be explained shortly.

Our third prescription is almost identical to the first prescription. We applied Type-I migration to all planets simultaneously and we included an inner disk edge. The only difference is that the planets were initially placed further out in the disk ($>1$AU). This prescription specifically investigate the {\it ex-situ} formation of the TOI-1136 planets followed by large-scale migration.

Fig. \ref{fig:edge_migration} shows the time evolution of the period ratios, orbital eccentricities, and resonant angles in a successful disk simulation using the first prescription. The planets were locked into their observed MMR on 10s-kyr timescales. Once in resonance, the resonant angle changed from a state of circulation to libration. Even though the planets started on circular orbits, resonant interaction can pump up the eccentricity. Another well-known result of convergent disk migration is that the deviation from MMR $\Delta \equiv \frac{P_{out}/P_{in}}{p/q}-1$ and the equilibrium orbital eccentricity $e$ are inversely related \citep[e.g.,][]{Ramos}. The inverse relation is determined by the ratio between semi-major axis and eccentricity damping timescales $K\equiv\tau_a/\tau_e$. During disk migration, $e$ is damped down by the disk and is pumped up by resonant interaction. The equilibrium eccentricity is given by the balance of the $e$ pumping and $e$ damping \citep{TerquemPapaloizou2019}. In a Sessin-type resonant Hamiltonian \citep{Sessin}, if we ignore secular interaction and work in the limit of small $e$, the argument of pericenter precesses at a rate $\dot{\omega}\propto1/e$ for a planet in MMR \citep[see also][]{Laune2022}. For a pair of planets to remain in MMR, the period ratio has to deviate away from MMR ($\Delta$ increases) such that the conjunctions shift spatially in pace with the precession of pericenters: $\dot{\phi}_{\rm 12}=qn_1-pn_2+\dot{\varpi} \approx0$. Our disk migration simulations recovered this general behavior (bottom row of Fig. \ref{fig:migration_success}). A smaller $K\equiv\tau_a/\tau_e$, slower damping of orbital eccentricity, leads to larger equilibrium $e$ and smaller deviation from MMR $\Delta$ (Fig.~\ref{fig:migration_success} and Fig.~\ref{fig: repulsion_e_delta}). To reproduce the observed $\Delta$ of $\sim 10^{-4}$ (gray area in Fig.~\ref{fig:migration_success}), $K\equiv\tau_a/\tau_e$ has to be smaller than about 100.

We carried out about 200 simulations for each prescription. This was not an exact number as some realizations went unstable. The results are summarized in Fig.~\ref{fig:migration_success}. We consider a simulation successful if all six planets get locked into their observed MMR with no more than 0.1\% deviation; and the respective 2-body and 3-body resonant angles are all librating. The most common failure mode is that the planets e and f skip the weaker second-order 7:5 MMR and gets locked in the nearby stronger first-order MMR (4:3 and 3:2, see third row of Fig. \ref{fig:migration_success}).  Our simulations disfavored the third prescription: long-scale (from $1$AU to 0.05AU) Type-I migration. None of 200 simulations with this prescription managed to form a system like TOI-1136. \citet{Xu0217} found that the capture into second-order resonance is more likely with slower migration (see their Eqn.~44). There is a paradox here if the planets experienced long-scale migration, their migration rate must be high enough so that they can arrive at the observed 0.05AU separation before the disk dissipates after $\sim3$Myr. On the other hand, the weak 7:5 second-order resonance is easily skipped during fast migration. Even though in some realizations planet e and f get initially captured into 7:5 MMR, $1$AU to 0.05AU is such a long journey that perturbations form the other planets eventually disrupted the weak 7:5 MMR.

Our two short-scale (0.1AU) migration prescriptions both abundantly produce TOI-1136 analogs (Fig.~\ref{fig:migration_success}). However the second prescription, migrating only the outermost planet, seems less likely. To form analogs of TOI-1136, the second prescription often requires slower migration with timescales of several Myr that often exceeds typical disk lifetime (second row of Fig.~\ref{fig:migration_success}). One may argue that in transition disks, the gas surface density is low enough that Type-I migration is also significantly slower. However, transition disk is short-lived leaving it little time for the migration to deposit the planets deep in resonance. In particular, the innermost planets have to wait for the outer planets to be captured into resonance sequentially before resonant interaction starts acting on it. Our simulations very rarely deposit planet b and c to the observed 10$^{-4}$ level from perfect resonance (bottom row of Fig.~\ref{fig:migration_success}).

\begin{figure*}
\begin{center}
\includegraphics[width = 1.02\columnwidth]{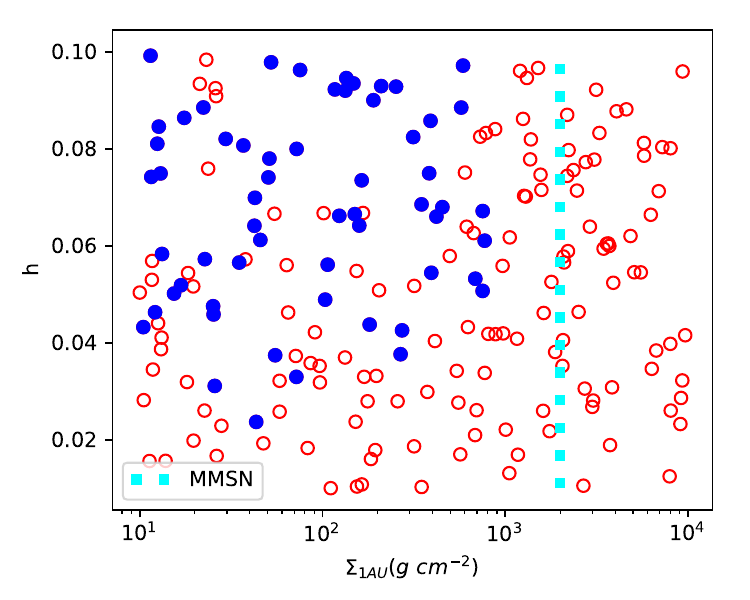}
\includegraphics[width = 0.98\columnwidth]{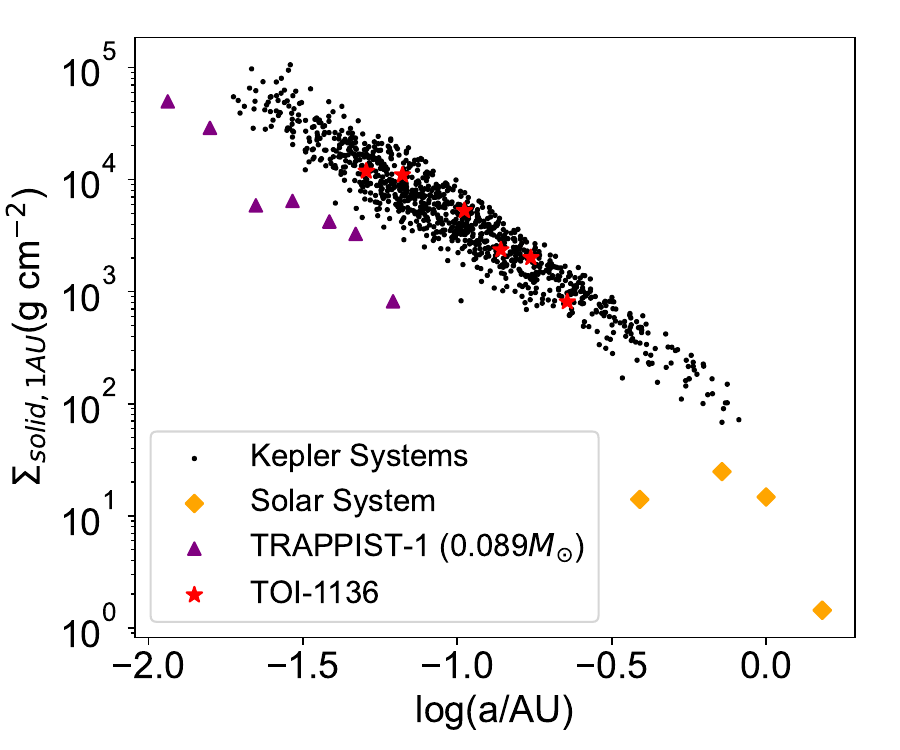}
\caption{Properties of the protoplanetary disk that formed TOI-1136. Left: the total disk surface density and the scale height (h$\equiv H/R$) of our disk migration simulation. The successful simulations (blue solid symbols) suggest that the TOI-1136 likely had a lower total surface density ($\Sigma_{\rm total, 1AU}\lesssim1000$g~cm$^{-2}$) than the Minimum-mass Solar Nebula \citep[MMSN,][]{Hayashi}. The slower migration in a lower density disk facilitated the capture into resonance particularly the 7:5 second-order resonance. Right: The Minimum-Mass Extrasolar Nebula of TOI-1136 constructed directly from the TTV-measured masses using the method in \citet{Dai_mmen}. TOI-1136 fall close to the {\it Kepler} multi-planet systems with an estimated solid surface density of $\Sigma_{\rm solid, 1AU}\approx50$g~cm$^{-2}$. The two panels together suggest an enhanced dust-to-gas ratio of at least $\gtrsim$0.05 within the innermost 1AU possibly due to radial drift and gas disk dispersal \citep[e.g.][]{Gorti,Cridland,Birnstiel}.}
\label{fig:disk_properties}
\end{center}
\end{figure*}

Our first prescription, short-scale (0.1AU) Type-I migration on all planets with a disk edge, seems to be the more likely scenario. As shown in Fig.~\ref{fig:migration_success}, the first prescription can produce systems like TOI-1136 (including 7:5 MMR) even with rapid Type-I migration of $\tau_a = 10^4-10^6$yr. This is thanks to the inner edge of the protoplanetary disk which slows down and even reverses the effective migration \citep{Masset_2006,Kretke2012}. The disk edge stalls the inner planets at the edge and thereby allows planets further out to catch up and join the resonant chain \citep{Izidoro}. As shown in the top panel of Fig.~\ref{fig:edge_migration}, even though some planet pairs initially underwent divergent migration, all planet pairs eventually switched to convergent migration and got locked into MMR. Moreover, since all planets migrated simultaneously and captured into resonance quickly, they are deposited deeper in resonance after the simulation ($\Delta$ can be as low as $10^{-5}$, bottom row in Fig. \ref{fig:migration_success}). Such deep resonances better match the observed TOI-1136 system. Within the limitations of Type-I migration prescription of \citet{Cresswell2006}, \citet{Baruteau2014}, and \citet{Pichierri}, our successful disk migration simulations translates to a protoplanetary disk no denser than $\sim 1000$~g~cm$^{-2}$ at 1AU ( Fig.~\ref{fig:disk_properties}). This is comparable but lower than the surface density of the MMSN \citep[$\approx1700$~g~cm$^{-2}$;][]{Hayashi}.

Another signpost of convergent disk migration is that neighboring planets in MMR have anti-aligned arguments of pericenters \citep[e.g.,][]{Batygin_repulsion}. This is a robust prediction of convergent disk migration as it does not depend on initial conditions. Anti-aligned pericenters have been observed in some of the known resonant chains \citep[e.g., TRAPPIST-1;][]{Agol2021}. For TOI-1136, our disk migration simulations produced anti-aligned pericenters for neighboring planets (see the top left panel of Fig.~\ref{fig:ttv_resonant_angle}). A significant fraction our TTV posterior samples are indeed consistent with an anti-aligned configuration (top center panel of Fig.~\ref{fig:ttv_resonant_angle}). However other TTV solution are not apsidall anti-aligned (top right panel of Fig.~\ref{fig:ttv_resonant_angle}). On a population level (Fig. \ref{fig: e_vector}), our TTV solutions are suggestive of the apsidal anti-alignment, however more TTV data is needed to confirm this trend.

\citet{Macdonald2018} proposed that post-formation eccentricity damping alone could also produce a resonant chain of planets. In our limited exploration of this possibility, we could only deposit the inner two or three planets of TOI-1136 into resonance. The other planets, which have much longer tidal timescale (see Section \ref{sec:repulsion}), showed negligible evolution within a 700-Myr age. We argue that post-formation eccentricity damping may explain pairs or triplets of resonant planets, however it struggles to explain a 6-planet resonant chain such as TOI-1136. Some other process, e.g. Type-I migration, is required to initialize the planets close to resonance. The observed orbital architecture of TOI-1136, particularly the depth of MMR and the 7:5 second-order MMR, is most consistent with the scenario of a short-scale (0.1 AU), Type-I migration with an inner disk edge.

\subsection{Resonant Repulsion}\label{sec:repulsion}
After the protoplanetary disk dissipates, a resonant chain of planets in a system like TOI-1136 may experience planetesimal scattering \citep[e.g.,][]{Chatterjee_2015}, orbital instabilities followed by giant impact collisions \citep[e.g.,][]{Izidoro,Goldberg2022}, secular chaos \citep[e.g.,][]{Petrovich}, and tidal dissipation \citep[e.g.,][]{Lithwick_repulsion}, all of which could move the system off resonance. If the system is lucky, it may evade giant impacts, planetesimal scattering, however some amount of tidal resonant repulsion \citep{Papaloizou2010,Lithwick_repulsion,Batygin_repulsion,Delisle2014,Pichierri2021} seems unavoidable. Resonant repulsion is well understood for a pair of planets in first-order MMR: tidal damping of both planets' eccentricities causes $\Delta$ to rise, taking the system further from MMR. This is not to be confused with a simple divergence of orbits due to the faster tidal orbital decay of the inner planet. In resonant repulsion, the outer planet moves outward. The underlying physics is almost identical to the $e$-$\Delta$-$K$ relationship we described in Section \ref{sec:migration}. Again, when the resonant interaction dominates and in the limit of small $e$, the argument of pericenter precesses at a rate $\dot{\omega} \propto1/e$. To stay in MMR, the period ratios of two resonant planets must positively deviate away from MMR to catch up with the ever faster precession of the pericenter. In short, as $e$ gets damped by tides,  $\dot{\omega}$ precesses faster and $\Delta$ has to increase to maintain the MMR. This process can continue until the resonance is broken. Again Kepler-221 is a great example \citep{Goldberg2021}.

Most of the {\it Kepler} multi-planet systems are not near first-order MMR. There is only a small overabundance just wide of first-order resonances and a lack of planets just short of them \citep{Fabrycky2014}. See also Fig.~\ref{closeness_MMR}. A number of works have explored whether this preponderance of wide-of-resonance systems could be produced by resonant repulsion \citep[e.g.,][]{LeeMH,Silburt}. The general conclusion is that with only eccentricity tides, resonant repulsion is too slow to explain the entire {\it Kepler} sample. \citet{Millholland_obliquity} pointed out that obliquity tides may solve this problem by enhancing the rate of tidal dissipation. Regardless of the source of dissipation, the long-term asymptotic behavior of resonant repulsion is the same, as long as the process does not break the resonance (or the Cassini state for the case of obliquity tides; \citet{Batygin_repulsion}). The long-term asymptotic behavior obeys a power-law relation: $\Delta \propto (t/\tau_e)^{1/3}$ \citep[e.g., Eqn.~26 of][]{Lithwick_repulsion}.

We simulated the resonant repulsion for TOI-1136. The initial conditions are our disk migration simulations that successfully locked all six planets of TOI-1136 into their observed MMR (Section \ref{sec:migration}). We integrated these systems forward in time in {\tt REBOUNDx} \citep{Tamayo_x}. We used the symplectic {\tt WHFAST} integrator \citep{1991AJ....102.1528W}. We included tidal damping on all planets using the {\tt modify\_orbits\_forces} routine in {\tt REBOUNDx}.  We parameterized $\tau_e$ using the equilibrium-tide expression \citep{Murray}
\begin{equation}\label{eqn:tides}
    \tau_e = \frac{2}{21n}\frac{Q}{k_2}\frac{m_p}{m_\star}\left(\frac{a}{r_p}\right)^5
\end{equation}
where $n$ is the mean motion of the planet; $Q$ is the tidal quality factor; $k_2$ is the tidal Love number; $m_\star$ is the stellar mass; and $m_p$, $r_p$, and $a$ are the planetary mass, radius, and semi-major axis, respectively.

To guide our discussion, we first examine the theoretical behavior of resonant repulsion for each pair of planets in TOI-1136 \citep{Lithwick_repulsion}. This calculation assumes the planets are only in pairwise first-order resonance. According to their Eqn.~26, $\Delta$ grows as $(t/\tau_e)^{1/3}$ with a proportionality that changes with planetary parameters and the relevant resonance. The process of resonant repulsion is independent of the absolute scale of $\tau_e$ as long as the system is maintained in resonance. This is why \citet{Goldberg2021} were able to use a $\tau_e$ of just 10 years to speed up their numerical investigation of Kepler-221. The situation is more complicated for a resonant chain of planets: the effect of tidal damping on individual planets will be transmitted other planets by their resonant interaction. Resonant repulsion could proceed for all resonantly-locked planets even though tides may only operate strongly on the inner, larger planets.

With six planets in TOI-1136, each of which may have a different reduced tidal quality factor $Q^\prime \equiv Q/k_2$, there
are too many possibilities to consider. For simplicity, we assumed that all planets have the same $Q^\prime$ but different $\tau_e$ given by Eqn.~\ref{eqn:tides}. In this case, planets b, c, and d have comparable $\tau_e$ $\sim5$ Gyr if the planets have Neptune-like $Q^\prime\approx3\times10^4$ \citep[e.g.,][]{Banfield,Zhang_neptune}. Planets e, f, and g have $\tau_e$ that are longer by at least two orders of magnitude. However, we also tried to decrease the $Q^\prime$ of planet b by two orders of magnitude than the other planets. This possibility was
entertained because planet b is plausibly rocky ($1.9R_\oplus$), which would make it much more dissipative than a gaseous planet. We also explored the possibility that planets d and f may have $Q^\prime$ smaller than the other planets by two orders of magnitude. The motivation is that d and f are the largest planets; perhaps their radii are inflated
by the heat of obliquity tidal dissipation.

\begin{figure*}
\begin{center}
        \includegraphics[width=1.8\columnwidth]{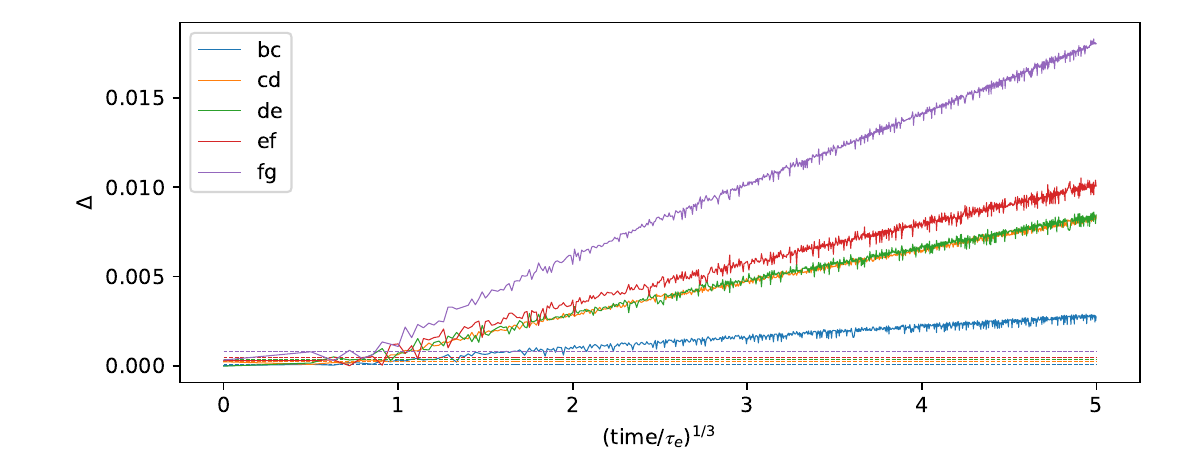}
\caption{The time evolution of the deviation from MMR $\Delta$ in our resonant repulsion simulations of TOI-1136 (Section \ref{sec:repulsion}). We dynamically evolved the resonant chains generated from our convergent disk migration simulations (Section \ref{sec:migration}) after including tidal dissipation. 
We note a characteristic behavior in which $\Delta$ grow with $(t/\tau_e)^{1/3}$ is seen for all planet pairs as long as they remain in a resonant chain. The x-axis is plotted with $\tau_e$ of the most dissipative planet. Even though tidal dissipation may be concentrated on the most dissipative planet (usually planet b), resonant repulsion occurs on all planet pairs as long as they remain a resonant chain. Depending on the masses of the planets, the rate of resonant repulsion are typically $\frac{\Delta}{(t/\tau_e)^{1/3}}$ of order $10^{-3}$ (solid lines). However, the observed deviations are of order $10^{-4}$ (horizontal dotted lines). The intersection between solid lines and horizontal lines tell us the number of $\tau_e$ cycle that have elapsed in the 700-Myr lifetime of TOI-1136. As shown in the plot, the intersection happened at small $(t/\tau_e)^{1/3}$ indicating minimum tidal evolution since formation. We described in the text that we rule out a few scenarios that would enhance the rate of tidal dissipation.}
\label{fig:repulsion}
\end{center}
\end{figure*}

 We integrated the TOI-1136 for about 100 $\tau_e$ of the most dissipative planet to determine the asymptotic behavior of resonant repulsion. The qualitative behavior is similar regardless of the choice of $Q^\prime$ and is shown in Fig.~\ref{fig:repulsion}. The theoretical $\Delta \propto (t/\tau_e)^{1/3}$ behavior held up well even though the TOI-1136 is in a resonant chain rather than a resonant pair, for which the theory was originally derived \citep{Batygin_repulsion,Lithwick_repulsion}. We experimented with $\tau_e$ between 10$^3$ to 10$^5$ yr, and confirmed that the qualitative behavior stayed the same. We computed using Eqn.~26 of \citet{Lithwick_repulsion} that planet pairs bc, cd, de and fg would deviate from MMR $\Delta$ by about $0.0005$ to $0.004$ after 1 $\tau_e$. $\Delta$ would double these amounts after 8$\tau_e$ given the $1/3$ power law. {\tt REBOUNDx} simulations revealed consistent rates of resonant repulsion of $0.0006$ to $0.004$ for various planet pairs (Fig.~\ref{fig:repulsion}). The precise values depend on the TTV-measured masses.

The observed deviations from MMR ($\Delta$) coupled with an age estimate for the system can be used to put constraints on the tidal quality factor $Q^\prime$ of the planets \citep{Lithwick_repulsion,Brasser2021}. Compared to the other known systems with resonant chains, TOI-1136 is young, with an estimated age of 700 Myr. In Fig.~\ref{fig:repulsion}, we plotted the evolution of $\Delta$ as a function of time; again note the the long-term asymptotic behavior is $\Delta \propto (t/\tau_e)^{1/3}$. In theory, the intersection of the currently observed $\Delta$ (horizontal dashed lines) and the resonant repulsion $\Delta \propto (t/\tau_e)^{1/3}$ power law (solid lines) could provide an empirical estimate of $\tau_e$. However, the $\Delta \propto (t/\tau_e)^{1/3}$ relation only holds asymptotically for long-term evolution (Fig.~\ref{fig:repulsion}). Due to other terms in the Hamiltonian,  the early-time behavior deviates significantly from a perfect $(t/\tau_e)^{1/3}$ relation. Nonetheless, we can see that intersection happened early on with $(t/\tau_e)^{1/3}\lesssim1$. In other words, the 700-Myr-old TOI-1136 has barely undergone a single $\tau_e$ of the most dissipative planet. We can hence rule out an Earth-like or Mars-like $Q^\prime$ for planet b (1.9$R_\oplus$). If b had a terrestrial $Q^\prime$ of 1000 \citep{Murray}, $\tau_e$ would have been $\sim 120$ Myr. About 5 $\tau_e$ cycles have elapsed in TOI-1136's lifetime, the $\Delta$ would have been significantly higher than the observed value. We summarize a few representative cases in Tab \ref{tab:q_comparison}.

\begin{deluxetable*}{cccc}
\tablecaption{Rate of Resonant Repulsion \label{tab:q_comparison}}
\tablehead{
\colhead{Scenario} & \colhead{$\tau_e$ of planet b} &  \colhead{Simulated $\Delta$ after 700 Myr $^1$} & \colhead{Largest Observed $\Delta$}}
\startdata
Earth-Like ($Q/k_2 \approx 100$) &$12$ Myr & $1.4\%$ & $\lesssim0.08\%$ \\
Mars-Like ($Q/k_2 \approx 1000$) &$120$ Myr & $0.7\%$ & $\lesssim0.08\%$  \\
Neptune-Like ($Q/k_2 \approx 30000$) &$3.8$ Gyr & $0.1\%$ & $\lesssim0.08\%$  \\
\enddata
\tablecomments{1. Deviation from MMR $\Delta$ after 700 Myr of resonant repulsion. Reported here is the planet pair that shows the fastest deviation. See text for detail.}
\end{deluxetable*}

\begin{figure*}
\begin{center}
\includegraphics[width = 1.9\columnwidth]{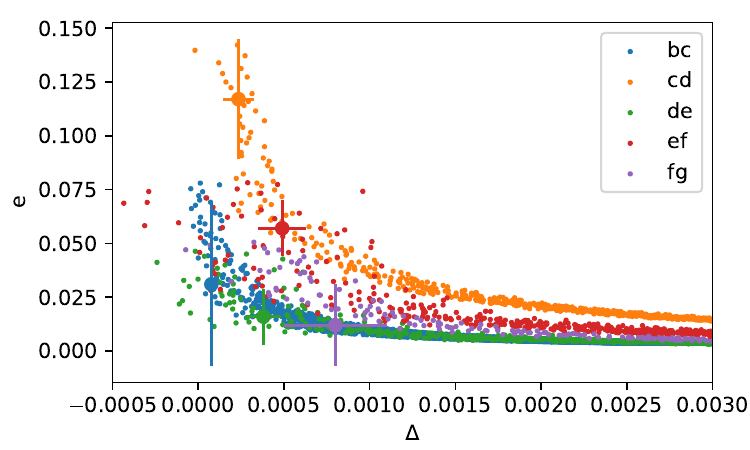}
\caption{The evolution of deviation from MMR $\Delta$ and the orbital eccentricity provides a different perspective on resonant repulsion (c.f. Fig. \ref{fig:repulsion}). A resonant planetary system starts with substantial eccentricity and deep in resonance right after disk migration (upper left corner of this $\Delta$-$e$ parameter space). Over time the orbital eccentricities are damped by tides, the deviation $\Delta$ from MMR increases. Qualitatively, this is because the precession of pericenter scales inversely with eccentricity in Sessin-type Hamiltonian \citep{Sessin}. To maintain resonance, the orbital period has to deviate from perfect integer ratio to catch up with the precession. The measured constraints on $\Delta$ and $e$ from our TTV analysis are shown by the errorbars. The high-e and low-$\Delta$ TOI-1136 system likely has not undergone much resonant repulsion since formation. \citet{Brasser2021} made a similar plot for TRAPPIST-1. With $e\lesssim 0.005$ and $\Delta\gtrsim$1\%, TRAPPIST-1 is a mature (a few Gyr old) system that has likely evolved to the bottom right of this $\Delta$-$e$ parameter space.}
\label{fig: repulsion_e_delta}
\end{center}
\end{figure*}

The constraints on orbital eccentricity from our TTV analysis also shed light on the progress of resonant repulsion. In $e$-$\Delta$ space (Fig.~\ref{fig: repulsion_e_delta}), each planet follows an evolution track that is anti-correlated in $e$ and $\Delta$. The underlying physics was explained at the start of this section. Soon after the convergent migration, the system was deep in resonance ($\Delta$ can be as low as $10^{-5}$ in our disk migration simulations) with large orbital eccentricities ($e\approx0.1$). As tides operate, eccentricities get damped and resonant repulsion drives the system towards larger $\Delta$. We plotted the measured $e$ and $\Delta$ constraints from our TTV analyses in Fig.~\ref{fig: repulsion_e_delta}. The relatively high $e$ and small $\Delta$ in our TTV solutions are consistent with the very end of disk migration or the very start of resonant repulsion. In other words, TOI-1136 has undergone very minimal resonant repulsion and still records the orbital architecture from disk migration. Even the most dissipative planet in TOI-1136 likely has $\tau_e$ that is at least $700$ Myr if not much longer. For example, if all of the planets in TOI-1136 have Neptune-like $Q^\prime\approx 3\times10^4$, $\tau_e$ would be at least 4 Gyr, and one would not expect to see significant resonant repulsion in its 700-Myr lifetime. In contrast, most {\it Kepler} near-resonant TTV systems, typically a few Gyr old, have $\Delta\approx1\%$, and damped eccentricity $e\approx0.02$ \citep[e.g.,][]{HaddenLithwick2014}. They have likely undergone many cycles of tidal damping thanks to perhaps obliquity tides \citep{Millholland_obliquity} or other mechanisms. 

Planetesimal scattering can also induce deviations from MMR \citep{Chatterjee_2015}. One can put an upper limit on the integrated amount of planetesimal scattering based on the extremely deep resonances observed in TOI-1136. For Kepler-223, \citet{Moore} found that there could not have been more than one Mars mass of orbit-crossing planetesimals or the systems would have been pulled out of resonance. Similarly, \citet{Raymond2021} investigated the same question for TRAPPIST-1. TOI-1136 may be amenable to a similar investigation, which is left for a future work.

\section{Discussion}\label{sec:discussion}
\subsection{A system deep in resonance }\label{sec:commensurability}
TOI-1136 is a deeply resonant planetary system. We now compare it with other known multi-planet systems. We only included planets discovered by the transit method in this comparison because orbital periods are much more precisely measured in transit surveys than in other types of surveys. We did not include the {\it TESS} Objects of Interest \citep[TOI; i.e.,][]{Guerrero} because TOIs typically have much shorter observational baselines, hence the orbital periods are not as precisely measured as in the {\it Kepler} mission. Moreover, many TOIs have not been confirmed yet. 

\begin{figure*}
\center
\includegraphics[width = 1.\columnwidth]{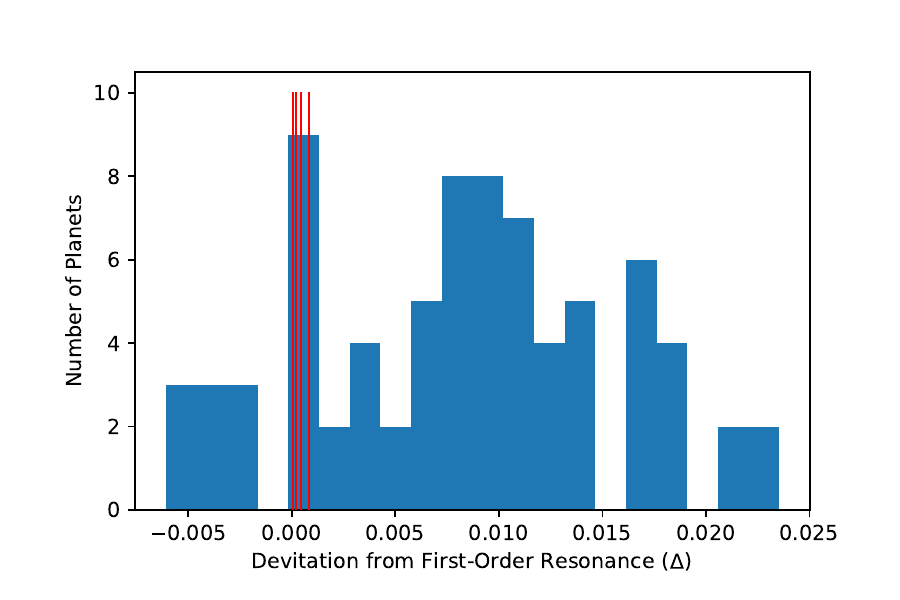}
\includegraphics[width = 1.\columnwidth]{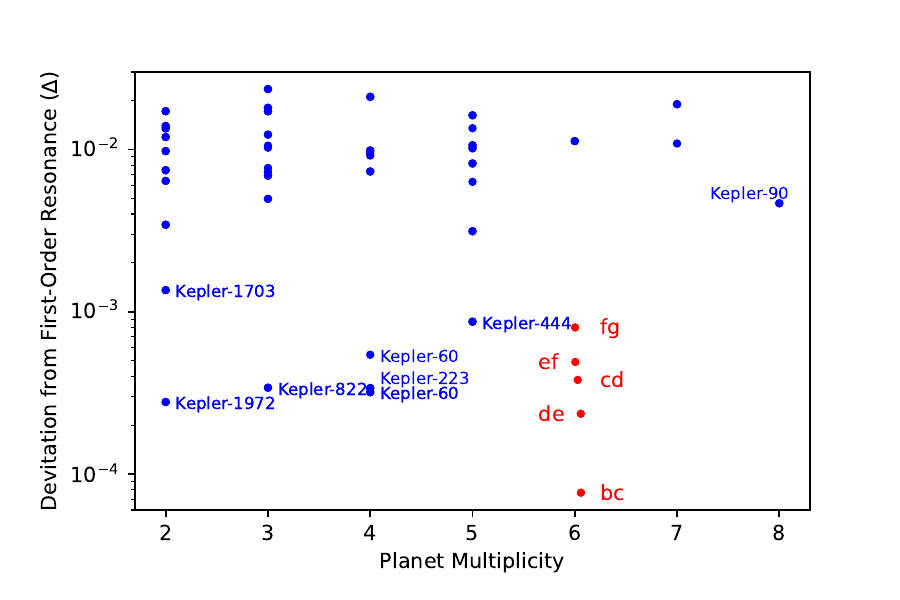}
\caption{The deviation from first order MMR ($\Delta \equiv \frac{P_{out}/P_{in}}{p/q}-1$) in TOI-1136 (red symbols) and the vetted {\it Kepler} multi-planet sample \citep[blue symbols,][]{CKS1}. Most near-resonant {\it Kepler} multi-planets have a $\Delta$ of $\sim 1\%$ from perfect integer ratio. TOI-1136 joins a small number of systems deep in resonance with a $\Delta$ $\lesssim10^{-3}$. Many other planets with similarly low $\Delta$ also have a resonant chain of planets.} 
\label{closeness_MMR}
\end{figure*}

Fig.~\ref{closeness_MMR} shows that TOI-1136 stands out as one of dozen planetary systems with orbital periods extremely close to MMR. Near-resonant {\it Kepler} multi-planets typically deviate positively from MMR by about $1$ to $2\%$ \citep{Fabrycky2014}. However, the planets orbiting TOI-1136 have $\Delta$ that are roughly two orders of magnitude smaller according to our analyses (Section \ref{sec:stability}). The other planetary systems with similarly low $\Delta$ are also resonant-chain systems, such as Kepler-60 \citep{Gozdziewski,JontofHutter2016} and Kepler-223 \citep{MillsNature}. 

\begin{figure*}
\center
\includegraphics[width = 1.6\columnwidth]{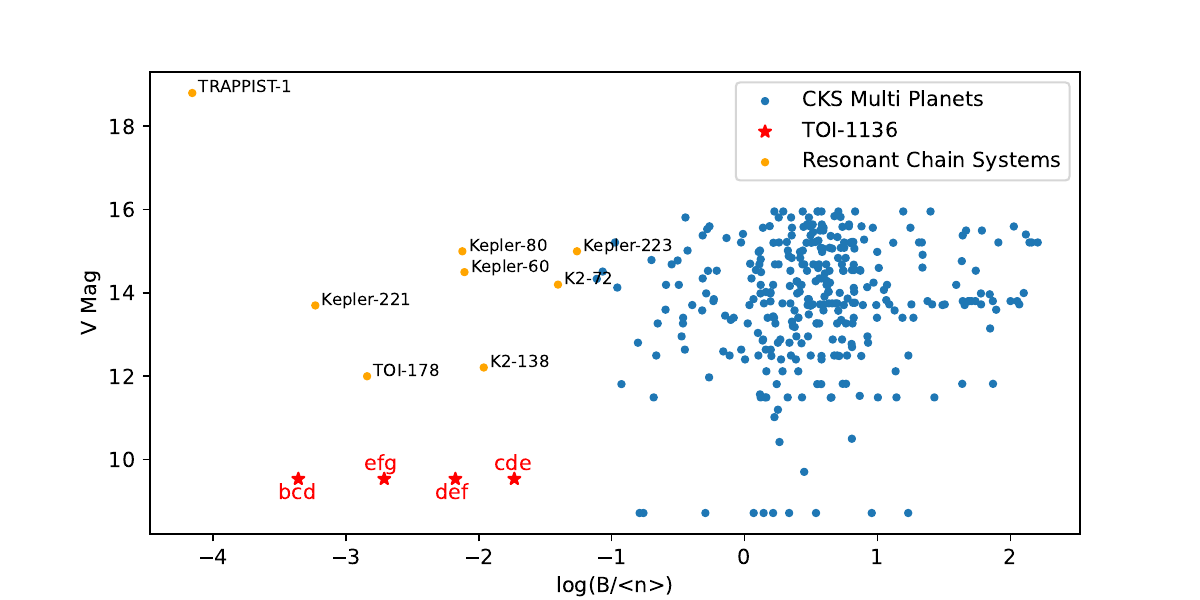}
\caption{$B$ is another metric for identifying multi-planet systems in Generalized Laplace Resonance \citep{Goldberg2021}. $B = |pn_1-(p+q)n_2+qn_3|$, where $n=2\pi/P$ is the mean motion, $p$ and $q$ are co-prime integers. For resonant systems, the resonant angle is librating, hence its time derivative $B$ should be small in magnitude. Plotted here is $B$ normalized by the average mean motion $<n>$ of the {\it Kepler} multi-planet systems (blue), TOI-1136 (red), and other known resonant chains (orange). $B$ values in TOI-1136 are similarly low as the other resonant-chain systems. TOI-1136 is by far the most observable resonant-chain system with a V-band magnitude of 9.5.} 
\label{B_scatter}
\end{figure*}

Another metric for identifying resonance was proposed by \citep{Goldberg2021}: $B = pn_1-(p+q)n_2+qn_3\approx\dot{\phi}_{123}$. It quantifies how fast the 3-body resonant angle changes with time. This metric is useful for picking out planetary systems that are in generalized Laplace resonance, in which case the resonant angle librates and $B$ is small in magnitude. In TOI-1136, the values of $B$ for the neighboring triplets bcd, cde, def, and efg are all smaller than in the general {\it Kepler} multi-planet systems by at least an order of magnitude (Fig.~\ref{B_scatter}). Again, the planetary systems with similarly small $B$ are those with resonant chains: Kepler-221 \citep{Goldberg2021}, Kepler-223 \citep{MillsNature}, Kepler-60 \citep{Gozdziewski} etc. Even without a TTV analysis, the depths of resonance among the six TOI-1136 planets seem extremely unlikely unless there is some underlying physical process that drove the planets into resonance. 

Our TTV and dynamical analyses (Section \ref{sec:transits} and \ref{sec:resonance}) provided further evidence for a resonant chain in TOI-1136. We showed that the planets of TOI-1136 display TTV on timescales that are more consistent with the libration period of the resonant angles ($700$ to $5000$ days) than the super-period or the circulation of the resonant angles ($\gtrsim10000$ days). Moreover, our stable TTV solutions predominantly showed the libration of the various resonant angles (Fig.~\ref{fig:ttv_resonant_angle}, \ref{fig:resonant_structure} and Table \ref{tab:libration}) with libration centers near the theoretically predicted values \citep{Siegel}.

\subsection{Planet ef (7:5 MMR) is the weakest link}\label{sec:ef}

The ef pair is the only second-order resonance in TOI-1136. Previous investigation has shown that second-order MMR is both much more difficult to form from disk migration and more easily disrupted than first-order MMR \citep{Xu0217}. This is because a second-order MMR, compared to first-order resonance, is suppressed by a factor of orbital eccentricity $e$. Moreover, the width of the second-order MMR in phase space is much thinner \citep[Fig. \ref{fig:resonant_structure}][]{Murray}. \citet{Mah} showed that planets b, c, and d of TRAPPIST-1 (second and third-order resonance) were often the first to be displaced from resonance in their dynamical simulations. Similarly, our dynamical modeling of TOI-1136 (Section \ref{sec:stability}) indicated that planets e and f are often the first to be dislodged from resonance. Dynamical instability often ensues after the ef pair is removed from the resonant chain.

We further experimented with the possibility that the ef pair are in the nearby 3:2 and 4:3 first-order MMR despite a period ratio that is close to 7:5 commensurability. One notable example is Kepler-221 \citep{Goldberg2021}, the planets are in a Laplace resonance even though their pair-wise period ratios deviated by >10\% from small integer ratio. For TOI-1136, we analyzed the resonant angles $\phi_{\rm ef}$, $\phi_{\rm def}$, and $\phi_{\rm efg}$ in 100 random draws of the stable TTV solutions. In all of these solutions, the resonant angles $\phi_{\rm ef}$, $\phi_{\rm def}$, and $\phi_{\rm efg}$ are circulating when computed with 3:2 or 4:3 MMR. A 7:5 second-order resonance for planet e and f is the simpler and preferred solution.

We also explored the possibility that there is an additional planet between planet e and f such that the planets are in a chain of 5:6:7 first-order MMR. The existence of such a planet would eliminate the need that planet e and f are in the much weaker 7:5 second-order MMR. In Appendix \ref{sec:additional_transits}, we showed that both systematic transit search and a careful visual inspection were not able to detect this hypothetical planet. Moreover, we calculated the mutual Hill radius for the 5:6:7 configuration. The planets are separated by only 6 mutual Hill radii even if the hypothetical middle planet is only about $1M_\oplus$. Such tight packing is seen in $<0.5\%$ of all Kepler multi-planet systems; and may compromise the overall stability of system \citep{PuWU}. Furthermore, including this hypothetical planet did not lead to an improved TTV solution. All of these results are against the possibility of another planet between planets e and f.

Therefore planets e and f are likely indeed in a 7:5 second-order MMR. This represents a weak link in the resonant chain of TOI-1136, and may threaten the overall stability as the 700-Myr-old system continues to mature. In the {\it Kepler} multi-planet sample, there is an over-abundance of planets just outside first-order resonance \citep{Fabrycky2014}. However there is no noticeable feature near second-order resonance \citep[except perhaps 5:3][]{Steffen2015}. The discovery of TOI-1136 shows that second-order MMR can be produced in at least some protoplanetary disks, as suggested by \citet{Xu0217}. If so, does it mean that the observed paucity of second-order MMR in mature planetary systems is due to the dynamical fragility of such a configuration? \citet{Izidoro} was puzzled that in order to reproduce the observed fraction of resonant systems in {\it Kepler}, at least 75\% (or even 95\%) of their simulated, initially resonant planetary systems must go unstable. However, only 50-60\% of their first-order MMR chains went unstable. Maybe the inclusion of the weaker second-order MMR could increase the rate of orbital instability. We note that in the revised models of \citet{Izidoro2021}, the fraction of unstable planetary systems could reach 95\%. Moreover in some of their simulated planetary systems contained second-order MMRs. 

\subsection{Comparison with Other Resonant Chains} \label{sec:comparison}
TOI-1136 joins a handful of known planetary systems with a resonant chain: GJ 876 \citep{Rivera,Millholland876}, TRAPPIST-1 \citep{Gillon,Luger2017,Wang_trappist,Agol2021}, TOI-178 \citep{Leleu2021}, Kepler-80 \citep{MacDonald2016}, Kepler-60 \citep{Gozdziewski}, K2-138 \citep{Christiansen2018}, Kepler-223 \citep{MillsNature}, and Kepler-221 \citep{Goldberg2021}. K2-72 \citep{Crossfield2016}, V1298 Tau \citep{David2019} as well as the {\it Kepler} systems labeled in Fig.~\ref{closeness_MMR} might also have resonant chains, pending
further analysis. \citet{Tejada} argued that V1298 Tau cannot be in resonance based on stability considerations.

TOI-1136 is the second known resonant-chain system with a well-established age as young as a few hundred million years. The other system is Kepler-221, with an age
of about 600 Myr \citep{Goldberg2021}. The rest of the resonant-chain systems are at least several Gyr old or have no precise age estimates. TOI-1136 and Kepler-221 seem to have had disparate evolution tracks despite similar ages. In Kepler-221, although the pairwise orbital period ratios (1.765 and 1.829) are farther from commensurability than in TOI-1136, the 3-body resonant angle changes so slowly (small $B$, Fig.~\ref{B_scatter}) that the resonant angle is most likely librating. The interpretation offered by \citet{Goldberg2021} is that Kepler-221 underwent rapid tidal resonant repulsion, possibly with the help of obliquity tides. \citet{Goldberg2021} estimated a total of 7000$\tau_e$ must have elapsed so that the system reached the current state of 10\% off resonance. On the other hand, TOI-1136 has barely moved from perfect orbital period commensurability. One possible explanation is that the conditions for capturing planets into a Cassini state \citep{Millholland_obliquity} were simply not available for TOI-1136. Its resonant repulsion has to proceed with the much slower eccentricity tides. We will return to this point in Section \ref{sec:obliquity_tides}. Based on the preceding argument, Kepler-223 \citep[not to be confused with Kepler-221,][]{MillsNature} may represent the future of TOI-1136. Kepler-223 is about 6 Gyr old, and its four transiting planets are likely in a 4-body resonant chain that only involves first-order MMR. Despite its 6-Gyr age, Kepler-223 seems to have avoided giant impact collisions, resonant repulsion, and planetesimal scattering, any of which could have induced deviations from MMR. Its orbital period ratios are still deep in resonance (1.3336, 1.5015, and 1.3339).

TOI-1136 also has the first known resonant chain that has a second-order MMR between neighboring first-order MMR. Kepler-29 b and c have a period ratio that deviates from a 9:7 MMR at a $10^{-4}$ level \citep{Fabrycky_Kepler29,JontofHutter2016}, however existing TTV could not determine if the system is in resonance nor its dynamical origin \citep{Migaszewski,Vissapragada29}. TOI-178 b is near a second-order 5:3 MMR with planet c \citep{Leleu2021}. However, the period ratio is shorter than expected if the system was resonant (1.95 day vs. 1.91 day). \citet{Leleu2021} suspected that tidal dissipation might have broken planet b away from resonance. In TRAPPIST-1, it is also the case that the inner three planets are close to third-order (8:5) and second-order (5:3) MMR; \citet{Agol2021} showed that the 3-body resonant angle involving b, c, and d is likely librating. However, they could not tell if the 2-body resonant angles were also librating. Nonetheless, the presence of a $490$-day super-period in the TTV of TRAPPIST-1 suggests the circulation of the 2-body resonant angle. One may be tempted to suggest that these innermost planets were initially in first-order MMR, but later disrupted by a crossing of the disk edge or tidal evolution \citep{Huang_Ormel}.  TOI-1136 is a very rare case --- possibly unique among the known systems --- a resonant chain with a second-order MMR between first-order resonances.

\subsection{The Disk that Formed TOI-1136}\label{sec:formation_environment}
The second-order resonance between TOI-1136 e and f allows us to place stringent constraints on TOI-1136's formation environment. Planets e and f most likely started with an initial period ratio close to 1.4 such that they did not get captured into the nearby, much stronger 3:2 first-order MMR. \citet{Xu0217} showed that the successful capture and stability of a second-order MMR is facilitated by lower initial orbital eccentricity, a planet mass ratio $m_2/m_1$ close to unity, and most importantly slower disk  migration. 

In our disk migration simulations (Section \ref{sec:migration}), the disks that successfully locked e and f into a 7:5 MMR all had lower total surface density (hence lower migration rate) compared to the MMSN ($\Sigma_{1AU}\lesssim1000$g~cm$^{-2}$). In comparison, \citet{Huhn} used a very similar prescription of disk migration with an inner disk edge to constrain the formation of Kepler-223, which only contains first-order MMR \citep{MillsNature}. \citet{Huhn} noted that Kepler-223 could form from convergent disk migration with a wider range of disk properties: the disk surface density can be a few times denser than the MMSN but still lock all planets of Kepler-223 into a resonant chain.

The rate of disk migration allowed us to constrain the total disk surface density of TOI-1136's protoplanetary disk. Our analyses in Section \ref{sec:migration} suggested that the TOI-1136 planets formed mostly {\it in-situ} followed by short-scale migration. If so, one can also constrain the solid surface density by spreading out the masses of the planets into their local feeding zones. We computed the Minimum-Mass Extrasolar Nebula of TOI-1136 using the TTV masses following the method in \citet{Dai_mmen}. TOI-1136 joined the other {\it Kepler} multi-planet systems with a similar solid surface density of $\Sigma_{\rm solid, 1AU}\approx50$g~cm$^{-2}$ (Fig. \ref{fig:disk_properties}). The total surface density $\Sigma_{1AU}\lesssim1000$g~cm$^{-2}$ and the solid surface density together suggest an enhanced dust-to-gas ratio of $\gtrsim$0.05 within the innermost 1AU of TOI-1136. This is higher than typically assumed 0.01 in the Interstellar Medium, and may suggest radial drift of dust and early gas disk dispersal \citep[e.g.][]{Gorti,Cridland,Birnstiel}.

Previous disk migration simulations placed meaningful constraints on disk turbulence \citep{Adams2008,Rein2009,Huhn}. Turbulence may increase the libration amplitudes of the planets captured in resonance and even disrupt the resonance if the turbulence is strong enough. We did not include turbulence in our convergent disk migration simulations in Section \ref{sec:migration}, because the libration amplitudes of TOI-1136 are still poorly constrained by the available TTV data.  Recent work by \citet{Jensen_Millholland} indicated that typical methods for inferring libration amplitudes of resonant planetary systems can be strongly biased by measurement uncertainties. We defer a discussion of the libration amplitudes to a future work where the libration amplitudes are better constrained.

TOI-1136 has a highly coplanar planetary system that is also well-aligned with the host star. The fact that all six (potentially seven) planets transit already hints at a low mutual inclination. According to our transit modeling (Section \ref{sec:transits}), assuming all planets transit the same hemisphere of the host star, the measured orbital inclination implies a mutual inclination of $1.1^\circ$. The planet with the most discrepant orbital inclination is planet b at  $86.44_{-0.21}^{+0.27}~^\circ$. The other five planets all have orbital inclinations around 89.5$^\circ$.  If we assume that the planets have the same longitudes of ascending node which can be tested with future transit duration variation analyses, the dispersion of orbital inclinations ($0.15^\circ$) is proxy for their mutual inclination. Previous works have also found that the innermost planet of a multi-planet system often has the largest orbital inclination, likely due to an equipartitioning of the angular momentum deficit \citep{Steffen2016,Dai2018,Weiss2018,Petrovich}. For comparison, the TRAPPIST-1 planets have even lower mutual inclinations of about 0.04$^\circ$ \citep{Agol2021}. 

Our RM measurement of TOI-1136 revealed a planetary system that is well-aligned with the rotation of the host star. The sky-projected stellar obliquity $\lambda$ is $5\pm5^\circ$ and the stellar obliquity is less than 28$^\circ$ with a 95\% credible level. \citet{Hirano2020} measured the stellar obliquity of TRAPPIST-1, and they also found evidence for a well-aligned planetary system. \citet{Spalding} pointed out that planets may couple with the oblateness of their host star, the differential nodal precession may induce a mutual inclination of $\approx 2\Psi$. Hence, for TOI-1136, the measured low mutual inclination and low stellar obliquity corroborate each other: if the stellar obliquity were high, a large mutual inclination would have been generated by the differential precession. \citet{Batygin2015_capture} suggested that if a protoplanetary disk has substantial axial asymmetry, capturing planets into MMR during disk migration is much more difficult. If TOI-1136 had a stellar companion, a stellar fly-by event \citep[e.g][]{Xiang-Gruess} or the perturbation from the companion \citep{Batygin_nature} may induce disk warp, axial asymmetry and primordial misalignment that are detrimental to the formation of a resonant chain like TOI-1136. No spectroscopic, visual, blended, or comoving stellar companions were found for TOI-1136 (Section \ref{sec:stellar_para}). We suggest that TOI-1136, at 700-Myr-old, still preserves the pristine orbital architecture formed by a slow migration in an isolated disk with no primordial misalignment or disk asymmetry.

\subsection{Mass and radius}

\begin{figure*}
\center
\includegraphics[width = 1.7\columnwidth]{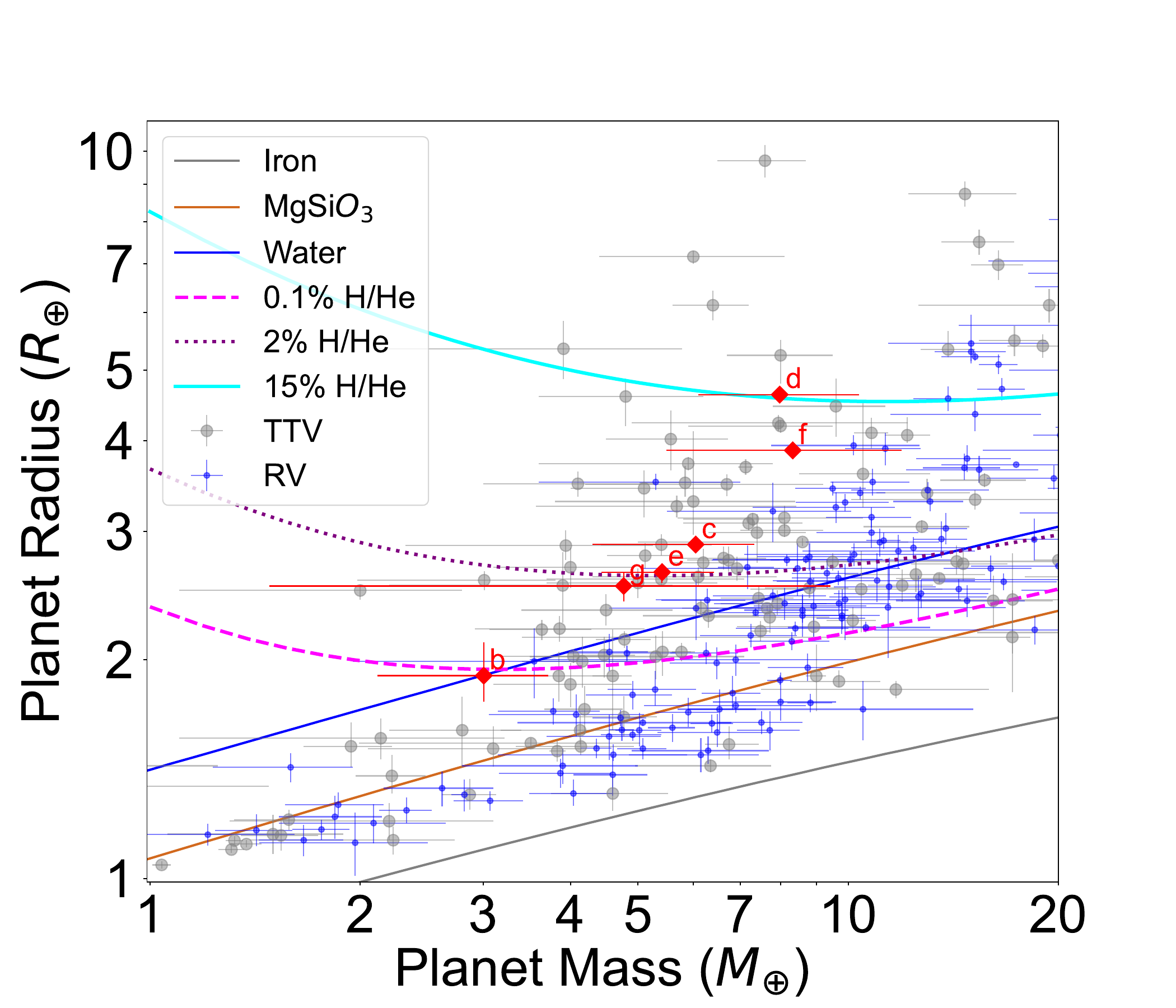}
\caption{The measured masses and radii of TOI-1136 planets and other planets on NASA Exoplanet Archive. We distinguished mass measurements from RV (blue) and TTV analyses (gray). It was pointed out that TTV planets may have systematically lower masses than RV planets of the same radii \citep[e.g.,][]{Steffen_ttv,Hadden2017,MillsMazeh}. Although there is still large mass uncertainty based on the existing TTV dataset, TOI-1136 planets tend to have lower densities than the RV planets even though the e-mass degeneracy does not affect this in-resonant system \citep{Nesvorny}. The theoretical mass radius relationships are from \citet{Zeng2016} and \citet{Chen_Rogers}. TOI-1136 b, being the innermost, lowest-mass planet, seems to have lost substantial H/He after 700 Myr of photoevaporation. The outer, more massive planets are generally consistent with having $\sim2$ to $15\%$ their mass in H/He.}
\label{fig:mass_radius}
\end{figure*}

Fig.~\ref{fig:mass_radius} shows the masses and radii of the TOI-1136 planets along with the theoretical mass-radius relationships from  \citet{Zeng2016} and \citet{Chen_Rogers}. We also plot archival mass measurements from the NASA Exoplanet Archive\footnote{\url{https://exoplanetarchive.ipac.caltech.edu}}. Using the model by \citet{Chen_Rogers} which takes into account the age, mass, composition and insolation level of a planet, the required mass of the H/He envelope increases from 0.1\% for the smallest planet b up to about 15\% in mass for the largest planet d. 

TOI-1136 is about 700 Myr old. The innermost planet b should have experienced extensive photoevaporation for hundreds of Myr \citep{OwenWu,Fulton,Zhang2022}. With a core mass of about $3M_\oplus$, planet b does not have a deep enough gravitational potential well to prevent photoevaporation \citep[see the self-consistent hydrodynamic simulations of photoevaporation by][]{WangDai2018}. On the other hand, the more massive and more distant planets in TOI-1136 should experience sequentially weaker photoevaporation. A dynamically quiet, multi-planet system like TOI-1136 is a good testbed for photoevaporation theory. Given the delicate orbital architecture, the orbital distances likely did not change significantly since formation. There has not been any giant impact collisions that could have removed the gaseous envelopes \citep{Inamdar}. The planets are subject to the same XUV spectrum other than scaling with their orbital distances. By comparing the extent of the mass loss for each planet, TOI-1136 offers an opportunity to probe the variance of the efficiency of photoevaporation \citep{OwenCampos}. Future observations of outflowing material via Lyman-$\alpha$ or metastable He observations \citep[e.g.,][]{Spake,Zhang2022} coupled with hydrodynamic modeling \citep[e.g.,][]{Wang_2021_wasp107} may shed light on this issue.

 In Fig.~\ref{fig:mass_radius}, we color-coded mass measurements from TTV and RV analyses separately. TOI-1136 conforms to the previously suggested trend that TTV planets tend to have lower masses than RV planets of the same radii \citep[e.g.,][]{Steffen_ttv}. One possible explanation is that for near-resonant systems (majority of the TTV sample), strong $e$-mass degeneracy may bias the TTV  measurements toward lower values \citep{Lithwick_ttv}. However, for TOI-1136, a resonant TTV case, such a degeneracy is minimal \citep{Nesvorny}. Instead, both the amplitude and periodicity of resonant TTV contain information on the masses of the planets independently. Hence the $e$-mass degeneracy does not seem to be a convincing explanation for TTV planets' lower densities. Alternatively, TTV planets might have originated from beyond the water snow line where conditions are more conducive for accreting a thick atmosphere \citep[e.g.][]{Lee2016}. Another possibility is obliquity tides: \citet{Millholland_obliquity} argued that obliquity tides might inflate the radii of near resonant systems as the tidal dissipation goes into heating the planets. We discuss the obliquity tides in more detail in the next section.

\subsection{Obliquity tides inflating d and f?} \label{sec:obliquity_tides}

 Strong tidal dissipation due to obliquity tides may offer enough tidal damping to explain both the observed resonant repulsion in {\it Kepler} multi-planet systems \citep{Millholland_obliquity} and a possible radius inflation of near-resonant planets \citep{Millholland2019}. Obliquity tides can be much more dissipative than eccentricity tides. When a planet maintains a non-zero obliquity \citep[Cassini State 2 being the most favorable,][]{Colombo,Peale_cassini}, the tidal bulge will move in the co-rotating frame and leads to significant dissipation. The capture of a system into a Cassini state (secular spin-orbit resonance) both excites and sustains a non-zero planetary obliquity. This resonance happens when the nodal precession frequency matches the spin precession frequency. Resonant planetary systems are more likely to be in a Cassini state. This is because during their convergent disk migration, the nodal precession frequency sweeps through a range of frequencies as the semi-major axes of the planets change, allowing a crossing of the frequencies.

TOI-1136 d and f have larger radii ($>4R_\oplus$) than the other planets ($<3R_\oplus$), to some extent discrepant with the previously noted trend of intra-system uniformity of multi-planet systems \citep{Weiss_peas,Millholland_peas,Wang_uniform}. Could their larger radii be caused by obliquity tides? As \citet{Millholland_obliquity} and \citet{Millholland2019} suggested, obliquity tides can lead to both resonant repulsion and radius inflation of resonant planets. The TOI-1136 planets are currently still deep in resonance with deviations from MMR $\Delta$ of the order $10^{-4}$. The system has undergone minimal resonant repulsion since formation (Section \ref{sec:repulsion}). Even assuming the planets have moved by a $\Delta$ of $10^{-3}$, the corresponding tidal heating luminosity spread out in the system's age of 700 Myr is about 10$^{16}$W. This only accounts for $10^{-4}$ of the bolometric insolation planet d receives from the star (10$^{20}$W). According to the MESA simulation by \citet{Millholland2019}, this level of additional heating due to tides could not inflate the planetary radius by more than 10\%. This is not sufficient to inflate the radii of d and f to $>4R_\oplus$ if they were initially similar to the other planets with radii $<3R_\oplus$. The radii of d and f may require another explanation such as slower photoevaporation, or dusty outflows \citep{WangDai,Gao} which is testable with a near infrared transmission spectrum from the James Webb Space Telescope \citep{Gardner}. In this hypothesis, the radius inflation of planets d and f is temporary. As the system matures, the radii may drop down to conform to the intra-system uniformity of mature multi-planet systems.

\subsection{A Precursor of Kepler Multi-planet Systems?}

\begin{figure*}
\center
\includegraphics[width = 1.8\columnwidth]{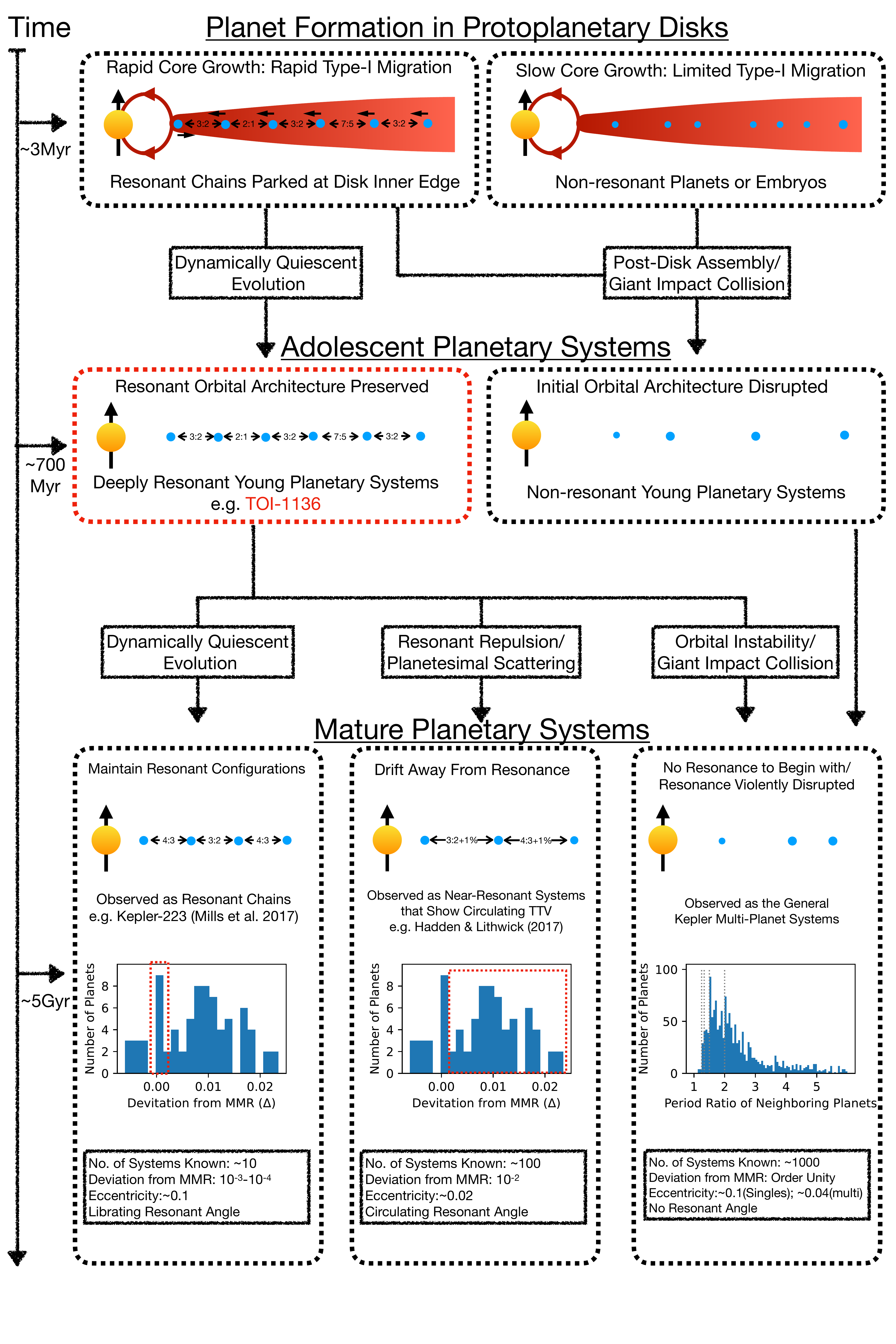}
\caption{A schematic showing how TOI-1136 fits into the broader picture of planet formation. It provides an adolescent planetary system that still records the initial condition from convergent disk migration. Depending on whether its future dynamical evolution is quiescent or violent, it may stay as a resonant chain like Kepler-223 \citep[bottom left, ][]{Mills2016}, mildly evolve into a near-resonant system (bottom center), or has its resonant structure violently disrupted and become a non-resonant {\it Kepler} multi-planet system (bottom right).}
\label{fig:formation_pathway}
\end{figure*}

Finally, we place TOI-1136 in the broader context of the formation and dynamical evolution of close-in, sub-Neptune planets. Fig. \ref{fig:formation_pathway} shows our understanding of where the field stands and how TOI-1136 fits in. Planet embryos grow in protoplanetary disks. The rate of Type-I migration is proportional to the masses of the cores \citep{Kley_2012}. Therefore, depending on the rate of core growth, Type-I migration may or may not play a significant role in the formation of close-in sub-Neptune planets. For systems where planetary cores assembled quickly, Type-I migration may routinely generate a chain of resonant planets parked at the inner edge of the disk. TOI-1136 is an example of this scenario: it is an 700-Myr-old adolescent planetary system that still records the deeply resonant configuration from disk migration.  On the other hand, in system with slower core growth, planet embryos undergo limited migration and are generally non-resonant when the disk dissipates. Or if the disk is turbulent or axially asymmetric \citep{Adams2008,Batygin_nature}, one would also expect a non-resonant configuration. Post-disk assembly of non-resonant planetary systems will likely remain non-resonant (right column of Fig. \ref{fig:formation_pathway}).

Fast forwarding to a $\sim$5-Gyr-old mature planetary system, TOI-1136 may remain deeply resonant if it only experiences negligible dynamical evolution such as orbital instability \citep{Izidoro} and planetesimal scattering \citep{Chatterjee_2015}. The deeply resonant, 6-Gyr-old Kepler-223 \citep{MillsNature} may be the future of TOI-1136. So far there are about $\sim10$ resonant-chain systems. If TOI-1136 continues to undergo mild resonant repulsion and planetesimal scattering, it may join the population of near-resonant ($\Delta$ = 1-2\%), multi-planet {\it Kepler} systems that show circulating TTV. There are about 100-200 such systems that have been discovered by {\it Kepler} \citep[e.g.][]{JontofHutter2016,Hadden2017}. If the future dynamical evolution of TOI-1136 is more violent, orbital instability and giant impact collision \citep{Izidoro,Goldberg2022} may totally disrupt the resonance in TOI-1136. It will end up as a non-resonant planetary systems that dominates the mature {\it Kepler} multi-planet sample.

\section{Summary}\label{sec:summary}
Disk migration may be a common stage of planet formation \citep{Ward1997,Kley_2012}. If so, many close-in, tightly-packed, multi-planet systems as observed by {\it Kepler} should host planets in a chain of mean motion resonances \citep[e.g.,][]{Kley_2012,Lee2002,Cresswell}. In reality, only a small subset of {\it Kepler} multi-planets are observed near resonance with a typical deviation $\Delta$ of 1 to 2\% \citep{Fabrycky2014}. Over billions of years of dynamical evolution, a combination of effects including resonant repulsion \citep{Lithwick_repulsion}, dynamical instability \citep{Goldberg_stability}, secular chaos \citep{WuLithwick}, planetesimal scattering \citep{Chatterjee_2015} could lead to the slow deviation or the disruption of the migration-induced resonance.

To complete this picture, we present TOI-1136 a young planetary system with a resonant chain of six planets. The system is so deep in resonance that it probably still preserves a "pristine" orbital architecture from convergent disk migration. It may be a precursor of many of the {\it Kepler} near-resonant multi-planets before dynamical evolution eventually dislodged the planets from perfect resonance over Gyr timescale. Our observations and dynamical modeling revealed the following characteristics of TOI-1136:

\begin{itemize}

 \item TOI-1136 is about 700-Myr-old based on gyrochronology, activity indicators, and Li absorption.
 
  \item A Rossiter-McLaughlin measurement of planet d revealed a planetary system well-aligned with the host's rotation with sky-projected stellar obliquity of $5\pm5^\circ$. All six planets transit, which implies a low mutual inclination between the planets: $1.1^\circ$ or just $0.15^\circ$ after excluding the most inclined planet b.
  
  \item No spectroscopic, AO, visual, or comoving stellar companion was detected for TOI-1136. The low stellar obliquity, coupled with the coplanarity, and dynamical fragility of a resonant chain of planets, point to the formation of TOI-1136 in an isolated disk with no stellar fly-by, disk warp or significant axial asymmetry.
 
 \item There are six transiting planets with each neighboring pair showing anti-correlated TTVs. The TTVs are most likely driven by the libration of resonant angles (libration periods) rather than by the circulation of resonant angles (super-periods).
 
  \item Our TTV analysis revealed the masses of the planets. The mass and radius of the innermost and the light planet b suggests only a 0.1\%-by-mass H/He envelope. This is consistent with the expectation of 700 Myr of photoevaporation.
 
 \item The orbital period ratios are extremely close to ratios of small integers, with a deviation $\Delta \equiv \frac{P_{out}/P_{in}}{p/q}-1$ of the order $10^{-4}$.

 \item The closeness to MMR and the libration of the various resonant angles suggest that TOI-1136 planets are in resonance rather than near resonance.
 
 \item Planets e and f are close to a 7:5 second-order MMR. TOI-1136 is the first known resonant chain with a second-order MMR between first-order MMR. The weaker and more delicate second-order MMR is much more difficult to form in disk migration and more easily dislodged from resonance later on.

 \item Our disk migration simulations favor Type-I migration with an inner disk edge for TOI-1136. The edge helps to halt the migration the planets and converts divergent encounters into convergent ones. To lock the ef pair into a 7:5 second-order MMR, the disk has to be less dense than than the MMSN with $\Sigma_{\rm 1AU}\lesssim 1000$g~cm$^{-2}$. 

 \item Our resonant repulsion simulations indicate that TOI-1136 has undergone minimum tidal dissipation since its formation. Strong tidal dissipation due to a rocky planet b or obliquity tides on planets d and f seems unlikely.
 
\end{itemize}

We encourage additional photometric follow-up observation of this system using space-based and ground-based facilities in the next few years to refine dynamical constraints on this system. TOI-1136 is also amenable to metastable Helium observation and transmission spectroscopy that will help better understand this young planetary system.

\software{Isoclassify \citep{Huber17}, COMOVE \citep{Tofflemire}, lmfit \citep{LM}, Batman \citep{Kreidberg2015}, REBOUND \citep{Rein}, emcee \citep{emcee}, JAX \citep{jax2018github}}

\acknowledgements
We thank Doug Lin, Eric Agol, Dan Fabrycky, Sarah Millhollland, Jared Siegel, Ji-Wei Xie, Wenrui Xu, Wei Zhu, Shuo Huang, Jason Steffen, and André Izidoro for insightful comments. We than Armaan Goyal for computing the Ginni Index for TOI-1136 and the Kepler sample. We thank Cicero Lu for investigating the SED of this system.

We acknowledge the use of public TESS data from pipelines at the TESS Science Office and at the TESS Science Processing Operations Center. Resources supporting this work were provided by the NASA High-End Computing (HEC) Program through the NASA Advanced Supercomputing (NAS) Division at Ames Research Center for the production of the SPOC data products.

J.L-B. acknowledges financial support received from "la Caixa" Foundation (ID 100010434) and from the European Unions Horizon 2020 research and innovation programme under the Marie Slodowska-Curie grant agreement No 847648, with fellowship code LCF/BQ/PI20/11760023. This research has also been partly funded by the Spanish State Research Agency (AEI) Projects No.PID2019-107061GB-C61. 

J.M.A.M. is supported by the National Science Foundation Graduate Research Fellowship Program under Grant No. DGE-1842400. J.M.A.M. acknowledges the LSSTC Data Science Fellowship Program, which is funded by LSSTC, NSF Cybertraining Grant No. 1829740, the Brinson Foundation, and the Moore Foundation; his participation in the program has benefited this work.

Based on observations (programme ID: A43/TAC11) made with the Italian Telescopio Nazionale Galileo (TNG) operated on the island of La Palma by the Fundación Galileo Galilei of the INAF (Istituto Nazionale di Astrofisica) at the Spanish Observatorio del Roque de los Muchachos of the Instituto de Astrofisica de Canarias. Funding for the Stellar Astrophysics Centre is provided by The Danish National Research Foundation (Grant agreement no.: DNRF106).

\bibliography{main}

\begin{thebibliography}{}
\expandafter\ifx\csname natexlab\endcsname\relax\def\natexlab#1{#1}\fi
\providecommand{\url}[1]{\href{#1}{#1}}
\providecommand{\dodoi}[1]{doi:~\href{http://doi.org/#1}{\nolinkurl{#1}}}
\providecommand{\doeprint}[1]{\href{http://ascl.net/#1}{\nolinkurl{http://ascl.net/#1}}}
\providecommand{\doarXiv}[1]{\href{https://arxiv.org/abs/#1}{\nolinkurl{https://arxiv.org/abs/#1}}}

\bibitem[{Adams {et~al.}(2008)Adams, Laughlin, \& Bloch}]{Adams2008}
Adams, F.~C., Laughlin, G., \& Bloch, A.~M. 2008, \apj, 683, 1117,
  \dodoi{10.1086/589986}

\bibitem[{{Agol} {et~al.}(2021){Agol}, {Hernandez}, \&
  {Langford}}]{2021MNRAS.507.1582A}
{Agol}, E., {Hernandez}, D.~M., \& {Langford}, Z. 2021, \mnras, 507, 1582,
  \dodoi{10.1093/mnras/stab2044}

\bibitem[{Agol {et~al.}(2005)Agol, Steffen, Sari, \& Clarkson}]{Agol2005}
Agol, E., Steffen, J., Sari, R., \& Clarkson, W. 2005, Monthly Notices of the
  Royal Astronomical Society, 359, 567,
  \dodoi{10.1111/j.1365-2966.2005.08922.x}

\bibitem[{Agol {et~al.}(2021)Agol, Dorn, Grimm, Turbet, Ducrot, Delrez, Gillon,
  Demory, Burdanov, Barkaoui, Benkhaldoun, Bolmont, Burgasser, Carey, Wit,
  Fabrycky, Foreman-Mackey, Haldemann, Hernandez, Ingalls, Jehin, Langford,
  Leconte, Lederer, Luger, Malhotra, Meadows, Morris, Pozuelos, Queloz,
  Raymond, Selsis, Sestovic, Triaud, \& Grootel}]{Agol2021}
Agol, E., Dorn, C., Grimm, S.~L., {et~al.} 2021, Planetary Science Journal, 2,
  \dodoi{10.3847/PSJ/abd022}

\bibitem[{{Albrecht} {et~al.}(2021){Albrecht}, {Marcussen}, {Winn}, {Dawson},
  \& {Knudstrup}}]{Albrecht2021}
{Albrecht}, S.~H., {Marcussen}, M.~L., {Winn}, J.~N., {Dawson}, R.~I., \&
  {Knudstrup}, E. 2021, \apjl, 916, L1, \dodoi{10.3847/2041-8213/ac0f03}

\bibitem[{{Andrews}(2020)}]{Andrews_review}
{Andrews}, S.~M. 2020, \araa, 58, 483,
  \dodoi{10.1146/annurev-astro-031220-010302}

\bibitem[{{Ataiee} \& {Kley}(2020)}]{Ataiee}
{Ataiee}, S., \& {Kley}, W. 2020, \aap, 635, A204,
  \dodoi{10.1051/0004-6361/201936390}

\bibitem[{{Bailey} {et~al.}(2022){Bailey}, {Gilbert}, \& {Fabrycky}}]{Bailey}
{Bailey}, N., {Gilbert}, G., \& {Fabrycky}, D. 2022, \aj, 163, 13,
  \dodoi{10.3847/1538-3881/ac2f46}

\bibitem[{{Banfield} \& {Murray}(1992)}]{Banfield}
{Banfield}, D., \& {Murray}, N. 1992, \icarus, 99, 390,
  \dodoi{10.1016/0019-1035(92)90155-Z}

\bibitem[{Barnes {et~al.}(2015)Barnes, Weingrill, Granzer, Spada, \&
  Strassmeier}]{barnes_2015}
Barnes, S.~A., Weingrill, J., Granzer, T., Spada, F., \& Strassmeier, K.~G.
  2015, \aap, 583, A73, \dodoi{10.1051/0004-6361/201526129}

\bibitem[{{Baruteau} {et~al.}(2014){Baruteau}, {Crida}, {Paardekooper},
  {Masset}, {Guilet}, {Bitsch}, {Nelson}, {Kley}, \&
  {Papaloizou}}]{Baruteau2014}
{Baruteau}, C., {Crida}, A., {Paardekooper}, S.~J., {et~al.} 2014, in
  Protostars and Planets VI, ed. H.~{Beuther}, R.~S. {Klessen}, C.~P.
  {Dullemond}, \& T.~{Henning}, 667,
  \dodoi{10.2458/azu_uapress_9780816531240-ch029}

\bibitem[{Batygin(2012)}]{Batygin_nature}
Batygin, K. 2012, Nature, 491, 418, \dodoi{10.1038/nature11560}

\bibitem[{{Batygin}(2015)}]{Batygin2015_capture}
{Batygin}, K. 2015, \mnras, 451, 2589, \dodoi{10.1093/mnras/stv1063}

\bibitem[{{Batygin} \& {Morbidelli}(2013{\natexlab{a}})}]{Batygin_repulsion}
{Batygin}, K., \& {Morbidelli}, A. 2013{\natexlab{a}}, \aj, 145, 1,
  \dodoi{10.1088/0004-6256/145/1/1}

\bibitem[{{Batygin} \& {Morbidelli}(2013{\natexlab{b}})}]{Batygin_resonance}
---. 2013{\natexlab{b}}, \aap, 556, A28, \dodoi{10.1051/0004-6361/201220907}

\bibitem[{{Beauge}(1994)}]{Beauge}
{Beauge}, C. 1994, Celestial Mechanics and Dynamical Astronomy, 60, 225,
  \dodoi{10.1007/BF00693323}

\bibitem[{Beaugé {et~al.}(2006)Beaugé, Michtchenko, \&
  Ferraz-Mello}]{Beauge2006}
Beaugé, C., Michtchenko, T.~A., \& Ferraz-Mello, S. 2006, Monthly Notices of
  the Royal Astronomical Society, 365, 1160,
  \dodoi{10.1111/j.1365-2966.2005.09779.x}

\bibitem[{{Berger} {et~al.}(2018){Berger}, {Howard}, \&
  {Boesgaard}}]{Berger_li}
{Berger}, T.~A., {Howard}, A.~W., \& {Boesgaard}, A.~M. 2018, \apj, 855, 115,
  \dodoi{10.3847/1538-4357/aab154}

\bibitem[{{Betancourt}(2017)}]{2017arXiv170102434B}
{Betancourt}, M. 2017, arXiv e-prints, arXiv:1701.02434.
\newblock \doarXiv{1701.02434}

\bibitem[{Bingham {et~al.}(2018)Bingham, Chen, Jankowiak, Obermeyer, Pradhan,
  Karaletsos, Singh, Szerlip, Horsfall, \& Goodman}]{bingham2018pyro}
Bingham, E., Chen, J.~P., Jankowiak, M., {et~al.} 2018, arXiv preprint
  arXiv:1810.09538

\bibitem[{{Birnstiel} {et~al.}(2010){Birnstiel}, {Dullemond}, \&
  {Brauer}}]{Birnstiel}
{Birnstiel}, T., {Dullemond}, C.~P., \& {Brauer}, F. 2010, \aap, 513, A79,
  \dodoi{10.1051/0004-6361/200913731}

\bibitem[{{Bouma} {et~al.}(2021){Bouma}, {Curtis}, {Hartman}, {Winn}, \&
  {Bakos}}]{Bouma_li}
{Bouma}, L.~G., {Curtis}, J.~L., {Hartman}, J.~D., {Winn}, J.~N., \& {Bakos},
  G.~{\'A}. 2021, \aj, 162, 197, \dodoi{10.3847/1538-3881/ac18cd}

\bibitem[{{Bouma} {et~al.}(2022){Bouma}, {Kerr}, {Curtis}, {Isaacson},
  {Hillenbrand}, {Howard}, {Kraus}, {Bieryla}, {Latham}, {Petigura}, \&
  {Huber}}]{Bouma_2022}
{Bouma}, L.~G., {Kerr}, R., {Curtis}, J.~L., {et~al.} 2022, arXiv e-prints,
  arXiv:2205.01112.
\newblock \doarXiv{2205.01112}

\bibitem[{Bradbury {et~al.}(2018)Bradbury, Frostig, Hawkins, Johnson, Leary,
  Maclaurin, Necula, Paszke, Vander{P}las, Wanderman-{M}ilne, \&
  Zhang}]{jax2018github}
Bradbury, J., Frostig, R., Hawkins, P., {et~al.} 2018, {JAX}: composable
  transformations of {P}ython+{N}um{P}y programs, 0.2.5.
\newblock \url{http://github.com/google/jax}

\bibitem[{Brandt \& Huang(2015)}]{brandt_age_2015}
Brandt, T.~D., \& Huang, C.~X. 2015, \apj, 807, 24,
  \dodoi{10.1088/0004-637X/807/1/24}

\bibitem[{{Brasser} {et~al.}(2022){Brasser}, {Pichierri}, {Dobos}, \&
  {Barr}}]{Brasser2021}
{Brasser}, R., {Pichierri}, G., {Dobos}, V., \& {Barr}, A.~C. 2022, \mnras,
  515, 2373, \dodoi{10.1093/mnras/stac1907}

\bibitem[{{Bruntt} {et~al.}(2010){Bruntt}, {Bedding}, {Quirion}, {Lo Curto},
  {Carrier}, {Smalley}, {Dall}, {Arentoft}, {Bazot}, \& {Butler}}]{Bruntt2010}
{Bruntt}, H., {Bedding}, T.~R., {Quirion}, P.~O., {et~al.} 2010, \mnras, 405,
  1907, \dodoi{10.1111/j.1365-2966.2010.16575.x}

\bibitem[{{Buchhave} {et~al.}(2010){Buchhave}, {Bakos}, {Hartman}, {Torres},
  {Kov{\'a}cs}, {Latham}, {Noyes}, {Esquerdo}, {Everett}, {Howard}, {Marcy},
  {Fischer}, {Johnson}, {Andersen}, {F{\H{u}}r{\'e}sz}, {Perumpilly},
  {Sasselov}, {Stefanik}, {B{\'e}ky}, {L{\'a}z{\'a}r}, {Papp}, \&
  {S{\'a}ri}}]{buchhave2010}
{Buchhave}, L.~A., {Bakos}, G.~{\'A}., {Hartman}, J.~D., {et~al.} 2010, \apj,
  720, 1118, \dodoi{10.1088/0004-637X/720/2/1118}

\bibitem[{Buchhave {et~al.}(2012)Buchhave, Latham, Johansen, Bizzarro, Torres,
  Rowe, Batalha, Borucki, Brugamyer, Caldwell, Bryson, Ciardi, Cochran, Endl,
  Esquerdo, Ford, Geary, Gilliland, Hansen, Isaacson, Laird, Lucas, Marcy,
  Morse, Robertson, Shporer, Stefanik, Still, \& Quinn}]{buchhave2012}
Buchhave, L.~A., Latham, D., Johansen, A., {et~al.} 2012, Nature, 486, 375

\bibitem[{{Carter} {et~al.}(2012){Carter}, {Agol}, {Chaplin}, {Basu},
  {Bedding}, {Buchhave}, {Christensen-Dalsgaard}, {Deck}, {Elsworth},
  {Fabrycky}, {Ford}, {Fortney}, {Hale}, {Handberg}, {Hekker}, {Holman},
  {Huber}, {Karoff}, {Kawaler}, {Kjeldsen}, {Lissauer}, {Lopez}, {Lund},
  {Lundkvist}, {Metcalfe}, {Miglio}, {Rogers}, {Stello}, {Borucki}, {Bryson},
  {Christiansen}, {Cochran}, {Geary}, {Gilliland}, {Haas}, {Hall}, {Howard},
  {Jenkins}, {Klaus}, {Koch}, {Latham}, {MacQueen}, {Sasselov}, {Steffen},
  {Twicken}, \& {Winn}}]{Carter}
{Carter}, J.~A., {Agol}, E., {Chaplin}, W.~J., {et~al.} 2012, Science, 337,
  556, \dodoi{10.1126/science.1223269}

\bibitem[{{Chambers}(1999)}]{Chambers}
{Chambers}, J.~E. 1999, \mnras, 304, 793,
  \dodoi{10.1046/j.1365-8711.1999.02379.x}

\bibitem[{Chatterjee \& Ford(2015)}]{Chatterjee_2015}
Chatterjee, S., \& Ford, E.~B. 2015, The Astrophysical Journal, 803, 33,
  \dodoi{10.1088/0004-637x/803/1/33}

\bibitem[{{Chen} \& {Rogers}(2016)}]{Chen_Rogers}
{Chen}, H., \& {Rogers}, L.~A. 2016, \apj, 831, 180,
  \dodoi{10.3847/0004-637X/831/2/180}

\bibitem[{{Chiang} \& {Laughlin}(2013)}]{Chiang}
{Chiang}, E., \& {Laughlin}, G. 2013, \mnras, 431, 3444,
  \dodoi{10.1093/mnras/stt424}

\bibitem[{{Chontos} {et~al.}(2022){Chontos}, {Murphy}, {MacDougall},
  {Fetherolf}, {Van Zandt}, {Rubenzahl}, {Beard}, {Huber}, {Batalha},
  {Crossfield}, {Dressing}, {Fulton}, {Howard}, {Isaacson}, {Kane}, {Petigura},
  {Robertson}, {Roy}, {Weiss}, {Behmard}, {Dai}, {Dalba}, {Giacalone}, {Hill},
  {Lubin}, {Mayo}, {Mo{\v{c}}nik}, {Polanski}, {Rosenthal}, {Scarsdale},
  {Turtelboom}, {Ricker}, {Vanderspek}, {Latham}, {Seager}, {Winn}, {Jenkins},
  {Quinn}, {Guerrero}, {Collins}, {Ciardi}, {Shporer}, {Goeke}, {Levine},
  {Ting}, {Bieryla}, {Collins}, {Kielkopf}, {Barkaoui}, {Benni},
  {Esparza-Borges}, {Conti}, {Hooton}, {Kagetani}, {Laloum}, {Marino},
  {Massey}, {Murgas}, {Papini}, {Schwarz}, {Srdoc}, {Stockdale}, {Wang},
  {Wittrock}, \& {Zou}}]{Chontos}
{Chontos}, A., {Murphy}, J. M.~A., {MacDougall}, M.~G., {et~al.} 2022, \aj,
  163, 297, \dodoi{10.3847/1538-3881/ac6266}

\bibitem[{Christiansen {et~al.}(2018)Christiansen, Crossfield, Barentsen,
  Lintott, Barclay, Simmons, Petigura, Schlieder, Dressing, Vanderburg, Ciardi,
  Allen, McMaster, Miller, Veldthuis, Allen, Wolfenbarger, Cox, Zemiro, Howard,
  Livingston, Sinukoff, Catron, Grey, Kusch, Terentev, Vales, \&
  Kristiansen}]{Christiansen2018}
Christiansen, J.~L., Crossfield, I. J.~M., Barentsen, G., {et~al.} 2018,
  \dodoi{10.3847/1538-3881/aa9be0}

\bibitem[{{Ciardi} {et~al.}(2015){Ciardi}, {Beichman}, {Horch}, \&
  {Howell}}]{ciardi2015}
{Ciardi}, D.~R., {Beichman}, C.~A., {Horch}, E.~P., \& {Howell}, S.~B. 2015,
  \apj, 805, 16, \dodoi{10.1088/0004-637X/805/1/16}

\bibitem[{{Collins} {et~al.}(2017){Collins}, {Kielkopf}, {Stassun}, \&
  {Hessman}}]{Collins:2017}
{Collins}, K.~A., {Kielkopf}, J.~F., {Stassun}, K.~G., \& {Hessman}, F.~V.
  2017, \aj, 153, 77, \dodoi{10.3847/1538-3881/153/2/77}

\bibitem[{{Colombo}(1966)}]{Colombo}
{Colombo}, G. 1966, \aj, 71, 891, \dodoi{10.1086/109983}

\bibitem[{{Cosentino} {et~al.}(2012){Cosentino}, {Lovis}, {Pepe}, {Collier
  Cameron}, {Latham}, {Molinari}, {Udry}, {Bezawada}, {Black}, {Born},
  {Buchschacher}, {Charbonneau}, {Figueira}, {Fleury}, {Galli}, {Gallie},
  {Gao}, {Ghedina}, {Gonzalez}, {Gonzalez}, {Guerra}, {Henry}, {Horne},
  {Hughes}, {Kelly}, {Lodi}, {Lunney}, {Maire}, {Mayor}, {Micela}, {Ordway},
  {Peacock}, {Phillips}, {Piotto}, {Pollacco}, {Queloz}, {Rice}, {Riverol},
  {Riverol}, {San Juan}, {Sasselov}, {Segransan}, {Sozzetti}, {Sosnowska},
  {Stobie}, {Szentgyorgyi}, {Vick}, \& {Weber}}]{Cosentino2012}
{Cosentino}, R., {Lovis}, C., {Pepe}, F., {et~al.} 2012, in Society of
  Photo-Optical Instrumentation Engineers (SPIE) Conference Series, Vol. 8446,
  Ground-based and Airborne Instrumentation for Astronomy IV, ed. I.~S.
  {McLean}, S.~K. {Ramsay}, \& H.~{Takami}, 84461V, \dodoi{10.1117/12.925738}

\bibitem[{{Cresswell} \& {Nelson}(2006)}]{Cresswell2006}
{Cresswell}, P., \& {Nelson}, R.~P. 2006, \aap, 450, 833,
  \dodoi{10.1051/0004-6361:20054551}

\bibitem[{{Cresswell} \& {Nelson}(2008)}]{Cresswell}
---. 2008, \aap, 482, 677, \dodoi{10.1051/0004-6361:20079178}

\bibitem[{Cridland {et~al.}(2016)Cridland, Pudritz, \& Birnstiel}]{Cridland}
Cridland, A.~J., Pudritz, R.~E., \& Birnstiel, T. 2016, Monthly Notices of the
  Royal Astronomical Society, 465, 3865, \dodoi{10.1093/mnras/stw2946}

\bibitem[{{Crossfield} {et~al.}(2016){Crossfield}, {Ciardi}, {Petigura},
  {Sinukoff}, {Schlieder}, {Howard}, {Beichman}, {Isaacson}, {Dressing},
  {Christiansen}, {Fulton}, {L{\'e}pine}, {Weiss}, {Hirsch}, {Livingston},
  {Baranec}, {Law}, {Riddle}, {Ziegler}, {Howell}, {Horch}, {Everett}, {Teske},
  {Martinez}, {Obermeier}, {Benneke}, {Scott}, {Deacon}, {Aller}, {Hansen},
  {Mancini}, {Ciceri}, {Brahm}, {Jord{\'a}n}, {Knutson}, {Henning}, {Bonnefoy},
  {Liu}, {Crepp}, {Lothringer}, {Hinz}, {Bailey}, {Skemer}, \&
  {Defrere}}]{Crossfield2016}
{Crossfield}, I. J.~M., {Ciardi}, D.~R., {Petigura}, E.~A., {et~al.} 2016,
  \apjs, 226, 7, \dodoi{10.3847/0067-0049/226/1/7}

\bibitem[{Curtis {et~al.}(2019)Curtis, Agüeros, Douglas, \&
  Meibom}]{curtis_2019}
Curtis, J.~L., Agüeros, M.~A., Douglas, S.~T., \& Meibom, S. 2019, \apj, 879,
  49, \dodoi{10.3847/1538-4357/ab2393}

\bibitem[{{Dai} {et~al.}(2018){Dai}, {Masuda}, \& {Winn}}]{Dai2018}
{Dai}, F., {Masuda}, K., \& {Winn}, J.~N. 2018, \apjl, 864, L38,
  \dodoi{10.3847/2041-8213/aadd4f}

\bibitem[{{Dai} {et~al.}(2020){Dai}, {Winn}, {Schlaufman}, {Wang}, {Weiss},
  {Petigura}, {Howard}, \& {Fang}}]{Dai_mmen}
{Dai}, F., {Winn}, J.~N., {Schlaufman}, K., {et~al.} 2020, \aj, 159, 247,
  \dodoi{10.3847/1538-3881/ab88b8}

\bibitem[{{Dai} {et~al.}(2021){Dai}, {Howard}, {Batalha}, {Beard}, {Behmard},
  {Blunt}, {Brinkman}, {Chontos}, {Crossfield}, {Dalba}, {Dressing}, {Fulton},
  {Giacalone}, {Hill}, {Huber}, {Isaacson}, {Kane}, {Lubin}, {Mayo},
  {Mo{\v{c}}nik}, {Akana Murphy}, {Petigura}, {Rice}, {Robertson}, {Rosenthal},
  {Roy}, {Rubenzahl}, {Weiss}, {Zandt}, {Beichman}, {Ciardi}, {Collins},
  {Gonzales}, {Howell}, {Matson}, {Matthews}, {Schlieder}, {Schwarz}, {Ricker},
  {Vanderspek}, {Latham}, {Seager}, {Winn}, {Jenkins}, {Caldwell}, {Colon},
  {Dragomir}, {Lund}, {McLean}, {Rudat}, \& {Shporer}}]{Dai_1444}
{Dai}, F., {Howard}, A.~W., {Batalha}, N.~M., {et~al.} 2021, \aj, 162, 62,
  \dodoi{10.3847/1538-3881/ac02bd}

\bibitem[{{David} {et~al.}(2019){David}, {Petigura}, {Luger}, {Foreman-Mackey},
  {Livingston}, {Mamajek}, \& {Hillenbrand}}]{David2019}
{David}, T.~J., {Petigura}, E.~A., {Luger}, R., {et~al.} 2019, \apjl, 885, L12,
  \dodoi{10.3847/2041-8213/ab4c99}

\bibitem[{{Deck} \& {Agol}(2015)}]{Deck_chopping}
{Deck}, K.~M., \& {Agol}, E. 2015, \apj, 802, 116,
  \dodoi{10.1088/0004-637X/802/2/116}

\bibitem[{{Deck} {et~al.}(2014){Deck}, {Agol}, {Holman}, \&
  {Nesvorn{\'y}}}]{2014ApJ...787..132D}
{Deck}, K.~M., {Agol}, E., {Holman}, M.~J., \& {Nesvorn{\'y}}, D. 2014, \apj,
  787, 132, \dodoi{10.1088/0004-637X/787/2/132}

\bibitem[{Delisle(2017)}]{Delisle2017}
Delisle, J.~B. 2017, Astronomy and Astrophysics, 605,
  \dodoi{10.1051/0004-6361/201730857}

\bibitem[{{Delisle} \& {Laskar}(2014)}]{Delisle2014}
{Delisle}, J.~B., \& {Laskar}, J. 2014, \aap, 570, L7,
  \dodoi{10.1051/0004-6361/201424227}

\bibitem[{Douglas {et~al.}(2017)Douglas, Agüeros, Covey, \&
  Kraus}]{douglas_2017}
Douglas, S.~T., Agüeros, M.~A., Covey, K.~R., \& Kraus, A. 2017, \apj, 842,
  83, \dodoi{10.3847/1538-4357/aa6e52}

\bibitem[{{Doyle} {et~al.}(2014){Doyle}, {Davies}, {Smalley}, {Chaplin}, \&
  {Elsworth}}]{Doyle2014}
{Doyle}, A.~P., {Davies}, G.~R., {Smalley}, B., {Chaplin}, W.~J., \&
  {Elsworth}, Y. 2014, \mnras, 444, 3592, \dodoi{10.1093/mnras/stu1692}

\bibitem[{Duane {et~al.}(1987)Duane, Kennedy, Pendleton, \&
  Roweth}]{DUANE1987216}
Duane, S., Kennedy, A., Pendleton, B.~J., \& Roweth, D. 1987, Physics Letters
  B, 195, 216 , \dodoi{https://doi.org/10.1016/0370-2693(87)91197-X}

\bibitem[{{Eastman} {et~al.}(2013){Eastman}, {Gaudi}, \& {Agol}}]{Eastman2013}
{Eastman}, J., {Gaudi}, B.~S., \& {Agol}, E. 2013, \pasp, 125, 83,
  \dodoi{10.1086/669497}

\bibitem[{{Espaillat} {et~al.}(2014){Espaillat}, {Muzerolle}, {Najita},
  {Andrews}, {Zhu}, {Calvet}, {Kraus}, {Hashimoto}, {Kraus}, \&
  {D'Alessio}}]{Espaillat}
{Espaillat}, C., {Muzerolle}, J., {Najita}, J., {et~al.} 2014, in Protostars
  and Planets VI, ed. H.~{Beuther}, R.~S. {Klessen}, C.~P. {Dullemond}, \&
  T.~{Henning}, 497, \dodoi{10.2458/azu_uapress_9780816531240-ch022}

\bibitem[{{Fabrycky}(2010)}]{2010arXiv1006.3834F}
{Fabrycky}, D.~C. 2010, ArXiv e-prints.
\newblock \doarXiv{1006.3834}

\bibitem[{{Fabrycky} {et~al.}(2012){Fabrycky}, {Ford}, {Steffen}, {Rowe},
  {Carter}, {Moorhead}, {Batalha}, {Borucki}, {Bryson}, {Buchhave},
  {Christiansen}, {Ciardi}, {Cochran}, {Endl}, {Fanelli}, {Fischer}, {Fressin},
  {Geary}, {Haas}, {Hall}, {Holman}, {Jenkins}, {Koch}, {Latham}, {Li},
  {Lissauer}, {Lucas}, {Marcy}, {Mazeh}, {McCauliff}, {Quinn}, {Ragozzine},
  {Sasselov}, \& {Shporer}}]{Fabrycky_Kepler29}
{Fabrycky}, D.~C., {Ford}, E.~B., {Steffen}, J.~H., {et~al.} 2012, \apj, 750,
  114, \dodoi{10.1088/0004-637X/750/2/114}

\bibitem[{{Fabrycky} {et~al.}(2014){Fabrycky}, {Lissauer}, {Ragozzine}, {Rowe},
  {Steffen}, {Agol}, {Barclay}, {Batalha}, {Borucki}, {Ciardi}, {Ford},
  {Gautier}, {Geary}, {Holman}, {Jenkins}, {Li}, {Morehead}, {Morris},
  {Shporer}, {Smith}, {Still}, \& {Van Cleve}}]{Fabrycky2014}
{Fabrycky}, D.~C., {Lissauer}, J.~J., {Ragozzine}, D., {et~al.} 2014, \apj,
  790, 146, \dodoi{10.1088/0004-637X/790/2/146}

\bibitem[{F\H{u}r\'esz(2008)}]{fureszTRES}
F\H{u}r\'esz, G. 2008, PhD thesis, University of Szeged, Hungary

\bibitem[{{Foreman-Mackey} {et~al.}(2013){Foreman-Mackey}, {Hogg}, {Lang}, \&
  {Goodman}}]{emcee}
{Foreman-Mackey}, D., {Hogg}, D.~W., {Lang}, D., \& {Goodman}, J. 2013, \pasp,
  125, 306, \dodoi{10.1086/670067}

\bibitem[{{Fulton} \& {Petigura}(2018)}]{CKS7}
{Fulton}, B.~J., \& {Petigura}, E.~A. 2018, \aj, 156, 264,
  \dodoi{10.3847/1538-3881/aae828}

\bibitem[{{Fulton} {et~al.}(2017){Fulton}, {Petigura}, {Howard}, {Isaacson},
  {Marcy}, {Cargile}, {Hebb}, {Weiss}, {Johnson}, {Morton}, {Sinukoff},
  {Crossfield}, \& {Hirsch}}]{Fulton}
{Fulton}, B.~J., {Petigura}, E.~A., {Howard}, A.~W., {et~al.} 2017, \aj, 154,
  109, \dodoi{10.3847/1538-3881/aa80eb}

\bibitem[{{Gagn{\'e}} {et~al.}(2018){Gagn{\'e}}, {Mamajek}, {Malo}, {Riedel},
  {Rodriguez}, {Lafreni{\`e}re}, {Faherty}, {Roy-Loubier}, {Pueyo}, {Robin}, \&
  {Doyon}}]{Gagne}
{Gagn{\'e}}, J., {Mamajek}, E.~E., {Malo}, L., {et~al.} 2018, \apj, 856, 23,
  \dodoi{10.3847/1538-4357/aaae09}

\bibitem[{{Gaia Collaboration} {et~al.}(2018){Gaia Collaboration}, {Brown},
  {Vallenari}, {Prusti}, {de Bruijne}, {Babusiaux}, \& {Bailer-Jones}}]{Gaia}
{Gaia Collaboration}, {Brown}, A.~G.~A., {Vallenari}, A., {et~al.} 2018, ArXiv
  e-prints.
\newblock \doarXiv{1804.09365}

\bibitem[{{Gao} \& {Zhang}(2020)}]{Gao}
{Gao}, P., \& {Zhang}, X. 2020, \apj, 890, 93, \dodoi{10.3847/1538-4357/ab6a9b}

\bibitem[{{Gardner} {et~al.}(2006){Gardner}, {Mather}, {Clampin}, {Doyon},
  {Greenhouse}, {Hammel}, {Hutchings}, {Jakobsen}, {Lilly}, {Long}, {Lunine},
  {McCaughrean}, {Mountain}, {Nella}, {Rieke}, {Rieke}, {Rix}, {Smith},
  {Sonneborn}, {Stiavelli}, {Stockman}, {Windhorst}, \& {Wright}}]{Gardner}
{Gardner}, J.~P., {Mather}, J.~C., {Clampin}, M., {et~al.} 2006, \ssr, 123,
  485, \dodoi{10.1007/s11214-006-8315-7}

\bibitem[{Gelman {et~al.}(2014)Gelman, Carlin, Stern, Dunson, Vehtari, \&
  Rubin}]{BB13945229}
Gelman, A., Carlin, J.~B., Stern, H.~S., {et~al.} 2014, Bayesian data analysis,
  3rd edn., Texts in statistical science (CRC Press).
\newblock \url{https://ci.nii.ac.jp/ncid/BB13945229}

\bibitem[{{Ghosh} \& {Lamb}(1979)}]{Ghosh}
{Ghosh}, P., \& {Lamb}, F.~K. 1979, \apj, 234, 296, \dodoi{10.1086/157498}

\bibitem[{{Gillon} {et~al.}(2017){Gillon}, {Triaud}, {Demory}, {Jehin}, {Agol},
  {Deck}, {Lederer}, {de Wit}, {Burdanov}, {Ingalls}, {Bolmont}, {Leconte},
  {Raymond}, {Selsis}, {Turbet}, {Barkaoui}, {Burgasser}, {Burleigh}, {Carey},
  {Chaushev}, {Copperwheat}, {Delrez}, {Fernandes}, {Holdsworth}, {Kotze}, {Van
  Grootel}, {Almleaky}, {Benkhaldoun}, {Magain}, \& {Queloz}}]{Gillon}
{Gillon}, M., {Triaud}, A. H.~M.~J., {Demory}, B.-O., {et~al.} 2017, \nat, 542,
  456, \dodoi{10.1038/nature21360}

\bibitem[{Goldberg \& Batygin(2021)}]{Goldberg2021}
Goldberg, M., \& Batygin, K. 2021, \dodoi{10.3847/1538-3881/abfb78}

\bibitem[{Goldberg \& Batygin(2022)}]{Goldberg2022}
---. 2022, \dodoi{10.3847/1538-3881/ac5961}

\bibitem[{{Goldberg} {et~al.}(2022){Goldberg}, {Batygin}, \&
  {Morbidelli}}]{Goldberg_stability}
{Goldberg}, M., {Batygin}, K., \& {Morbidelli}, A. 2022, arXiv e-prints,
  arXiv:2207.13833.
\newblock \doarXiv{2207.13833}

\bibitem[{{Goldreich} \& {Schlichting}(2014)}]{Goldreich2014}
{Goldreich}, P., \& {Schlichting}, H.~E. 2014, \aj, 147, 32,
  \dodoi{10.1088/0004-6256/147/2/32}

\bibitem[{{Goldreich} \& {Tremaine}(1979)}]{Goldreich1979}
{Goldreich}, P., \& {Tremaine}, S. 1979, \apj, 233, 857, \dodoi{10.1086/157448}

\bibitem[{{Gorti} {et~al.}(2015){Gorti}, {Hollenbach}, \& {Dullemond}}]{Gorti}
{Gorti}, U., {Hollenbach}, D., \& {Dullemond}, C.~P. 2015, \apj, 804, 29,
  \dodoi{10.1088/0004-637X/804/1/29}

\bibitem[{{Go{\'z}dziewski} {et~al.}(2016){Go{\'z}dziewski}, {Migaszewski},
  {Panichi}, \& {Szuszkiewicz}}]{Gozdziewski}
{Go{\'z}dziewski}, K., {Migaszewski}, C., {Panichi}, F., \& {Szuszkiewicz}, E.
  2016, \mnras, 455, L104, \dodoi{10.1093/mnrasl/slv156}

\bibitem[{{Guerrero} {et~al.}(2021){Guerrero}, {Seager}, {Huang}, {Vanderburg},
  {Garcia Soto}, {Mireles}, {Hesse}, {Fong}, {Glidden}, {Shporer}, {Latham},
  {Collins}, {Quinn}, {Burt}, {Dragomir}, {Crossfield}, {Vanderspek},
  {Fausnaugh}, {Burke}, {Ricker}, {Daylan}, {Essack}, {G{\"u}nther}, {Osborn},
  {Pepper}, {Rowden}, {Sha}, {Villanueva}, {Yahalomi}, {Yu}, {Ballard},
  {Batalha}, {Berardo}, {Chontos}, {Dittmann}, {Esquerdo}, {Mikal-Evans},
  {Jayaraman}, {Krishnamurthy}, {Louie}, {Mehrle}, {Niraula}, {Rackham},
  {Rodriguez}, {Rowden}, {Sousa-Silva}, {Watanabe}, {Wong}, {Zhan},
  {Zivanovic}, {Christiansen}, {Ciardi}, {Swain}, {Lund}, {Mullally},
  {Fleming}, {Rodriguez}, {Boyd}, {Quintana}, {Barclay}, {Col{\'o}n},
  {Rinehart}, {Schlieder}, {Clampin}, {Jenkins}, {Twicken}, {Caldwell},
  {Coughlin}, {Henze}, {Lissauer}, {Morris}, {Rose}, {Smith}, {Tenenbaum},
  {Ting}, {Wohler}, {Bakos}, {Bean}, {Berta-Thompson}, {Bieryla}, {Bouma},
  {Buchhave}, {Butler}, {Charbonneau}, {Doty}, {Ge}, {Holman}, {Howard},
  {Kaltenegger}, {Kane}, {Kjeldsen}, {Kreidberg}, {Lin}, {Minsky}, {Narita},
  {Paegert}, {P{\'a}l}, {Palle}, {Sasselov}, {Spencer}, {Sozzetti}, {Stassun},
  {Torres}, {Udry}, \& {Winn}}]{Guerrero}
{Guerrero}, N.~M., {Seager}, S., {Huang}, C.~X., {et~al.} 2021, \apjs, 254, 39,
  \dodoi{10.3847/1538-4365/abefe1}

\bibitem[{{Hadden}(2019)}]{Hadden_resonance}
{Hadden}, S. 2019, \aj, 158, 238, \dodoi{10.3847/1538-3881/ab5287}

\bibitem[{{Hadden} \& {Lithwick}(2014)}]{HaddenLithwick2014}
{Hadden}, S., \& {Lithwick}, Y. 2014, \apj, 787, 80,
  \dodoi{10.1088/0004-637X/787/1/80}

\bibitem[{{Hadden} \& {Lithwick}(2017)}]{Hadden2017}
---. 2017, \aj, 154, 5, \dodoi{10.3847/1538-3881/aa71ef}

\bibitem[{{Hayashi}(1981)}]{Hayashi}
{Hayashi}, C. 1981, Progress of Theoretical Physics Supplement, 70, 35,
  \dodoi{10.1143/PTPS.70.35}

\bibitem[{{Henrard}(1982)}]{Henrard}
{Henrard}, J. 1982, Celestial Mechanics, 27, 3, \dodoi{10.1007/BF01228946}

\bibitem[{{Henrard} {et~al.}(1986){Henrard}, {Lemaitre}, {Milani}, \&
  {Murray}}]{Henrard1986}
{Henrard}, J., {Lemaitre}, A., {Milani}, A., \& {Murray}, C.~D. 1986, Celestial
  Mechanics, 38, 335, \dodoi{10.1007/BF01238924}

\bibitem[{{Hirano} {et~al.}(2011{\natexlab{a}}){Hirano}, {Suto}, {Winn},
  {Taruya}, {Narita}, {Albrecht}, \& {Sato}}]{Hirano}
{Hirano}, T., {Suto}, Y., {Winn}, J.~N., {et~al.} 2011{\natexlab{a}}, \apj,
  742, 69, \dodoi{10.1088/0004-637X/742/2/69}

\bibitem[{{Hirano} {et~al.}(2011{\natexlab{b}}){Hirano}, {Suto}, {Winn},
  {Taruya}, {Narita}, {Albrecht}, \& {Sato}}]{Hirano2011}
---. 2011{\natexlab{b}}, \apj, 742, 69, \dodoi{10.1088/0004-637X/742/2/69}

\bibitem[{{Hirano} {et~al.}(2012){Hirano}, {Narita}, {Sato}, {Takahashi},
  {Masuda}, {Takeda}, {Aoki}, {Tamura}, \& {Suto}}]{Hirano_planet_planet}
{Hirano}, T., {Narita}, N., {Sato}, B., {et~al.} 2012, \apjl, 759, L36,
  \dodoi{10.1088/2041-8205/759/2/L36}

\bibitem[{{Hirano} {et~al.}(2020){Hirano}, {Gaidos}, {Winn}, {Dai}, {Fukui},
  {Kuzuhara}, {Kotani}, {Tamura}, {Hjorth}, {Albrecht}, {Huber}, {Bolmont},
  {Harakawa}, {Hodapp}, {Ishizuka}, {Jacobson}, {Konishi}, {Kudo}, {Kurokawa},
  {Nishikawa}, {Omiya}, {Serizawa}, {Ueda}, \& {Weiss}}]{Hirano2020}
{Hirano}, T., {Gaidos}, E., {Winn}, J.~N., {et~al.} 2020, \apjl, 890, L27,
  \dodoi{10.3847/2041-8213/ab74dc}

\bibitem[{{Hodapp} {et~al.}(2003){Hodapp}, {Jensen}, {Irwin}, {Yamada},
  {Chung}, {Fletcher}, {Robertson}, {Hora}, {Simons}, {Mays}, {Nolan}, {Bec},
  {Merrill}, \& {Fowler}}]{Hodapp}
{Hodapp}, K.~W., {Jensen}, J.~B., {Irwin}, E.~M., {et~al.} 2003, \pasp, 115,
  1388, \dodoi{10.1086/379669}

\bibitem[{{Hormuth} {et~al.}(2008){Hormuth}, {Brandner}, {Hippler}, \&
  {Henning}}]{hormuth08}
{Hormuth}, F., {Brandner}, W., {Hippler}, S., \& {Henning}, T. 2008, Journal of
  Physics Conference Series, 131, 012051,
  \dodoi{10.1088/1742-6596/131/1/012051}

\bibitem[{{Howard} {et~al.}(2010){Howard}, {Johnson}, {Marcy}, {Fischer},
  {Wright}, {Bernat}, {Henry}, {Peek}, {Isaacson}, {Apps}, {Endl}, {Cochran},
  {Valenti}, {Anderson}, \& {Piskunov}}]{Howard}
{Howard}, A.~W., {Johnson}, J.~A., {Marcy}, G.~W., {et~al.} 2010, \apj, 721,
  1467, \dodoi{10.1088/0004-637X/721/2/1467}

\bibitem[{{Howard} {et~al.}(2013){Howard}, {Sanchis-Ojeda}, {Marcy}, {Johnson},
  {Winn}, {Isaacson}, {Fischer}, {Fulton}, {Sinukoff}, \& {Fortney}}]{Howard78}
{Howard}, A.~W., {Sanchis-Ojeda}, R., {Marcy}, G.~W., {et~al.} 2013, \nat, 503,
  381, \dodoi{10.1038/nature12767}

\bibitem[{{Huang} \& {Ormel}(2022)}]{Huang_Ormel}
{Huang}, S., \& {Ormel}, C.~W. 2022, \mnras, 511, 3814,
  \dodoi{10.1093/mnras/stac288}

\bibitem[{{Huber}(2017)}]{Huber17}
{Huber}, D. 2017, {Isoclassify: V1.2}, v1.2,  Zenodo,
  \dodoi{10.5281/zenodo.573372}

\bibitem[{{Huber} {et~al.}(2017){Huber}, {Zinn}, {Bojsen-Hansen},
  {Pinsonneault}, {Sahlholdt}, {Serenelli}, {Silva Aguirre}, {Stassun},
  {Stello}, {Tayar}, {Bastien}, {Bedding}, {Buchhave}, {Chaplin}, {Davies},
  {Garc{\'\i}a}, {Latham}, {Mathur}, {Mosser}, \& {Sharma}}]{Huber}
{Huber}, D., {Zinn}, J., {Bojsen-Hansen}, M., {et~al.} 2017, \apj, 844, 102,
  \dodoi{10.3847/1538-4357/aa75ca}

\bibitem[{{H{\"u}hn} {et~al.}(2021){H{\"u}hn}, {Pichierri}, {Bitsch}, \&
  {Batygin}}]{Huhn}
{H{\"u}hn}, L.~A., {Pichierri}, G., {Bitsch}, B., \& {Batygin}, K. 2021, \aap,
  656, A115, \dodoi{10.1051/0004-6361/202142176}

\bibitem[{{Inamdar} \& {Schlichting}(2016)}]{Inamdar}
{Inamdar}, N.~K., \& {Schlichting}, H.~E. 2016, \apjl, 817, L13,
  \dodoi{10.3847/2041-8205/817/2/L13}

\bibitem[{{Isaacson} \& {Fischer}(2010)}]{Isaacson}
{Isaacson}, H., \& {Fischer}, D. 2010, \apj, 725, 875,
  \dodoi{10.1088/0004-637X/725/1/875}

\bibitem[{{Izidoro} {et~al.}(2021){Izidoro}, {Bitsch}, {Raymond}, {Johansen},
  {Morbidelli}, {Lambrechts}, \& {Jacobson}}]{Izidoro2021}
{Izidoro}, A., {Bitsch}, B., {Raymond}, S.~N., {et~al.} 2021, \aap, 650, A152,
  \dodoi{10.1051/0004-6361/201935336}

\bibitem[{{Izidoro} {et~al.}(2017){Izidoro}, {Ogihara}, {Raymond},
  {Morbidelli}, {Pierens}, {Bitsch}, {Cossou}, \& {Hersant}}]{Izidoro}
{Izidoro}, A., {Ogihara}, M., {Raymond}, S.~N., {et~al.} 2017, \mnras, 470,
  1750, \dodoi{10.1093/mnras/stx1232}

\bibitem[{{Jenkins} {et~al.}(2020){Jenkins}, {Tenenbaum}, {Seader}, {Burke},
  {McCauliff}, {Smith}, {Twicken}, \& {Chandrasekaran}}]{Jenkins2020J}
{Jenkins}, J.~M., {Tenenbaum}, P., {Seader}, S., {et~al.} 2020, {Kepler Data
  Processing Handbook: Transiting Planet Search}, Kepler Science Document
  KSCI-19081-003, id. 9. Edited by Jon M. Jenkins.

\bibitem[{{Jenkins} {et~al.}(2016){Jenkins}, {Twicken}, {McCauliff},
  {Campbell}, {Sanderfer}, {Lung}, {Mansouri-Samani}, {Girouard}, {Tenenbaum},
  {Klaus}, {Smith}, {Caldwell}, {Chacon}, {Henze}, {Heiges}, {Latham},
  {Morgan}, {Swade}, {Rinehart}, \& {Vanderspek}}]{jenkinsSPOC2016}
{Jenkins}, J.~M., {Twicken}, J.~D., {McCauliff}, S., {et~al.} 2016, in
  \procspie, Vol. 9913, Software and Cyberinfrastructure for Astronomy IV,
  99133E, \dodoi{10.1117/12.2233418}

\bibitem[{{Jensen} \& {Millholland}(2022)}]{Jensen_Millholland}
{Jensen}, D., \& {Millholland}, S.~C. 2022, arXiv e-prints, arXiv:2208.05423.
\newblock \doarXiv{2208.05423}

\bibitem[{{Jontof-Hutter} {et~al.}(2016){Jontof-Hutter}, {Ford}, {Rowe},
  {Lissauer}, {Fabrycky}, {Van Laerhoven}, {Agol}, {Deck}, {Holczer}, \&
  {Mazeh}}]{JontofHutter2016}
{Jontof-Hutter}, D., {Ford}, E.~B., {Rowe}, J.~F., {et~al.} 2016, \apj, 820,
  39, \dodoi{10.3847/0004-637X/820/1/39}

\bibitem[{{Kervella} {et~al.}(2022){Kervella}, {Arenou}, \&
  {Th{\'e}venin}}]{Gaia_dr3_companion}
{Kervella}, P., {Arenou}, F., \& {Th{\'e}venin}, F. 2022, \aap, 657, A7,
  \dodoi{10.1051/0004-6361/202142146}

\bibitem[{{Kipping}(2013)}]{Kipping}
{Kipping}, D.~M. 2013, \mnras, 435, 2152, \dodoi{10.1093/mnras/stt1435}

\bibitem[{{Kley} {et~al.}(2005){Kley}, {Lee}, {Murray}, \& {Peale}}]{Kley_2005}
{Kley}, W., {Lee}, M.~H., {Murray}, N., \& {Peale}, S.~J. 2005, \aap, 437, 727,
  \dodoi{10.1051/0004-6361:20052656}

\bibitem[{{Kley} \& {Nelson}(2012)}]{Kley_2012}
{Kley}, W., \& {Nelson}, R.~P. 2012, \araa, 50, 211,
  \dodoi{10.1146/annurev-astro-081811-125523}

\bibitem[{{Kokubo} \& {Makino}(2004)}]{2004PASJ...56..861K}
{Kokubo}, E., \& {Makino}, J. 2004, \pasj, 56, 861,
  \dodoi{10.1093/pasj/56.5.861}

\bibitem[{{Kov{\'a}cs} {et~al.}(2002){Kov{\'a}cs}, {Zucker}, \&
  {Mazeh}}]{Kovac2002}
{Kov{\'a}cs}, G., {Zucker}, S., \& {Mazeh}, T. 2002, \aap, 391, 369,
  \dodoi{10.1051/0004-6361:20020802}

\bibitem[{{Kreidberg}(2015)}]{Kreidberg2015}
{Kreidberg}, L. 2015, \pasp, 127, 1161, \dodoi{10.1086/683602}

\bibitem[{{Kretke} \& {Lin}(2012)}]{Kretke2012}
{Kretke}, K.~A., \& {Lin}, D.~N.~C. 2012, \apj, 755, 74,
  \dodoi{10.1088/0004-637X/755/1/74}

\bibitem[{Laune {et~al.}(2022)Laune, Rodet, \& Lai}]{Laune2022}
Laune, J., Rodet, L., \& Lai, D. 2022.
\newblock \url{http://arxiv.org/abs/2206.04810}

\bibitem[{{Lee} \& {Chiang}(2015)}]{Lee}
{Lee}, E.~J., \& {Chiang}, E. 2015, \apj, 811, 41,
  \dodoi{10.1088/0004-637X/811/1/41}

\bibitem[{{Lee} \& {Chiang}(2016)}]{Lee2016}
---. 2016, \apj, 817, 90, \dodoi{10.3847/0004-637X/817/2/90}

\bibitem[{{Lee} {et~al.}(2013){Lee}, {Fabrycky}, \& {Lin}}]{LeeMH}
{Lee}, M.~H., {Fabrycky}, D., \& {Lin}, D.~N.~C. 2013, \apj, 774, 52,
  \dodoi{10.1088/0004-637X/774/1/52}

\bibitem[{{Lee} \& {Peale}(2002)}]{Lee2002}
{Lee}, M.~H., \& {Peale}, S.~J. 2002, \apj, 567, 596, \dodoi{10.1086/338504}

\bibitem[{Leleu {et~al.}(2021)Leleu, Alibert, Hara, Hooton, Wilson, Robutel,
  Delisle, Laskar, Hoyer, Lovis, Bryant, Ducrot, Cabrera, Delrez, Acton,
  Adibekyan, Allart, Prieto, Alonso, Alves, Anderson, Angerhausen, Escudé,
  Asquier, Barrado, Barros, Baumjohann, Bayliss, Beck, Beck, Bekkelien, Benz,
  Billot, Bonfanti, Bonfils, Bouchy, Bourrier, Boué, Brandeker, Broeg, Buder,
  Burdanov, Burleigh, Bárczy, Cameron, Chamberlain, Charnoz, Cooke, Damme,
  Correia, Cristiani, Damasso, Davies, Deleuil, Demangeon, Demory, Marcantonio,
  Persio, Dumusque, Ehrenreich, Erikson, Figueira, Fortier, Fossati, Fridlund,
  Futyan, Gandolfi, Muñoz, Garcia, Gill, Gillen, Gillon, Goad, Hernández,
  Guedel, Günther, Haldemann, Henderson, Heng, Hogan, Isaak, Jehin, Jenkins,
  Jordán, Kiss, Kristiansen, Lam, Lavie, des Etangs, Lendl, Lillo-Box, Curto,
  Magrin, Martins, Maxted, McCormac, Mehner, Micela, Molaro, Moyano, Murray,
  Nascimbeni, Nunes, Olofsson, Osborn, Oshagh, Ottensamer, Pagano, Pallé,
  Pedersen, Pepe, Persson, Peter, Piotto, Polenta, Pollacco, Poretti, Pozuelos,
  Queloz, Ragazzoni, Rando, Ratti, Rauer, Raynard, Rebolo, Reimers, Ribas,
  Santos, Scandariato, Schneider, Sebastian, Sestovic, Simon, Smith, Sousa,
  Sozzetti, Steller, Mascareño, Szabó, Ségransan, Thomas, Thompson,
  Tilbrook, Triaud, Turner, Udry, Grootel, Venus, Verrecchia, Vines, Walton,
  West, Wheatley, Wolter, \& Osorio}]{Leleu2021}
Leleu, A., Alibert, Y., Hara, N.~C., {et~al.} 2021,
  \dodoi{10.1051/0004-6361/202039767}

\bibitem[{{Lin} \& {Papaloizou}(1986)}]{Lin1986}
{Lin}, D.~N.~C., \& {Papaloizou}, J. 1986, \apj, 309, 846,
  \dodoi{10.1086/164653}

\bibitem[{Lithwick \& Wu(2012)}]{Lithwick_repulsion}
Lithwick, Y., \& Wu, Y. 2012, The Astrophysical Journal, 756, L11,
  \dodoi{10.1088/2041-8205/756/1/L11}

\bibitem[{{Lithwick} {et~al.}(2012){Lithwick}, {Xie}, \& {Wu}}]{Lithwick_ttv}
{Lithwick}, Y., {Xie}, J., \& {Wu}, Y. 2012, \apj, 761, 122,
  \dodoi{10.1088/0004-637X/761/2/122}

\bibitem[{Lomb(1976)}]{Lomb1976}
Lomb, N.~R. 1976, Astrophysics and Space Science, 39, 447,
  \dodoi{10.1007/BF00648343}

\bibitem[{Luger {et~al.}(2017)Luger, Sestovic, Kruse, Grimm, Demory, Agol,
  Bolmont, Fabrycky, Fernandes, Grootel, Burgasser, Gillon, Ingalls, Jehin,
  Raymond, Selsis, Triaud, Barclay, Barentsen, Howell, Delrez, de~Wit,
  Foreman-Mackey, Holdsworth, Leconte, Lederer, Turbet, Almleaky, Benkhaldoun,
  Magain, Morris, Heng, \& Queloz}]{Luger2017}
Luger, R., Sestovic, M., Kruse, E., {et~al.} 2017,
  \dodoi{10.1038/s41550-017-0129}

\bibitem[{Macdonald \& Dawson(2018)}]{Macdonald2018}
Macdonald, M.~G., \& Dawson, R.~I. 2018

\bibitem[{MacDonald {et~al.}(2016)MacDonald, Ragozzine, Fabrycky, Ford, Holman,
  Isaacson, Lissauer, Lopez, Mazeh, Rogers, Rowe, Steffen, \&
  Torres}]{MacDonald2016}
MacDonald, M.~G., Ragozzine, D., Fabrycky, D.~C., {et~al.} 2016,
  \dodoi{10.3847/0004-6256/152/4/105}

\bibitem[{Mah(2018)}]{Mah}
Mah, J. 2018, Formation and dynamics of the resonant chain in the trappist-1
  exoplanet system.
\newblock \url{http://hdl.handle.net/10722/265344}

\bibitem[{{Mamajek} \& {Hillenbrand}(2008)}]{Mamajek}
{Mamajek}, E.~E., \& {Hillenbrand}, L.~A. 2008, \apj, 687, 1264,
  \dodoi{10.1086/591785}

\bibitem[{Masset {et~al.}(2006)Masset, Morbidelli, Crida, \&
  Ferreira}]{Masset_2006}
Masset, F.~S., Morbidelli, A., Crida, A., \& Ferreira, J. 2006, The
  Astrophysical Journal, 642, 478, \dodoi{10.1086/500967}

\bibitem[{{Masuda} \& {Winn}(2020)}]{Masuda_vsini}
{Masuda}, K., \& {Winn}, J.~N. 2020, \aj, 159, 81,
  \dodoi{10.3847/1538-3881/ab65be}

\bibitem[{{Mayor} {et~al.}(2003){Mayor}, {Pepe}, {Queloz}, {Bouchy},
  {Rupprecht}, {Lo Curto}, {Avila}, {Benz}, {Bertaux}, {Bonfils}, {Dall},
  {Dekker}, {Delabre}, {Eckert}, {Fleury}, {Gilliotte}, {Gojak}, {Guzman},
  {Kohler}, {Lizon}, {Longinotti}, {Lovis}, {Megevand}, {Pasquini}, {Reyes},
  {Sivan}, {Sosnowska}, {Soto}, {Udry}, {van Kesteren}, {Weber}, \&
  {Weilenmann}}]{Mayor2003}
{Mayor}, M., {Pepe}, F., {Queloz}, D., {et~al.} 2003, The Messenger, 114, 20

\bibitem[{{McNeil} {et~al.}(2005){McNeil}, {Duncan}, \& {Levison}}]{McNeil}
{McNeil}, D., {Duncan}, M., \& {Levison}, H.~F. 2005, \aj, 130, 2884,
  \dodoi{10.1086/497687}

\bibitem[{{Migaszewski} {et~al.}(2017){Migaszewski}, {Go{\'z}dziewski}, \&
  {Panichi}}]{Migaszewski}
{Migaszewski}, C., {Go{\'z}dziewski}, K., \& {Panichi}, F. 2017, \mnras, 465,
  2366, \dodoi{10.1093/mnras/stw2866}

\bibitem[{{Millholland}(2019)}]{Millholland2019}
{Millholland}, S. 2019, \apj, 886, 72, \dodoi{10.3847/1538-4357/ab4c3f}

\bibitem[{{Millholland} \& {Laughlin}(2019)}]{Millholland_obliquity}
{Millholland}, S., \& {Laughlin}, G. 2019, Nature Astronomy, 3, 424,
  \dodoi{10.1038/s41550-019-0701-7}

\bibitem[{{Millholland} {et~al.}(2017){Millholland}, {Wang}, \&
  {Laughlin}}]{Millholland_peas}
{Millholland}, S., {Wang}, S., \& {Laughlin}, G. 2017, \apjl, 849, L33,
  \dodoi{10.3847/2041-8213/aa9714}

\bibitem[{{Millholland} {et~al.}(2018{\natexlab{a}}){Millholland}, {Laughlin},
  {Teske}, {Butler}, {Burt}, {Holden}, {Vogt}, {Crane}, {Shectman}, \&
  {Thompson}}]{Millholland_2018_resonance}
{Millholland}, S., {Laughlin}, G., {Teske}, J., {et~al.} 2018{\natexlab{a}},
  \aj, 155, 106, \dodoi{10.3847/1538-3881/aaa894}

\bibitem[{{Millholland} {et~al.}(2018{\natexlab{b}}){Millholland}, {Laughlin},
  {Teske}, {Butler}, {Burt}, {Holden}, {Vogt}, {Crane}, {Shectman}, \&
  {Thompson}}]{Millholland876}
---. 2018{\natexlab{b}}, \aj, 155, 106, \dodoi{10.3847/1538-3881/aaa894}

\bibitem[{Mills \& Fabrycky(2017)}]{Mills2017}
Mills, S.~M., \& Fabrycky, D.~C. 2017, \dodoi{10.3847/2041-8213/aa6543}

\bibitem[{{Mills} {et~al.}(2016){Mills}, {Fabrycky}, {Migaszewski}, {Ford},
  {Petigura}, \& {Isaacson}}]{MillsNature}
{Mills}, S.~M., {Fabrycky}, D.~C., {Migaszewski}, C., {et~al.} 2016, \nat, 533,
  509, \dodoi{10.1038/nature17445}

\bibitem[{Mills {et~al.}(2016)Mills, Fabrycky, Migaszewski, Ford, Petigura, \&
  Isaacson}]{Mills2016}
Mills, S.~M., Fabrycky, D.~C., Migaszewski, C., {et~al.} 2016, Nature, 533,
  509, \dodoi{10.1038/nature17445}

\bibitem[{{Mills} \& {Mazeh}(2017)}]{MillsMazeh}
{Mills}, S.~M., \& {Mazeh}, T. 2017, \apjl, 839, L8,
  \dodoi{10.3847/2041-8213/aa67eb}

\bibitem[{{Moore} {et~al.}(2013){Moore}, {Hasan}, \& {Quillen}}]{Moore}
{Moore}, A., {Hasan}, I., \& {Quillen}, A.~C. 2013, \mnras, 432, 1196,
  \dodoi{10.1093/mnras/stt535}

\bibitem[{{Morris} {et~al.}(2020){Morris}, {Twicken}, {Smith}, {Clarke},
  {Jenkins}, {Bryson}, {Girouard}, \& {Klaus}}]{Morris2020}
{Morris}, R.~L., {Twicken}, J.~D., {Smith}, J.~C., {et~al.} 2020, {Kepler Data
  Processing Handbook: Photometric Analysis}, Kepler Science Document
  KSCI-19081-003, id. 6. Edited by Jon M. Jenkins.

\bibitem[{{Murray} \& {Dermott}(1999)}]{Murray}
{Murray}, C.~D., \& {Dermott}, S.~F. 1999, {Solar system dynamics}

\bibitem[{{Nelson}(2018)}]{Nelson2018}
{Nelson}, R.~P. 2018, in Handbook of Exoplanets, ed. H.~J. {Deeg} \& J.~A.
  {Belmonte}, 139, \dodoi{10.1007/978-3-319-55333-7_139}

\bibitem[{{Nesvorn{\'y}} \& {Vokrouhlick{\'y}}(2016)}]{Nesvorny}
{Nesvorn{\'y}}, D., \& {Vokrouhlick{\'y}}, D. 2016, \apj, 823, 72,
  \dodoi{10.3847/0004-637X/823/2/72}

\bibitem[{Newville {et~al.}(2014)Newville, Stensitzki, Allen, \&
  Ingargiola}]{LM}
Newville, M., Stensitzki, T., Allen, D.~B., \& Ingargiola, A. 2014, {LMFIT:
  Non-Linear Least-Square Minimization and Curve-Fitting for Python}, 0.8.0,
  Zenodo, \dodoi{10.5281/zenodo.11813}

\bibitem[{{Ogihara} \& {Ida}(2009)}]{Ogihara}
{Ogihara}, M., \& {Ida}, S. 2009, \apj, 699, 824,
  \dodoi{10.1088/0004-637X/699/1/824}

\bibitem[{{Ostriker} \& {Shu}(1995)}]{Ostriker}
{Ostriker}, E.~C., \& {Shu}, F.~H. 1995, \apj, 447, 813, \dodoi{10.1086/175920}

\bibitem[{{Owen} \& {Campos Estrada}(2020)}]{OwenCampos}
{Owen}, J.~E., \& {Campos Estrada}, B. 2020, \mnras, 491, 5287,
  \dodoi{10.1093/mnras/stz3435}

\bibitem[{{Owen} \& {Wu}(2017)}]{OwenWu}
{Owen}, J.~E., \& {Wu}, Y. 2017, \apj, 847, 29,
  \dodoi{10.3847/1538-4357/aa890a}

\bibitem[{{Papaloizou} \& {Terquem}(2010)}]{Papaloizou2010}
{Papaloizou}, J. C.~B., \& {Terquem}, C. 2010, \mnras, 405, 573,
  \dodoi{10.1111/j.1365-2966.2010.16477.x}

\bibitem[{{Peale}(1969)}]{Peale_cassini}
{Peale}, S.~J. 1969, \aj, 74, 483, \dodoi{10.1086/110825}

\bibitem[{{Petigura} {et~al.}(2017){Petigura}, {Howard}, {Marcy}, {Johnson},
  {Isaacson}, {Cargile}, {Hebb}, {Fulton}, {Weiss}, {Morton}, {Winn}, {Rogers},
  {Sinukoff}, {Hirsch}, \& {Crossfield}}]{CKS1}
{Petigura}, E.~A., {Howard}, A.~W., {Marcy}, G.~W., {et~al.} 2017, \aj, 154,
  107, \dodoi{10.3847/1538-3881/aa80de}

\bibitem[{{Petit} {et~al.}(2020){Petit}, {Petigura}, {Davies}, \&
  {Johansen}}]{Petit}
{Petit}, A.~C., {Petigura}, E.~A., {Davies}, M.~B., \& {Johansen}, A. 2020,
  \mnras, 496, 3101, \dodoi{10.1093/mnras/staa1736}

\bibitem[{{Petrovich} {et~al.}(2018){Petrovich}, {Deibert}, \&
  {Wu}}]{Petrovich}
{Petrovich}, C., {Deibert}, E., \& {Wu}, Y. 2018, ArXiv e-prints.
\newblock \doarXiv{1804.05065}

\bibitem[{Phan {et~al.}(2019)Phan, Pradhan, \& Jankowiak}]{phan2019composable}
Phan, D., Pradhan, N., \& Jankowiak, M. 2019, arXiv preprint arXiv:1912.11554

\bibitem[{Pichierri {et~al.}(2021)Pichierri, Batygin, \&
  Morbidelli}]{Pichierri2021}
Pichierri, G., Batygin, K., \& Morbidelli, A. 2021.
\newblock \url{https://exoplanetarchive.ipac.}

\bibitem[{{Pichierri} {et~al.}(2018){Pichierri}, {Morbidelli}, \&
  {Crida}}]{Pichierri}
{Pichierri}, G., {Morbidelli}, A., \& {Crida}, A. 2018, Celestial Mechanics and
  Dynamical Astronomy, 130, 54, \dodoi{10.1007/s10569-018-9848-2}

\bibitem[{{Pu} \& {Wu}(2015)}]{PuWU}
{Pu}, B., \& {Wu}, Y. 2015, \apj, 807, 44, \dodoi{10.1088/0004-637X/807/1/44}

\bibitem[{{Ramos} {et~al.}(2017){Ramos}, {Charalambous},
  {Ben{\'\i}tez-Llambay}, \& {Beaug{\'e}}}]{Ramos}
{Ramos}, X.~S., {Charalambous}, C., {Ben{\'\i}tez-Llambay}, P., \&
  {Beaug{\'e}}, C. 2017, \aap, 602, A101, \dodoi{10.1051/0004-6361/201629642}

\bibitem[{Raymond {et~al.}(2021)Raymond, Izidoro, Bolmont, Dorn, Selsis,
  Turbet, Agol, Barth, Carone, Dasgupta, Gillon, \& Grimm}]{Raymond2021}
Raymond, S.~N., Izidoro, A., Bolmont, E., {et~al.} 2021,
  \dodoi{10.1038/s41550-021-01518-6}

\bibitem[{{Rein} \& {Liu}(2012)}]{Rein}
{Rein}, H., \& {Liu}, S.~F. 2012, \aap, 537, A128,
  \dodoi{10.1051/0004-6361/201118085}

\bibitem[{{Rein} \& {Papaloizou}(2009)}]{Rein2009}
{Rein}, H., \& {Papaloizou}, J.~C.~B. 2009, \aap, 497, 595,
  \dodoi{10.1051/0004-6361/200811330}

\bibitem[{{Rein} \& {Spiegel}(2015)}]{IAS15}
{Rein}, H., \& {Spiegel}, D.~S. 2015, \mnras, 446, 1424,
  \dodoi{10.1093/mnras/stu2164}

\bibitem[{{Rein} \& {Tamayo}(2015)}]{2015MNRAS.452..376R}
{Rein}, H., \& {Tamayo}, D. 2015, \mnras, 452, 376,
  \dodoi{10.1093/mnras/stv1257}

\bibitem[{{Ricker} {et~al.}(2014){Ricker}, {Winn}, {Vanderspek}, {Latham},
  {Bakos}, {Bean}, {Berta-Thompson}, {Brown}, {Buchhave}, {Butler}, {Butler},
  {Chaplin}, {Charbonneau}, {Christensen-Dalsgaard}, {Clampin}, {Deming},
  {Doty}, {De Lee}, {Dressing}, {Dunham}, {Endl}, {Fressin}, {Ge}, {Henning},
  {Holman}, {Howard}, {Ida}, {Jenkins}, {Jernigan}, {Johnson}, {Kaltenegger},
  {Kawai}, {Kjeldsen}, {Laughlin}, {Levine}, {Lin}, {Lissauer}, {MacQueen},
  {Marcy}, {McCullough}, {Morton}, {Narita}, {Paegert}, {Palle}, {Pepe},
  {Pepper}, {Quirrenbach}, {Rinehart}, {Sasselov}, {Sato}, {Seager},
  {Sozzetti}, {Stassun}, {Sullivan}, {Szentgyorgyi}, {Torres}, {Udry}, \&
  {Villasenor}}]{Ricker}
{Ricker}, G.~R., {Winn}, J.~N., {Vanderspek}, R., {et~al.} 2014, Society of
  Photo-Optical Instrumentation Engineers (SPIE) Conference Series, Vol. 9143,
  {Transiting Exoplanet Survey Satellite (TESS)}, 914320,
  \dodoi{10.1117/12.2063489}

\bibitem[{Rivera {et~al.}(2010)Rivera, Laughlin, Butler, Vogt, Haghighipour, \&
  Meschiari}]{Rivera}
Rivera, E.~J., Laughlin, G., Butler, R.~P., {et~al.} 2010, The Astrophysical
  Journal, 719, 890, \dodoi{10.1088/0004-637x/719/1/890}

\bibitem[{{Scargle}(1982)}]{Scargle1982}
{Scargle}, J.~D. 1982, \apj, 263, 835, \dodoi{10.1086/160554}

\bibitem[{{Schlaufman}(2010)}]{Schlaufman2010}
{Schlaufman}, K.~C. 2010, \apj, 719, 602, \dodoi{10.1088/0004-637X/719/1/602}

\bibitem[{Schwarz(1978)}]{Schwarz}
Schwarz, G. 1978, The Annals of Statistics, 6, 461 ,
  \dodoi{10.1214/aos/1176344136}

\bibitem[{{Scott} {et~al.}(2021){Scott}, {Howell}, {Gnilka}, {Stephens},
  {Salinas}, {Matson}, {Furlan}, {Horch}, {Everett}, {Ciardi}, {Mills}, \&
  {Quigley}}]{Scott2021}
{Scott}, N.~J., {Howell}, S.~B., {Gnilka}, C.~L., {et~al.} 2021, Frontiers in
  Astronomy and Space Sciences, 8, 138, \dodoi{10.3389/fspas.2021.716560}

\bibitem[{{Seager} \& {Mall{\'e}n-Ornelas}(2003)}]{Seager}
{Seager}, S., \& {Mall{\'e}n-Ornelas}, G. 2003, \apj, 585, 1038,
  \dodoi{10.1086/346105}

\bibitem[{{Sessin} \& {Ferraz-Mello}(1984)}]{Sessin}
{Sessin}, W., \& {Ferraz-Mello}, S. 1984, Celestial Mechanics, 32, 307,
  \dodoi{10.1007/BF01229087}

\bibitem[{{Siegel} \& {Fabrycky}(2021)}]{Siegel}
{Siegel}, J.~C., \& {Fabrycky}, D. 2021, \aj, 161, 290,
  \dodoi{10.3847/1538-3881/abf8a6}

\bibitem[{Silburt \& Rein(2015)}]{Silburt}
Silburt, A., \& Rein, H. 2015, Monthly Notices of the Royal Astronomical
  Society, 453, 4089, \dodoi{10.1093/mnras/stv1924}

\bibitem[{{Sinclair}(1975)}]{Sinclair}
{Sinclair}, A.~T. 1975, \mnras, 171, 59, \dodoi{10.1093/mnras/171.1.59}

\bibitem[{{Smith} {et~al.}(2012){Smith}, {Stumpe}, {Van Cleve}, {Jenkins},
  {Barclay}, {Fanelli}, {Girouard}, {Kolodziejczak}, {McCauliff}, {Morris}, \&
  {Twicken}}]{Smith2012}
{Smith}, J.~C., {Stumpe}, M.~C., {Van Cleve}, J.~E., {et~al.} 2012, \pasp, 124,
  1000, \dodoi{10.1086/667697}

\bibitem[{{Spake} {et~al.}(2018){Spake}, {Sing}, {Evans}, {Oklop{\v{c}}i{\'c}},
  {}, {Bourrier}, {Kreidberg}, {Rackham}, {Irwin}, {Ehrenreich}, {Wyttenbach},
  {Wakeford}, {Zhou}, {Chubb}, {Nikolov}, {Goyal}, {Henry}, {Williamson},
  {Blumenthal}, {Anderson}, {Hellier}, {Charbonneau}, {Udry}, \&
  {Madhusudhan}}]{Spake}
{Spake}, J.~J., {Sing}, D.~K., {Evans}, T.~M., {et~al.} 2018, \nat, 557, 68,
  \dodoi{10.1038/s41586-018-0067-5}

\bibitem[{{Spalding} \& {Batygin}(2016)}]{Spalding}
{Spalding}, C., \& {Batygin}, K. 2016, \apj, 830, 5,
  \dodoi{10.3847/0004-637X/830/1/5}

\bibitem[{{Stassun} {et~al.}(2019){Stassun}, {Oelkers}, {Paegert}, {Torres},
  {Pepper}, {De Lee}, {Collins}, {Latham}, {Muirhead}, {Chittidi},
  {Rojas-Ayala}, {Fleming}, {Rose}, {Tenenbaum}, {Ting}, {Kane}, {Barclay},
  {Bean}, {Brassuer}, {Charbonneau}, {Ge}, {Lissauer}, {Mann}, {McLean},
  {Mullally}, {Narita}, {Plavchan}, {Ricker}, {Sasselov}, {Seager}, {Sharma},
  {Shiao}, {Sozzetti}, {Stello}, {Vanderspek}, {Wallace}, \& {Winn}}]{Stassun}
{Stassun}, K.~G., {Oelkers}, R.~J., {Paegert}, M., {et~al.} 2019, \aj, 158,
  138, \dodoi{10.3847/1538-3881/ab3467}

\bibitem[{{Steffen}(2016)}]{Steffen_ttv}
{Steffen}, J.~H. 2016, \mnras, 457, 4384, \dodoi{10.1093/mnras/stw241}

\bibitem[{{Steffen} \& {Coughlin}(2016)}]{Steffen2016}
{Steffen}, J.~H., \& {Coughlin}, J.~L. 2016, Proceedings of the National
  Academy of Science, 113, 12023, \dodoi{10.1073/pnas.1606658113}

\bibitem[{{Steffen} \& {Hwang}(2015)}]{Steffen2015}
{Steffen}, J.~H., \& {Hwang}, J.~A. 2015, \mnras, 448, 1956,
  \dodoi{10.1093/mnras/stv104}

\bibitem[{{Stumpe} {et~al.}(2014){Stumpe}, {Smith}, {Catanzarite}, {Van Cleve},
  {Jenkins}, {Twicken}, \& {Girouard}}]{Stumpe2014}
{Stumpe}, M.~C., {Smith}, J.~C., {Catanzarite}, J.~H., {et~al.} 2014, \pasp,
  126, 100, \dodoi{10.1086/674989}

\bibitem[{{Stumpe} {et~al.}(2012){Stumpe}, {Smith}, {Van Cleve}, {Twicken},
  {Barclay}, {Fanelli}, {Girouard}, {Jenkins}, {Kolodziejczak}, {McCauliff}, \&
  {Morris}}]{Stumpe2012}
{Stumpe}, M.~C., {Smith}, J.~C., {Van Cleve}, J.~E., {et~al.} 2012, \pasp, 124,
  985, \dodoi{10.1086/667698}

\bibitem[{Tamayo {et~al.}(2017)Tamayo, Rein, Petrovich, \& Murray}]{Tamayo2017}
Tamayo, D., Rein, H., Petrovich, C., \& Murray, N. 2017, The Astrophysical
  Journal Letters, 840, L19, \dodoi{10.3847/2041-8213/AA70EA}

\bibitem[{{Tamayo} {et~al.}(2020){Tamayo}, {Rein}, {Shi}, \&
  {Hernandez}}]{Tamayo_x}
{Tamayo}, D., {Rein}, H., {Shi}, P., \& {Hernandez}, D.~M. 2020, \mnras, 491,
  2885, \dodoi{10.1093/mnras/stz2870}

\bibitem[{{Tayar} {et~al.}(2020){Tayar}, {Claytor}, {Huber}, \& {van
  Saders}}]{Tayar}
{Tayar}, J., {Claytor}, Z.~R., {Huber}, D., \& {van Saders}, J. 2020, arXiv
  e-prints, arXiv:2012.07957.
\newblock \doarXiv{2012.07957}

\bibitem[{{Tejada Arevalo} {et~al.}(2022){Tejada Arevalo}, {Tamayo}, \&
  {Cranmer}}]{Tejada}
{Tejada Arevalo}, R., {Tamayo}, D., \& {Cranmer}, M. 2022, \apjl, 932, L12,
  \dodoi{10.3847/2041-8213/ac70e0}

\bibitem[{Terquem \& Papaloizou(2007)}]{Terquem_2007}
Terquem, C., \& Papaloizou, J. C.~B. 2007, The Astrophysical Journal, 654,
  1110, \dodoi{10.1086/509497}

\bibitem[{{Terquem} \& {Papaloizou}(2019)}]{TerquemPapaloizou2019}
{Terquem}, C., \& {Papaloizou}, J. C.~B. 2019, \mnras, 482, 530,
  \dodoi{10.1093/mnras/sty2693}

\bibitem[{{Tofflemire} {et~al.}(2021){Tofflemire}, {Rizzuto}, {Newton},
  {Kraus}, {Mann}, {Vanderburg}, {Nelson}, {Hawkins}, {Wood}, {Zhou}, {Quinn},
  {Howell}, {Collins}, {Schwarz}, {Stassun}, {Bouma}, {Essack}, {Osborn},
  {Boyd}, {F{\H{u}}r{\'e}sz}, {Glidden}, {Twicken}, {Wohler}, {McLean},
  {Ricker}, {Vanderspek}, {Latham}, {Seager}, {Winn}, \&
  {Jenkins}}]{Tofflemire}
{Tofflemire}, B.~M., {Rizzuto}, A.~C., {Newton}, E.~R., {et~al.} 2021, \aj,
  161, 171, \dodoi{10.3847/1538-3881/abdf53}

\bibitem[{{Twicken} {et~al.}(2010){Twicken}, {Clarke}, {Bryson}, {Tenenbaum},
  {Wu}, {Jenkins}, {Girouard}, \& {Klaus}}]{Twicken2010}
{Twicken}, J.~D., {Clarke}, B.~D., {Bryson}, S.~T., {et~al.} 2010, in Society
  of Photo-Optical Instrumentation Engineers (SPIE) Conference Series, Vol.
  7740, Software and Cyberinfrastructure for Astronomy, ed. N.~M. {Radziwill}
  \& A.~{Bridger}, 774023, \dodoi{10.1117/12.856790}

\bibitem[{{Vissapragada} {et~al.}(2020){Vissapragada}, {Jontof-Hutter},
  {Shporer}, {Knutson}, {Liu}, {Thorngren}, {Lee}, {Chachan}, {Mawet},
  {Millar-Blanchaer}, {Nilsson}, {Tinyanont}, {Vasisht}, \&
  {Wright}}]{Vissapragada29}
{Vissapragada}, S., {Jontof-Hutter}, D., {Shporer}, A., {et~al.} 2020, \aj,
  159, 108, \dodoi{10.3847/1538-3881/ab65c8}

\bibitem[{{Vogt} {et~al.}(1994){Vogt}, {Allen}, {Bigelow}, {Bresee}, {Brown},
  {Cantrall}, {Conrad}, {Couture}, {Delaney}, {Epps}, {Hilyard}, {Hilyard},
  {Horn}, {Jern}, {Kanto}, {Keane}, {Kibrick}, {Lewis}, {Osborne},
  {Pardeilhan}, {Pfister}, {Ricketts}, {Robinson}, {Stover}, {Tucker}, {Ward},
  \& {Wei}}]{HIRES}
{Vogt}, S.~S., {Allen}, S.~L., {Bigelow}, B.~C., {et~al.} 1994, in Society of
  Photo-Optical Instrumentation Engineers (SPIE) Conference Series, Vol. 2198,
  Instrumentation in Astronomy VIII, ed. D.~L. {Crawford} \& E.~R. {Craine},
  362, \dodoi{10.1117/12.176725}

\bibitem[{{Wang} \& {Dai}(2018)}]{WangDai2018}
{Wang}, L., \& {Dai}, F. 2018, \apj, 860, 175, \dodoi{10.3847/1538-4357/aac1c0}

\bibitem[{{Wang} \& {Dai}(2019)}]{WangDai}
---. 2019, \apjl, 873, L1, \dodoi{10.3847/2041-8213/ab0653}

\bibitem[{{Wang} \& {Dai}(2021)}]{Wang_2021_wasp107}
---. 2021, \apj, 914, 99, \dodoi{10.3847/1538-4357/abf1ed}

\bibitem[{{Wang}(2017)}]{Wang_uniform}
{Wang}, S. 2017, Research Notes of the American Astronomical Society, 1, 26,
  \dodoi{10.3847/2515-5172/aa9be5}

\bibitem[{{Wang} {et~al.}(2017){Wang}, {Wu}, {Barclay}, \&
  {Laughlin}}]{Wang_trappist}
{Wang}, S., {Wu}, D.-H., {Barclay}, T., \& {Laughlin}, G.~P. 2017, arXiv
  e-prints, arXiv:1704.04290.
\newblock \doarXiv{1704.04290}

\bibitem[{Ward(1997)}]{Ward}
Ward, W.~R. 1997, Icarus, 126, 261,
  \dodoi{https://doi.org/10.1006/icar.1996.5647}

\bibitem[{{Ward}(1997)}]{Ward1997}
{Ward}, W.~R. 1997, \icarus, 126, 261, \dodoi{10.1006/icar.1996.5647}

\bibitem[{{Weidenschilling}(1977)}]{Weidenschilling}
{Weidenschilling}, S.~J. 1977, \apss, 51, 153, \dodoi{10.1007/BF00642464}

\bibitem[{{Weiss} {et~al.}(2018{\natexlab{a}}){Weiss}, {Marcy}, {Petigura},
  {Fulton}, {Howard}, {Winn}, {Isaacson}, {Morton}, {Hirsch}, {Sinukoff},
  {Cumming}, {Hebb}, \& {Cargile}}]{Weiss2018}
{Weiss}, L.~M., {Marcy}, G.~W., {Petigura}, E.~A., {et~al.} 2018{\natexlab{a}},
  \aj, 155, 48, \dodoi{10.3847/1538-3881/aa9ff6}

\bibitem[{{Weiss} {et~al.}(2018{\natexlab{b}}){Weiss}, {Isaacson}, {Marcy},
  {Howard}, {Petigura}, {Fulton}, {Winn}, {Hirsch}, {Sinukoff}, \&
  {Rowe}}]{Weiss_peas}
{Weiss}, L.~M., {Isaacson}, H.~T., {Marcy}, G.~W., {et~al.} 2018{\natexlab{b}},
  ArXiv e-prints.
\newblock \doarXiv{1808.03010}

\bibitem[{{Wisdom}(1986)}]{Wisdom1986}
{Wisdom}, J. 1986, Celestial Mechanics, 38, 175, \dodoi{10.1007/BF01230429}

\bibitem[{{Wisdom}(2006)}]{2006AJ....131.2294W}
---. 2006, \aj, 131, 2294, \dodoi{10.1086/500829}

\bibitem[{{Wisdom} \& {Holman}(1991)}]{1991AJ....102.1528W}
{Wisdom}, J., \& {Holman}, M. 1991, \aj, 102, 1528, \dodoi{10.1086/115978}

\bibitem[{{Wu} \& {Lithwick}(2011)}]{WuLithwick}
{Wu}, Y., \& {Lithwick}, Y. 2011, \apj, 735, 109,
  \dodoi{10.1088/0004-637X/735/2/109}

\bibitem[{{Xiang-Gruess}(2016)}]{Xiang-Gruess}
{Xiang-Gruess}, M. 2016, \mnras, 455, 3086, \dodoi{10.1093/mnras/stv2514}

\bibitem[{{Xu} \& {Lai}(2017)}]{Xu0217}
{Xu}, W., \& {Lai}, D. 2017, \mnras, 468, 3223, \dodoi{10.1093/mnras/stx668}

\bibitem[{{Zeng} {et~al.}(2016){Zeng}, {Sasselov}, \& {Jacobsen}}]{Zeng2016}
{Zeng}, L., {Sasselov}, D.~D., \& {Jacobsen}, S.~B. 2016, \apj, 819, 127,
  \dodoi{10.3847/0004-637X/819/2/127}

\bibitem[{{Zhang} \& {Hamilton}(2008)}]{Zhang_neptune}
{Zhang}, K., \& {Hamilton}, D.~P. 2008, \icarus, 193, 267,
  \dodoi{10.1016/j.icarus.2007.08.024}

\bibitem[{{Zhang} {et~al.}(2022){Zhang}, {Knutson}, {Dai}, {Wang}, {Ricker},
  {Schwarz}, {Mann}, \& {Collins}}]{Zhang2022}
{Zhang}, M., {Knutson}, H.~A., {Dai}, F., {et~al.} 2022, arXiv e-prints,
  arXiv:2207.13099.
\newblock \doarXiv{2207.13099}

\bibitem[{{Ziegler} {et~al.}(2020){Ziegler}, {Tokovinin}, {Brice{\~n}o},
  {Mang}, {Law}, \& {Mann}}]{ziegler2020}
{Ziegler}, C., {Tokovinin}, A., {Brice{\~n}o}, C., {et~al.} 2020, \aj, 159, 19,
  \dodoi{10.3847/1538-3881/ab55e9}

\end{thebibliography}

\newpage
\appendix
\section{TFOP Observations}
TOI-1136 received a number of follow-up observations from the TESS Follow-up Observing Program (TFOP). We refer the readers to the full list of observations on ExoFOP. We briefly summarize them here. 
As part of the standard process for validating transiting exoplanets and assessing the possible systematic errors in the planetary
radius due to light from bound or unbound companions  \citep{ciardi2015}, TOI~1136 was observed with higher-resolution instruments including near-infrared adaptive optics (AO) imaging at Palomar, Gemini-North, and Lick, optical speckle imaging at Gemini-North \citet{Scott2021}, and lucky imaging on the AstraLux instrument \citep{hormuth08} at the Calar Alto Observatory. The optical observations generally provided higher resolution than the NIR observations, while the NIR AO generally provided better sensitivity (especially to low-mass stars). The combination of the observations in multiple filters enables better characterization for any companions that may be detected.

Two reconnaissance spectra were obtained on UT 2019 December 3 and UT 2020 January 28 with the Tillinghast Reflector Echelle Spectrograph \citep[TRES][]{fureszTRES} located on the 1.5m telescope at the Fred Lawrence Whipple Observatory (FLWO) in Arizona. TRES is an echelle spectrograph that operates in the wavelength range 390-910nm and has a resolving power of 44,000. The TRES spectra were extracted using procedures outlined in \citet{buchhave2010}. The TRES spectra were also visually inspected. No signs of a composite spectrum (blended binary) were found. The TRES spectra were also used to derive stellar parameters using the Stellar Parameter Classification tool \citep[SPC][]{buchhave2012}. The resultant stellar parameters agreed well with our HIRES results presented in Section \ref{sec:stellar_para}. SPC gave $T_{\rm eff}=5775\pm50$K, log$g$=4.47$\pm$0.10, [m/H]=-0.02$\pm$0.08, $v$sin$i_\star$=6.7$\pm$0.5 km/s.

The KeplerCam on the 1.2m telescope at the FLWO was used to catch a transit of planet c on UT 2020 January 25 in the Sloan-z band. AstroImageJ \citep{Collins:2017}  was used to perform aperture photometry and model the predicted event. Unfortunately, this observation was performed before the team realized there is TTV in TOI-1136. We did not detect the transit event. 

\section{Search for Additional Planets}\label{sec:additional_transits}
We systematcially search for a seventh transiting planet in the {\it TESS} light curve. A BLS analysis did not detect another significant signal beyond the six known planets. We performed a visual inspection of the {\it TESS} light curve which revealed a possible seventh planet in this system. We saw a single transit-like event centered at BJD-2457000=2435.10 (Fig.~\ref{fig:single_transit}) with a duration of about 7.4 hours.  Assuming the planets are all on circular orbits, such a transit duration would imply an orbital period of 2.1 times that of planet g ($P_h = P_g(T_h/T_g)^3\approx 2.1$). However, after a thorough visual inspection, we could not identify another transit event of similar depth and duration in the existing {\it TESS} light curve. We analyzed this planet simultaneously with other planets in TOI-1136 following the procedure in Section \ref{sec:transits}. The transit depth implied a planetary radius near 2.5$R_\oplus$, although the data could accommodate a planet up to 5$R_\oplus$ if the planet is on a grazing orbit. We tried to add this seventh planet to our TTV model (Section \ref{sec:ttv}). We assumed that TOI-1136.07 followed the trend of the other planets and had an orbital period exactly twice that of planet g. However, adding this planet to our TTV model does not lead to an improvement of the fit. The fit looked almost identical visually and there is no improvement in the Bayesian Information Criterion \citep{Schwarz}. We did not include this planet candidate in our final analysis.

\section{HARPS-N Rossiter-McLaughlin Measurement}
We observed a spectroscopic transit of TOI-1136~d using the  High Accuracy Radial velocity Planetary Searcher North \citep[HARPS-N][]{Mayor2003,Cosentino2012} mounted on the 3.58~m Telescopio Nazionale Galileo (TNG) located on Roque de los Muchachos, La Palma, Spain. We observed a transit of TOI-1136 d on the night starting on UTC 2021 May 14 with observations between 21:30 UT annd 04:00 UT. The exposure time was set to 900~s and with an overhead of roughly 20~s, the sampling was approximately 920~s. Due to varying weather conditions the signal-to-noise ratio (in order 49) ranged from around 80 (in the beginning of the night) to around 40-50. 

We used our HARPS-N transit data to get an independent measure for the projected obliquity of TOI-1136 d. We sampled the posteriors using Markov Chain Monte Carlo (MCMC) sampling using the \texttt{emcee} \citep{emcee} package with the code by \citet{Hirano2011} to model the RM effect. We imposed Gaussian priors to $R_\mathrm{p}$, $a/R_\star$, and $i$ according to the values in Table~\ref{tab:planet_para}. Gaussian priors were imposed to the macro- and microturbulence with values of $3.13 \pm 1$ km~s$^{-1}$ \citep{Doyle2014} and $1.04 \pm 1$ km~s$^{-1}$ \citep{Bruntt2010} respectively. We let the mid-transit time and the systemic velocity, the $v \sin i_\star$, and the sky-projected obliquity $\lambda$ to vary freely. The posterior distribution indicates $\lambda$ of $6^{+28}_{-27}~^{\circ}$  consistent with the HIRES measurement. The mid-transit time of this event was at BJD of $2459349.525 \pm 0.005$.

\begin{figure*}
\begin{center}
\includegraphics[width = .4\columnwidth]{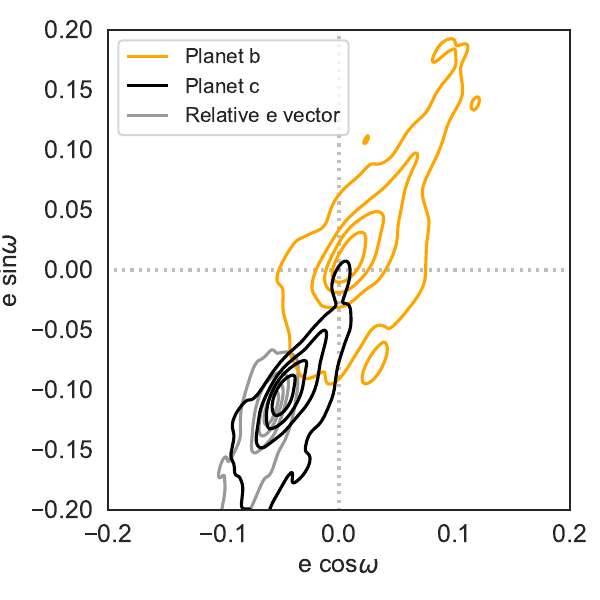}
\includegraphics[width = .4\columnwidth]{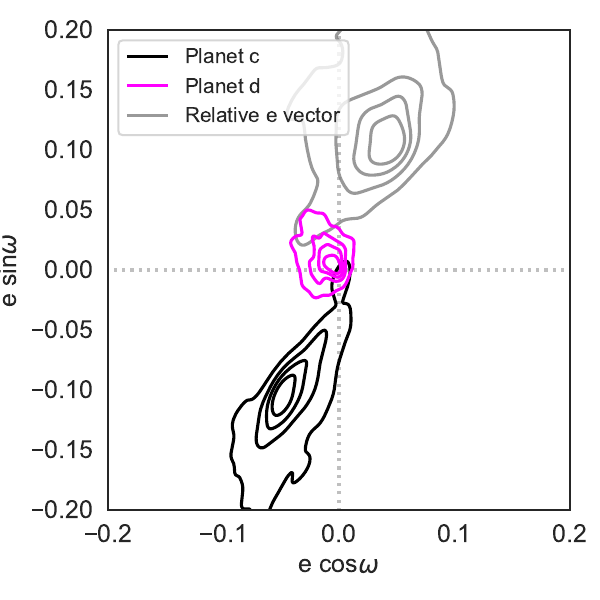}
\includegraphics[width = .4\columnwidth]{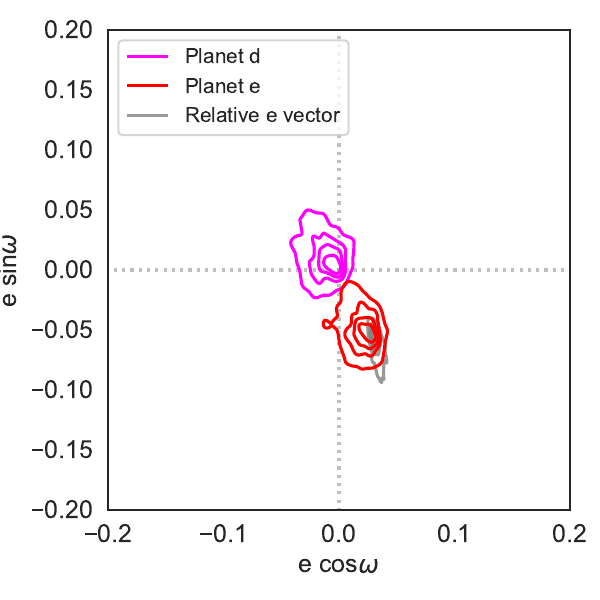}
\includegraphics[width = .4\columnwidth]{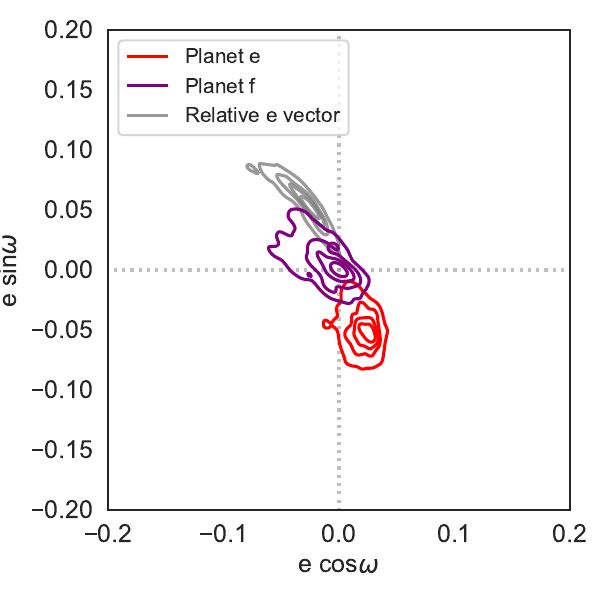}
\includegraphics[width = .4\columnwidth]{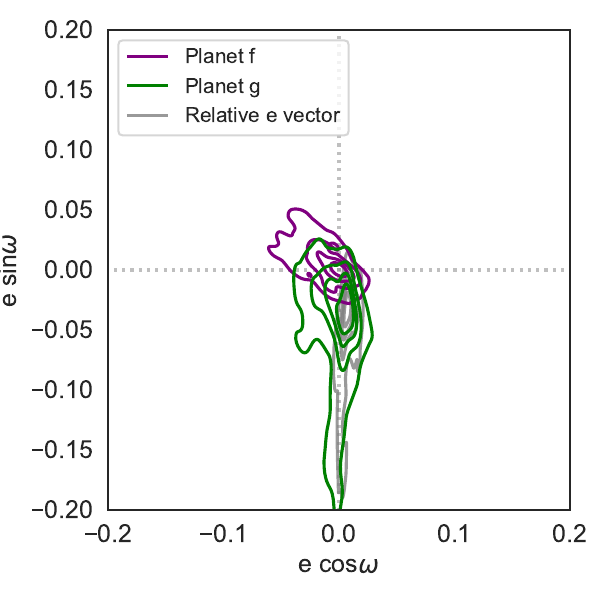}
\caption{The eccentricity vectors of adjacent planet pairs in TOI-1136. The colored contours represent successively higher posterior probability for each planet. The gray contours show the relative eccentricity vector between neighboring planets ($e_{\rm i+1}$cos$\omega_{\rm i+1}-e_{\rm i}$cos$\omega_{\rm i}$). The relative eccentricity vectors better capture any covariance. A classical prediction of convergent disk migration scenario is that adjacent planets should be apsidally anti-aligned (see Section \ref{sec:migration} for detail). For apsidally anti-aligned solutions, the gray contours tend to pushed away from the origin. Existing constraints on TOI-1136 may hint at apsidally anti-aligned configurations, however more data is required to confirm this trend.} 
\label{fig: e_vector}
\end{center}
\end{figure*}

\begin{figure*}
\center
\includegraphics[width = 0.5\columnwidth]{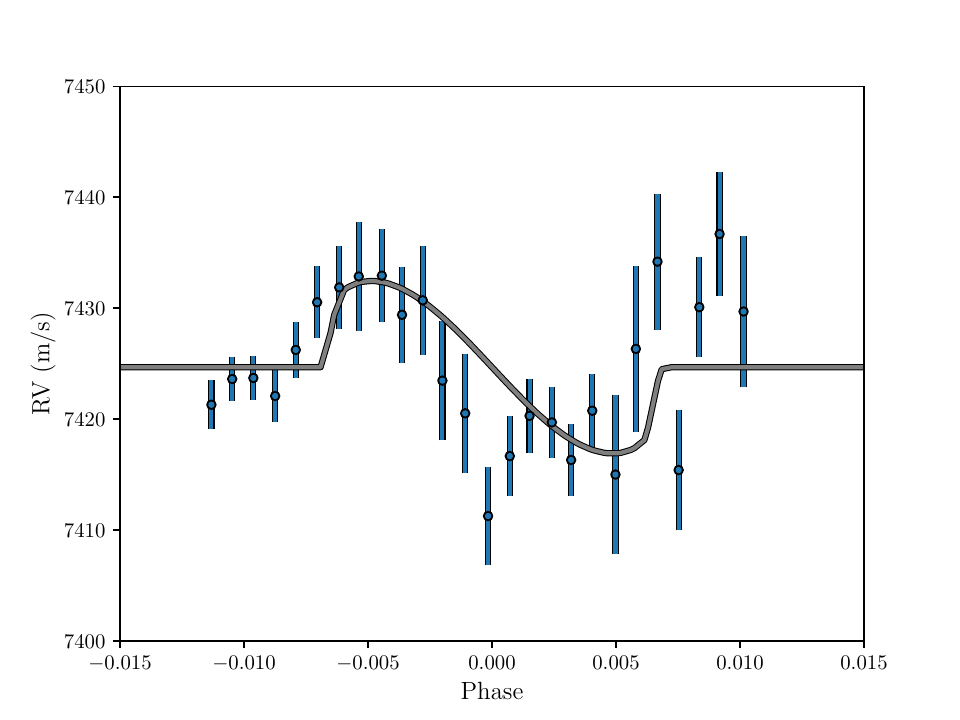}
\caption{A Rossiter-McLaughlin measurement of TOI-1136 d during a transit near BJD=2459349.525 with HARPS-N. We measured a stellar obliquity $\lambda$ of $6^{+28}_{-27}~^{\circ}$ which is consistent with the higher SNR HIRES measurement.}
\label{fig:harpsn_rm}
\end{figure*}

\begin{deluxetable}{ccc}
\tablecaption{Keck/HIRES Radial Velocities during a TOI-1136 d Transit near BJD=2459650.0310 \label{tab:rv}}
\tablehead{
\colhead{Time (BJD)} & \colhead{RV (m/s)} & \colhead{RV Unc. (m/s)}}
\startdata
2459649.865577 & 0.02 & 1.26\\
2459649.870924 & -2.86 & 1.32\\
2459649.876399 & 0.34 & 1.44\\
2459649.882382 & -1.91 & 1.51\\
2459649.888204 & -2.63 & 1.49\\
2459649.894153 & 0.31 & 1.28\\
2459649.90016 & -4.02 & 1.35\\
2459649.905831 & 2.21 & 1.16\\
2459649.911004 & -1.68 & 1.24\\
2459649.91597 & 1.32 & 1.19\\
2459649.92105 & -1.3 & 1.23\\
2459649.926027 & -0.95 & 1.2\\
2459649.930888 & -1.85 & 1.34\\
2459649.935772 & 0.87 & 1.24\\
2459649.940657 & 2.97 & 1.26\\
2459649.945529 & 0.4 & 1.37\\
2459649.950413 & 0.14 & 1.31\\
2459649.955355 & 8.23 & 1.27\\
2459649.960807 & 9.05 & 1.26\\
2459649.966779 & 8.41 & 1.18\\
2459649.971964 & 9.3 & 1.3\\
2459649.976848 & 8.87 & 1.21\\
2459649.98164 & 11.07 & 1.27\\
\enddata
\end{deluxetable}

\begin{deluxetable}{ccc}
\tablecaption{Keck/HIRES Radial Velocities during a TOI-1136 d Transit near BJD=2459650.0310 Continued}
\tablehead{
\colhead{Time (BJD)} & \colhead{RV (m/s)} & \colhead{RV Unc. (m/s)}}
\startdata
2459649.986466 & 8.68 & 1.19\\
2459649.991408 & 10.88 & 1.18\\
2459649.996327 & 9.78 & 1.24\\
2459650.001685 & 10.73 & 1.29\\
2459650.007449 & 7.21 & 1.22\\
2459650.013653 & -0.09 & 1.21\\
2459650.019509 & -2.09 & 1.36\\
2459650.025967 & 2.91 & 1.3\\
2459650.032345 & 0.92 & 1.31\\
2459650.038467 & 3.15 & 1.19\\
2459650.044567 & -5.95 & 1.35\\
2459650.050666 & -5.54 & 1.2\\
2459650.056233 & -8.68 & 1.25\\
2459650.061568 & -10.1 & 1.24\\
2459650.067182 & -10.54 & 1.2\\
2459650.073432 & -6.54 & 1.29\\
2459650.079751 & -5.96 & 1.26\\
2459650.085874 & -12.66 & 1.25\\
2459650.092135 & -10.78 & 1.28\\
2459650.098547 & -10.01 & 1.33\\
2459650.104531 & -8.63 & 1.2\\
2459650.110318 & -3.7 & 1.25\\
2459650.116023 & -0.24 & 1.26\\
2459650.121822 & -2.95 & 1.21\\
2459650.127343 & -0.08 & 1.32\\
2459650.132852 & -0.92 & 1.31\\
2459650.138465 & -2.98 & 1.28\\
2459650.144055 & -3.64 & 1.2\\
2459650.150016 & 1.72 & 1.29\\
2459650.156208 & -0.39 & 1.22\\
\enddata
\end{deluxetable}

\begin{deluxetable}{ccc}
\tablecaption{HARPS-N Radial Velocities during a TOI-1136 d Transit near BJD=2459349.525 \label{tab:rv_harpsn}}
\tablehead{
\colhead{Time (BJD)} & \colhead{RV (m/s)} & \colhead{RV Unc. (m/s)}}
\startdata
2459349.400949960109 &  7421.3 &  2.2\\
2459349.411447130144 &  7423.6 &  2.0\\
2459349.422129469924 &  7423.7 &  2.0\\
2459349.433135880157 &  7422.1 &  2.3\\
2459349.443552029785 &  7426.2 &  2.5\\
2459349.454326970037 &  7430.5 &  3.2\\
2459349.465472249780 &  7431.9 &  3.7\\
2459349.475332879927 &  7432.9 &  4.9\\
2459349.486998970155 &  7432.9 &  4.2\\
2459349.497172090225 &  7429.4 &  4.3\\
2459349.507634540088 &  7430.7 &  4.9\\
2459349.517553030048 &  7423.5 &  5.4\\
2459349.529068669770 &  7420.5 &  5.4\\
2459349.540619030129 &  7411.2 &  4.4\\
2459349.551579140127 &  7416.7 &  3.6\\
2459349.561578650028 &  7420.3 &  3.3\\
2459349.572816519998 &  7419.7 &  3.2\\
2459349.582538269926 &  7416.3 &  3.2\\
2459349.593209039886 &  7420.8 &  3.3\\
2459349.604898279998 &  7415.0 &  7.2\\
2459349.615210279822 &  7426.33 &  7.5\\
2459349.626124090049 &  7434.2 &  6.1\\
2459349.636806440074 &  7415.4 &  5.4\\
2459349.647245740052 &  7430.1 &  4.5\\
2459349.657476719934 &  7436.7 &  5.6\\
2459349.669582610019 &  7429.7 &  6.8\\
\enddata
\end{deluxetable}

\begin{deluxetable}{cccc}
\tablecaption{Measured Mid-Transit Times of TOI-1136 Planets \label{tab:ttv}}
\tablehead{
\colhead{Planet} & \colhead{Epoch} & \colhead{Mid-Transit Times (BJD-2457000)} & \colhead{Unc. (days)}}
\startdata
b  & 0 & 1684.2689 & 0.0128 \\
b  & 1 & 1688.4659 & 0.0153 \\
b  & 2 & 1692.6029 & 0.0085 \\
b  & 4 & 1700.9705 & 0.0193 \\
b  & 5 & 1705.1523 & 0.0162 \\
b  & 6 & 1709.3189 & 0.0118 \\
b  & 7 & 1713.4489 & 0.0213 \\
b  & 8 & 1717.6700 & 0.0124 \\
b  & 9 & 1721.8301 & 0.0203 \\
b  & 10 & 1726.0011 & 0.0109 \\
b  & 11 & 1730.1877 & 0.0170 \\
b  & 12 & 1734.3482 & 0.0108 \\
b  & 45 & 1872.0104 & 0.0072 \\
b  & 46 & 1876.2113 & 0.0148 \\
b  & 47 & 1880.3842 & 0.0182 \\
b  & 49 & 1888.7064 & 0.0152 \\
b  & 50 & 1892.8671 & 0.0162 \\
b  & 51 & 1897.0779 & 0.0112 \\
b  & 52 & 1901.2685 & 0.0096 \\
b  & 53 & 1905.4083 & 0.0182 \\
b  & 56 & 1917.9092 & 0.0198 \\
b  & 57 & 1922.1214 & 0.0164 \\
b  & 58 & 1926.2296 & 0.0173 \\
b  & 177 & 2422.6729 & 0.0102 \\
b  & 178 & 2426.8482 & 0.0080 \\
b  & 179 & 2431.0311 & 0.0086 \\
b  & 180 & 2435.1835 & 0.0089 \\
b  & 181 & 2439.3783 & 0.0135 \\
b  & 182 & 2443.5572 & 0.0112 \\
b  & 222 & 2610.4300 & 0.0108 \\
b  & 224 & 2618.8061 & 0.0100 \\
b  & 226 & 2627.1275 & 0.0166 \\
b  & 227 & 2631.3006 & 0.0106 \\
b  & 228 & 2635.4951 & 0.0137 \\
\enddata
\tablecomments{1. From HARPS RM Measurement (not included in our TTV modeling). 2. From HIRES RM measurement.}
\end{deluxetable}

\begin{deluxetable}{cccc}
\tablecaption{Measured Mid-Transit Times of TOI-1136 Planets Continued}
\tablehead{
\colhead{Planet} & \colhead{Epoch} & \colhead{Mid-Transit Times (BJD-2457000)} & \colhead{Unc. (days)}}
\startdata
c  & 0 & 1688.7211 & 0.0036 \\
c  & 1 & 1694.9699 & 0.0023 \\
c  & 2 & 1701.2284 & 0.0028 \\
c  & 3 & 1707.4861 & 0.0022 \\
c  & 4 & 1713.7520 & 0.0032 \\
c  & 5 & 1720.0034 & 0.0018 \\
c  & 6 & 1726.2605 & 0.0080 \\
c  & 7 & 1732.5187 & 0.0022 \\
c  & 30 & 1876.4569 & 0.0019 \\
c  & 33 & 1895.2336 & 0.0022 \\
c  & 34 & 1901.4896 & 0.0029 \\
c  & 35 & 1907.7538 & 0.0024 \\
c  & 37 & 1920.2739 & 0.0087 \\
c  & 117 & 2420.9969 & 0.0018 \\
c  & 118 & 2427.2600 & 0.0026 \\
c  & 120 & 2439.7757 & 0.0017 \\
c  & 121 & 2446.0284 & 0.0018 \\
c  & 148 & 2614.9957 & 0.0024 \\
c  & 149 & 2621.2509 & 0.0033 \\
c  & 150 & 2627.5170 & 0.0022 \\
c  & 151 & 2633.7730 & 0.0039 \\
d  & 0 & 1686.0671 & 0.0012 \\
d  & 1 & 1698.5858 & 0.0011 \\
d  & 3 & 1723.6219 & 0.0013 \\
d  & 4 & 1736.1413 & 0.0013 \\
d  & 15 & 1873.8428 & 0.0012 \\
d  & 16 & 1886.3601 & 0.0011 \\
d  & 18 & 1911.3936 & 0.0012 \\
d  & 19 & 1923.9092 & 0.0012 \\
d  & 53 & 2349.525 & 0.005$^1$ \\
d  & 59 & 2424.6430 & 0.0010 \\
d  & 60 & 2437.1649 & 0.0012 \\
d  & 74 & 2612.4673 & 0.0012 \\
d  & 77 & 2650.0310 & 0.0018$^2$\\
e  & 0 & 1697.7758 & 0.0022 \\
e  & 1 & 1716.5624 & 0.0099 \\
e  & 2 & 1735.3536 & 0.0097 \\
e  & 10 & 1885.7918 & 0.0044 \\
e  & 11 & 1904.5934 & 0.0070 \\
e  & 12 & 1923.4102 & 0.0112 \\
e  & 39 & 2430.9549 & 0.0034 \\
e  & 49 & 2618.8132 & 0.0044 \\
f  & 0 & 1699.3854 & 0.0018 \\
f  & 1 & 1725.7099 & 0.0019 \\
f  & 7 & 1883.6007 & 0.0020 \\
f  & 8 & 1909.9075 & 0.0025 \\
f  & 28 & 2436.2605 & 0.0015 \\
f  & 35 & 2620.5189 & 0.0014 \\
g  & 0 & 1711.9393 & 0.0071 \\
g  & 5 & 1909.6401 & 0.0054 \\
g  & 18 & 2423.6690 & 0.0040 \\
g  & 23 & 2621.3499 & 0.0027 \\
\enddata
\end{deluxetable}

\movetabledown=2.8in
\begin{rotatetable}
\begin{deluxetable}{llllllll}
\tablecaption{Planetary Parameters of TOI-1136}
\tablehead{
\colhead{Parameter}  & \colhead{Symbol} &  \colhead{planet b} &  \colhead{planet c}&  \colhead{planet d}&  \colhead{planet e}&  \colhead{planet f}&  \colhead{planet g}}
\startdata
From Transit Modeling\\
Planet/Star Radius Ratio & $R_p/R_\star$  
 &$0.0180_{-1.7e-3}^{+1.9e-3}$  
 & $0.02725_{-5.7e-4}^{+5.3e-4}$
 & $0.04379_{-6.6e-4}^{+7.1e-4}$
 & $0.02497_{-8.0e-4}^{+6.6e-4}$
 & $0.03671_{-9.8e-4}^{+9.9e-4}$
 & $0.0239_{-1.1e-3}^{+1.0e-3}$\\
Impact Parameter & $b$  
 & $0.705_{-5.6e-2}^{+4.6e-2}$
 & $0.15_{-1.0e-1}^{+1.4e-1}$
 & $0.24_{-1.1e-1}^{+1.1e-1}$
 & $0.37_{-1.4e-1}^{+9.3e-2}$
 & $0.42_{-1.5e-1}^{+1.1e-1}$
 & $0.31_{-1.6e-1}^{+1.1e-1}$\\
Scaled Semi-major Axis & $a/R_\star$ 
 & $11.29_{-2.4e-1}^{+2.8e-1}$
 & $14.80_{-3.2e-1}^{+3.7e-1}$
 & $23.50_{-5.1e-1}^{+5.9e-1}$
 & $30.81_{-6.7e-1}^{+7.7e-1}$
 & $38.56_{-8.3e-1}^{+9.6e-1}$
 & $50.6_{-1.1e+00}^{+1.3e+00}$\\
Transit Duration (hours)& $T_{\rm 14}$
 & $2.07_{-7.8e-2}^{+6.8e-2}$
 & $3.27_{-1.5e-1}^{+1.1e-2}$
 & $4.12_{-1.0e-1}^{+1.1e-2}$
 & $4.45_{-1.9e-1}^{+1.5e-1}$
 & $4.96_{-3.0e-1}^{+2.4e-1}$
 & $5.80_{-1.7e-1}^{+2.0e-1}$\\
Orbital Inclination (deg)   & $i$   &$86.44_{-2.1e-1}^{+2.7e-1}$
 & $89.42_{-5.5e-1}^{+3.9e-1}$
 & $89.41_{-2.8e-1}^{+2.8e-1}$
 & $89.31_{-1.8e-1}^{+2.6e-1}$
 & $89.38_{-1.7e-1}^{+2.2e-1}$
 & $89.65_{-1.3e-1}^{+1.8e-1}$\\
From Stable TTV Solutions$^1$\\
Mass Ratio & $m_p/m_\star$    
 & $0.00000885_{-2.6e-06}^{+2.1e-06}$
 & $0.00001780_{-5.1e-06}^{+3.8e-06}$
 & $0.00002349_{-5.5e-06}^{+7.0e-06}$
 & $0.00001594_{-2.9e-06}^{+3.0e-06}$
 & $0.00002452_{-8.3e-06}^{+1.1e-05}$
 & $0.00001404_{-9.7e-06}^{+1.4e-05}$\\
Orbital Period (days) & $P_{\rm orb}$   
&  $4.17278_{-1.8e-04}^{+2.4e-04}$
 & $6.25725_{-2.5e-04}^{+1.7e-04}$
 & $12.51937_{-4.1e-04}^{+3.7e-04}$
 & $18.7992_{-1.5e-03}^{+1.7e-03}$
 & $26.3162_{-1.3e-03}^{+1.7e-03}$
 & $39.5387_{-3.0e-03}^{+3.6e-03}$\\
Mean Anomaly (deg) & $M$    
 & $24_{-4.4e+01}^{+5.8e+01}$
 & $68_{-9.6e+00}^{+5.0e+00}$
 & $120_{-4.1e+01}^{+3.6e+01}$
 & $170_{-8.4e+00}^{+7.5e+00}$
 & $23_{-7.4e+01}^{+5.9e+01}$
 & $-113_{-1.7e+01}^{+4.1e+01}$\\
 Orbital Eccentricity & $e$  
 & $0.031_{-2.1e-02}^{+3.8e-02}$
 & $0.117_{-2.8e-02}^{+2.8e-02}$
 & $0.016_{-1.0e-02}^{+1.3e-02}$
 & $0.057_{-1.3e-02}^{+1.0e-02}$
 & $0.012_{-9.7e-03}^{+1.9e-02}$
 & $0.036_{-1.8e-02}^{+2.6e-02}$\\
 Argument of Pericenter (deg) & $\omega$  
 & $45_{-9.9e+01}^{+2.6e+01}$
 & $-113_{-4.0e+00}^{+7.3e+00}$
 & $118_{-4.4e+01}^{+3.6e+01}$
 & $-66_{-1.1e+01}^{+8.0e+00}$
 & $140_{-8.8e+01}^{+8.9e+01}$
 & $-87_{-3.1e+01}^{+1.7e+01}$\\
 Eccentricity Vector$^2$ & $e$cos$\omega$
 & $0.016_{-0.017}^{+0.029}$
 & $-0.046_{-0.013}^{+0.020}$
 & $-0.0091_{-0.0129}^{+0.0081}$
 & $0.022_{-0.011}^{+0.006}$
 & $-0.0033_{-0.019}^{+0.0075}$
 & $0.0021_{-0.013}^{+0.0076}$\\
 Eccentricity Vector& $e$sin$\omega$  
 & $0.012_{-0.020}^{+0.037}$
 & $-0.11_{-0.025}^{+0.023}$
 & $-0.0076_{-0.0097}^{+0.0146}$
 & $-0.051_{-0.011}^{+0.014}$
 & $-0.0012_{-0.0076}^{+0.0147}$
 & $-0.034_{-0.027}^{+0.021}$\\
Longitude of Ascending Node (deg) & $\Omega$  &  0 (fixed) &  0 (fixed) &  0 (fixed) &  0 (fixed) &  0 (fixed) &  0 (fixed)\\
Orbital Inclination (deg) & $i$  &  90 (fixed) &  90 (fixed) &  90 (fixed) &  90 (fixed) &  90 (fixed) &  90 (fixed)\\
\hline
From RM Modeling\\
Sky-Projected Stellar Obliquity (deg) & $\lambda$  &&&  5$\pm$5$^\circ$\\
Stellar Obliquity (deg) & $\Psi$  &&&  <28$^\circ$ (95\%)\\
Rotational Broadening (km/s) & $v$sin$i_\star$  &&&  6.7$\pm$0.6\\
\hline
Derived Parameters\\
Planetary Radius ($R_\oplus$)  & $R_{\rm p}$   &$1.90_{-0.15}^{+0.21}$
 & $2.879_{-0.062}^{+0.060}$
 & $4.627_{-0.072}^{+0.077}$
 & $2.639_{-0.088}^{+0.072}$
 & $3.88_{-0.11}^{+0.11}$
 & $2.53_{-0.12}^{+0.11}$ \\
Planetary Mass ($M_\oplus$)  & $M_{\rm p}$ 
 & $3.01_{-0.89}^{+0.71}$
 & $6.0_{-1.7}^{+1.3}$
 & $8.0_{-1.9}^{+2.4}$
 & $5.4_{-1.0}^{+1.0}$
 & $8.3_{-3.6}^{+2.8}$
 & $4.8_{-3.3}^{+4.7}$\\
\enddata
\tablecomments{1. Reported as osculating Keplerian elements at BJD=2458680 which is close to the first TESS observation. We note that the osculating orbital periods that depend sensitively on if there had just been a close encounter of planets. To compute the orbital period ratio between neighboring planets and the deviation from MMR $\Delta$, we used the average orbital period from N-body integration in Section \ref{sec:commensurability}. 2. We also report the eccentricity vectors directly because the arguments of pericenter often wrap around 2$\pi$.}
\label{tab:planet_para}
\end{deluxetable}
\end{rotatetable}

\begin{figure*}
\begin{center}
\includegraphics[width = 0.7\columnwidth]{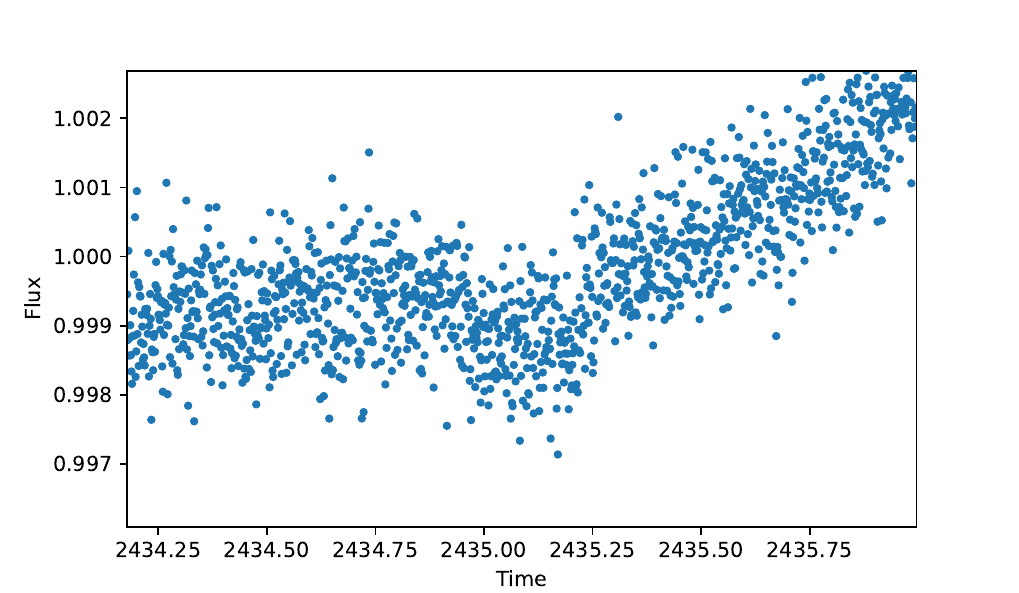}
\includegraphics[width = 0.7\columnwidth]{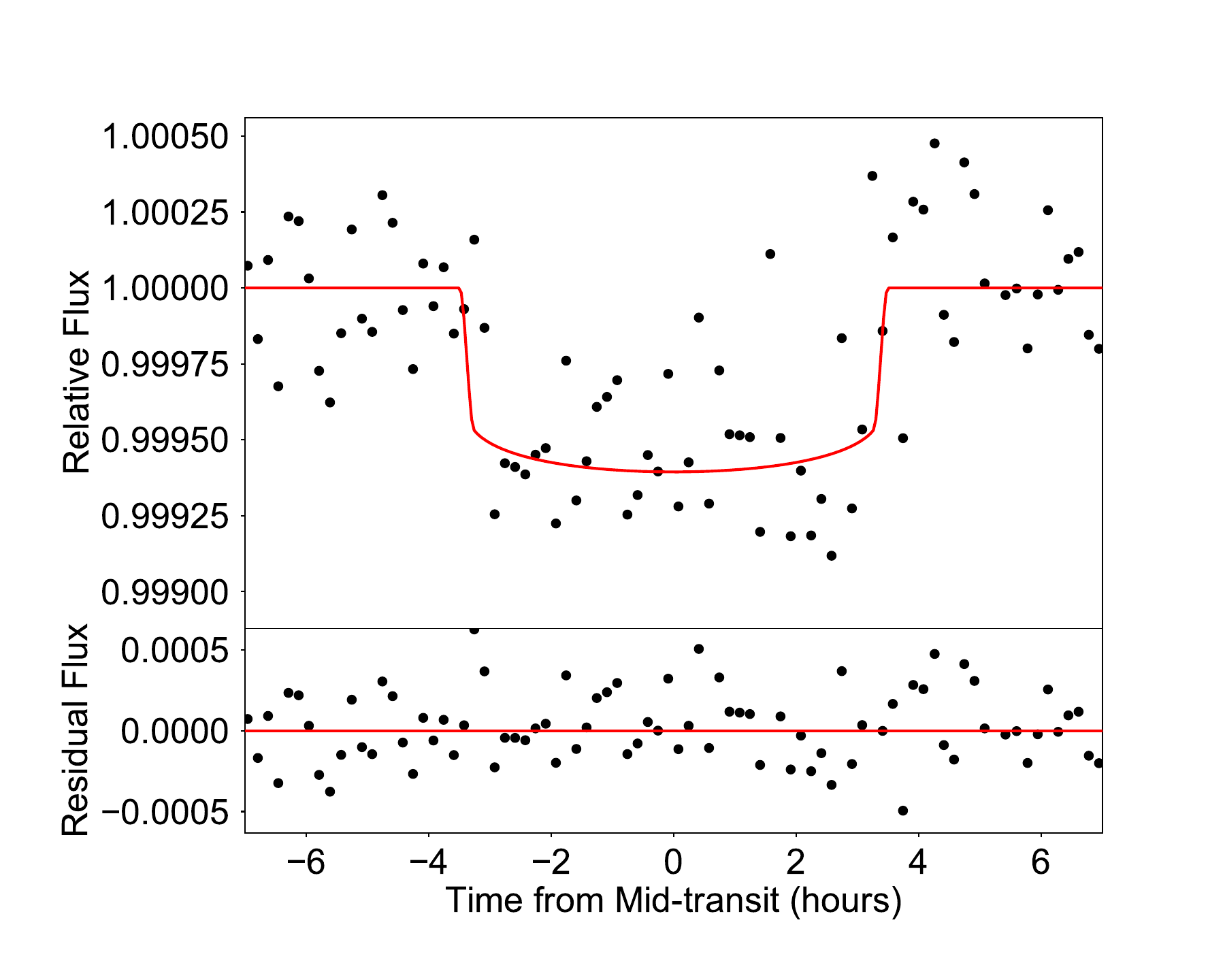}
\caption{Top: a possible single transit of a seventh planet in TOI-1136 was identified by visual inspection near BJD$-2457000=2435.10$ (left). We are unable to confirm this planet: no similarly shaped transit event was seen in the rest of the {\it TESS} light curve.  Bottome: the best-fit transit model of the detrended and binned light curve. The nominal transit depth suggests a planetary radius of about $2.5R_\oplus$. However, many posterior samples are also consistent with a larger planet ($5R_\oplus$) on a grazing orbit. The transit duration is about 7.4 hours. If the planets were on a circular, edge-on orbit. The implied orbital period is roughly twice the period of planet g ($P_h = P_g(T_h/T_g)^3\approx 2.1$) following the resonant pattern.}
\label{fig:single_transit}
\end{center}
\end{figure*}

\begin{figure*}
\begin{center}
\includegraphics[width = 1.\columnwidth]{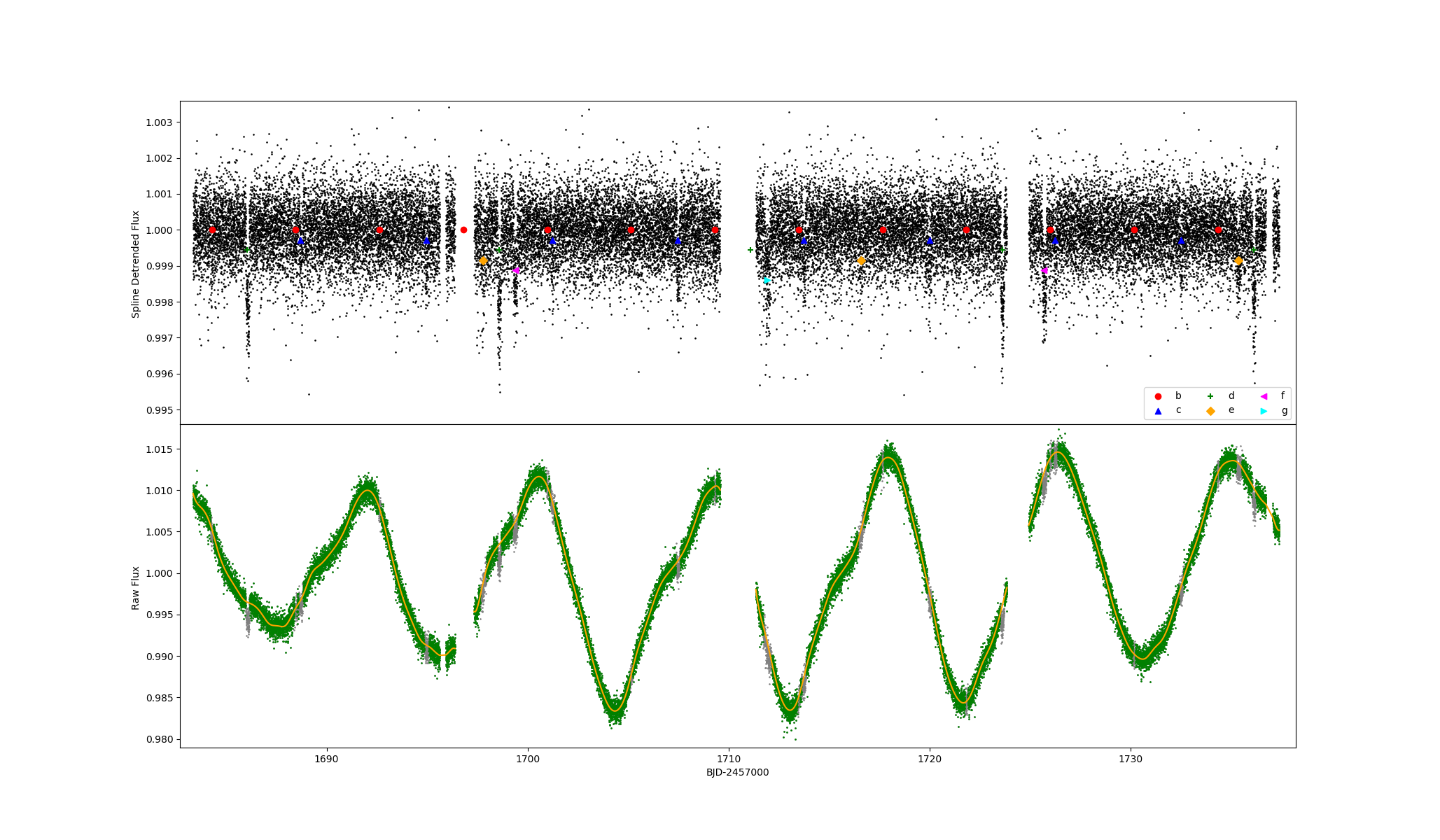}
\includegraphics[width = 1.\columnwidth]{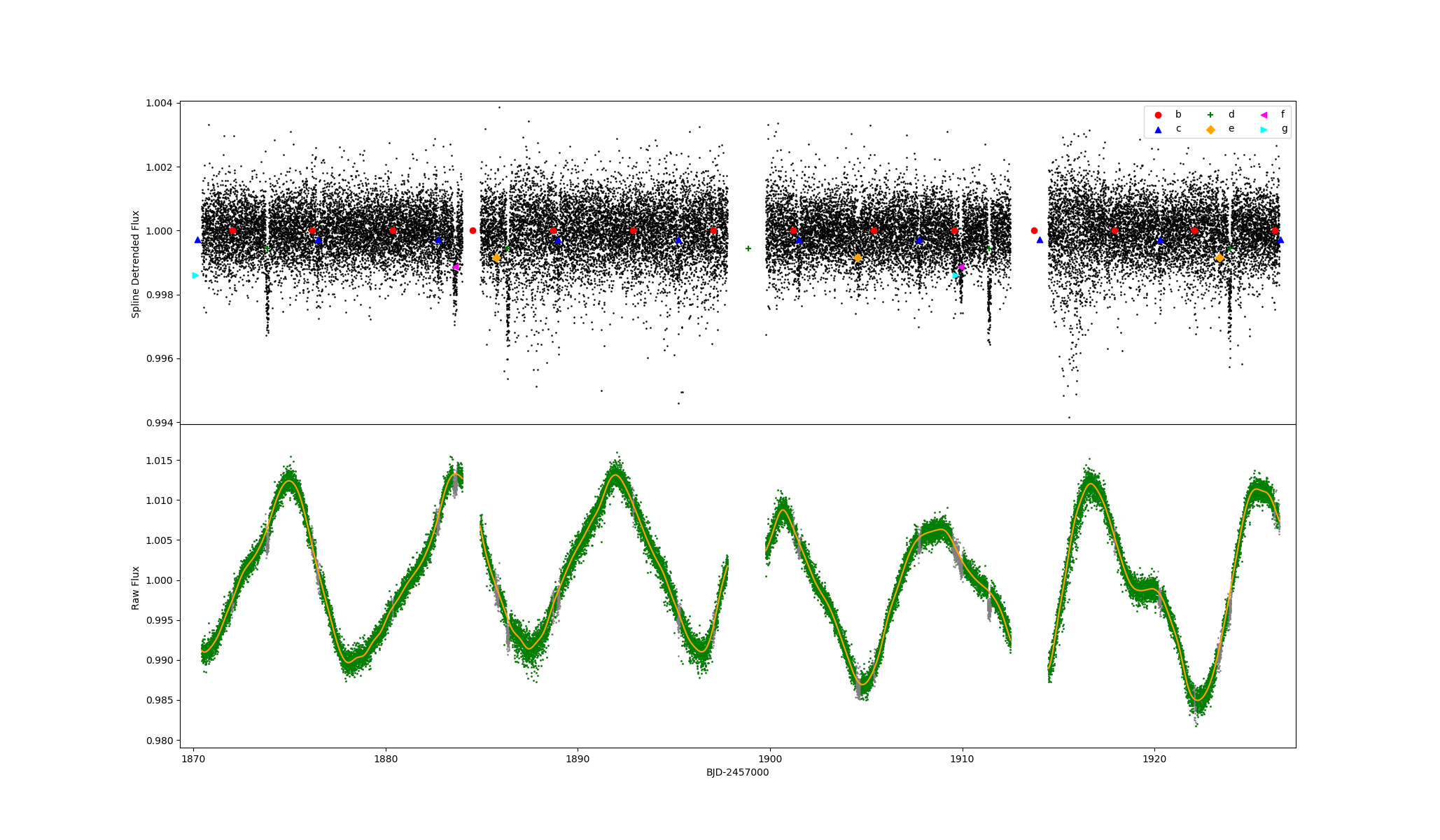}
\caption{The {\it TESS} light curves of TOI-1136 across different sectors (Sector 14, 15, 21 and 22). The quasi-periodic flux modulation due to stellar rotation is clearly visible. We removed these variations by fitting a cubic spline (orange curves) to the out-of-transit fluxes (green). The top panels shows the detrended light curve and mid-transit times of planets if there were no TTV.}
\label{fig:tess_raw_light_curve1}
\end{center}
\end{figure*}

\begin{figure*}
\begin{center}
\includegraphics[width = 1.\columnwidth]{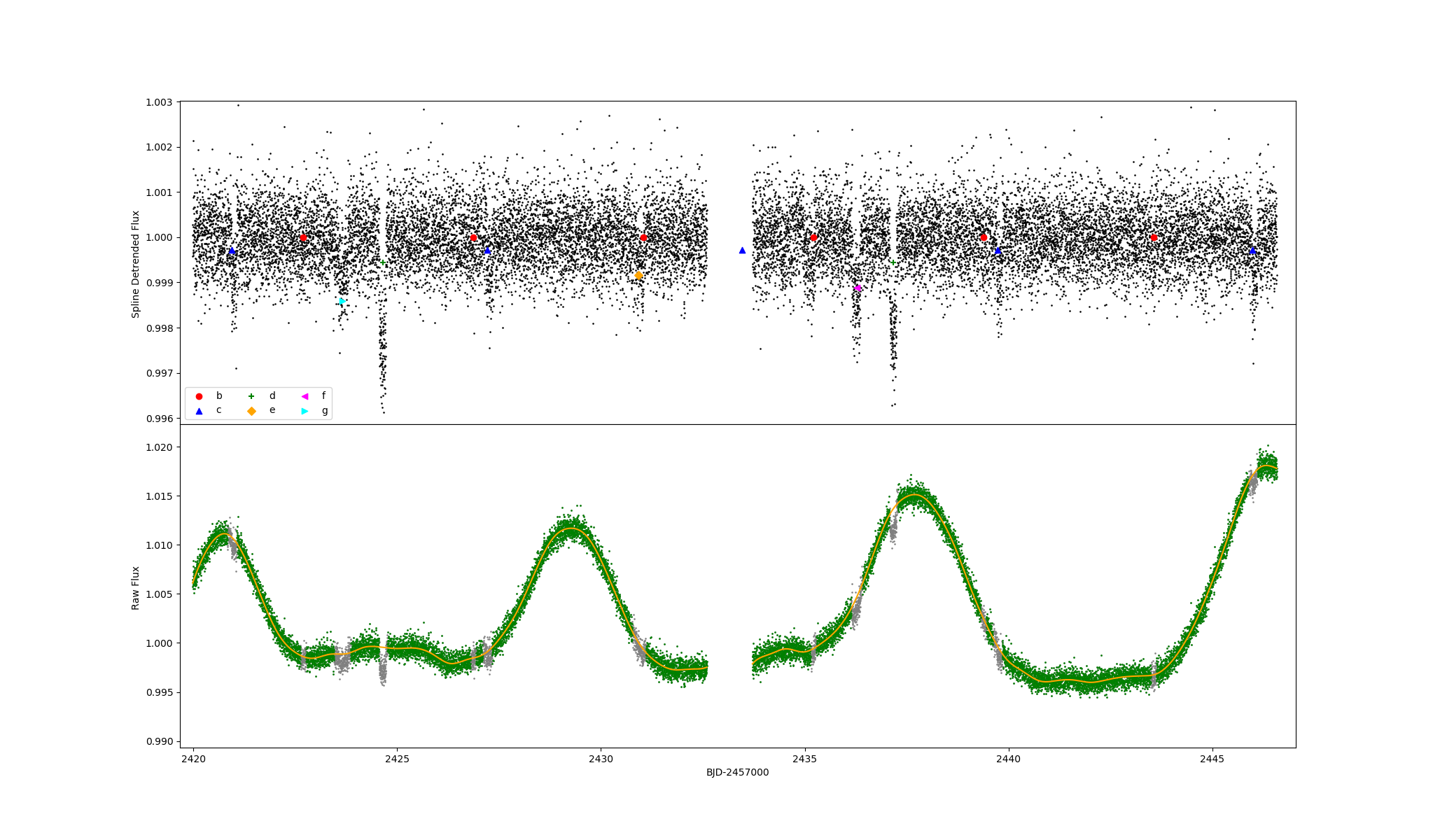}
\includegraphics[width = 1.\columnwidth]{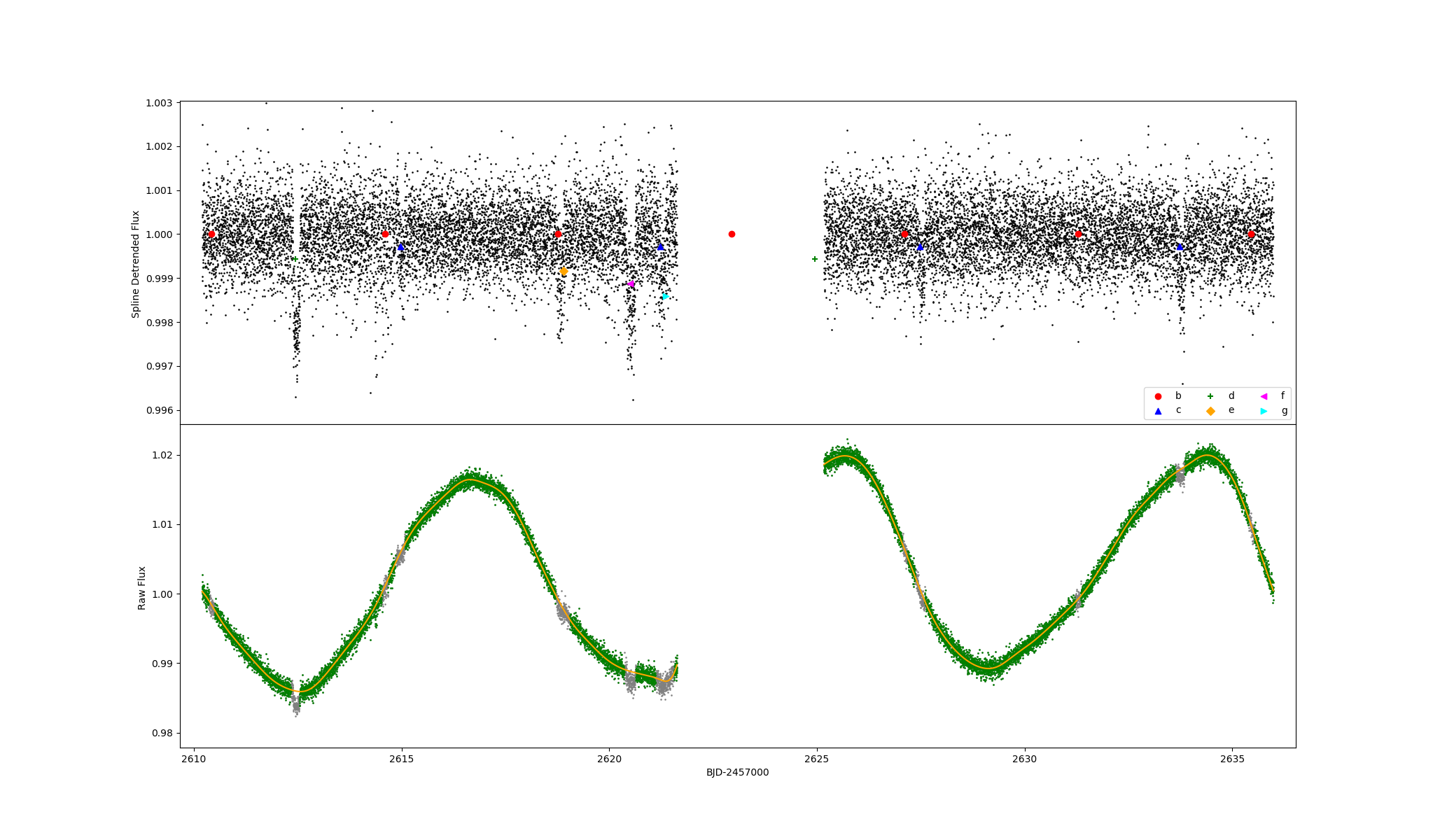}
\caption{Same as Fig.~\ref{fig:tess_raw_light_curve1} for Sectors 41 and 48.}
\label{fig:tess_raw_light_curve2}
\end{center}
\end{figure*}

\end{document}